\documentclass[sn-mathphys,Numbered]{sn-jnl}

\usepackage{graphicx}%
\usepackage{multirow}%
\usepackage{amsmath,amssymb,amsfonts}%
\usepackage{amsthm}%
\usepackage{mathrsfs,comment}%
\usepackage[title]{appendix}%
\usepackage{xcolor}%
\usepackage{textcomp}%
\usepackage{manyfoot}%
\usepackage{booktabs}%

\usepackage{listings}%
\usepackage{rotating}

\usepackage{bm}
\usepackage[ruled,linesnumbered]{algorithm2e}
\usepackage{cleveref}
\usepackage{textcomp}
\usepackage{xcolor}
\usepackage{caption}
\usepackage{subcaption}
\usepackage{lineno}
\begin{document}

\newcommand{\DONE}[1]{\textcolor{black}{\textbf{}#1}}
\newcommand{\REV}[1]{\textcolor{black}{\textbf{}#1}}

\newcommand{\ROUNDtwo}[1]{\textcolor{black}{\textbf{}#1}}
\newcommand{\SebaAle}[1]{\textcolor{black}{#1}}
\newcommand{\MeliVale}[1]{\textcolor{black}{#1}}

\title{\MeliVale{Unsupervised detection of Large-scale Weather Patterns in the Northern Hemisphere via Markov State Modelling: from Blockings to Teleconnections}}

\author[1]{\fnm{Sebastian} \sur{Springer}}\email{sebastian.springer@sissa.it}
\equalcont{These authors contributed equally to this work.}
\author*[1,5]{\fnm{Alessandro} \sur{Laio}}\email{laio@sissa.it}
\equalcont{These authors contributed equally to this work.}

\author[2]{\fnm{Vera } \sur{Melinda Galfi}}\email{v.m.galfi@vu.nl}
\equalcont{These authors contributed equally to this work.}

\author*[3,4]{\fnm{Valerio } \sur{Lucarini}}\email{v.lucarini@leicester.ac.uk}
\equalcont{These authors contributed equally to this work.}

\affil*[1]{
\orgdiv{Physics Department}, 
\orgname{International School for Advanced Studies (SISSA)}, \orgaddress{\street{Via Bonomea 265}, \city{Trieste}, \postcode{34136}, \country{Italy}}}

\affil[2]{
\orgdiv{ Institute for Environmental Studies}, 
\orgname{Vrije Universiteit Amsterdam}, \orgaddress{\street{De Boelelaan 1105}, \city{Amsterdam}, \postcode{081 HV}, \country{the Netherlands}}}

\affil[3]{
\orgdiv{School of Computing and Mathematical Sciences}, 
\orgname{University of Leicester}, \orgaddress{\street{University Road}, \city{Leicester}, \postcode{LE17RH}, \country{UK}}}

\affil[4]{
\orgdiv{Department of Mathematics and Statistics \& Centre for the Mathematics of Planet Earth}, 
\orgname{University of Reading}, \orgaddress{\street{Whiteknights Campus}, \city{Reading}, \postcode{RG66AX}, \country{UK}}}

\affil[5]{\orgname{International Center of Theoretical Physics (ICTP)}, \orgaddress{\street{Viale Miramare 11}, \city{Trieste}, \postcode{34151}, \country{Italy}}}

\abstract{
Detecting recurrent weather patterns and understanding the transitions between such regimes are key to advancing our knowledge on the low-frequency variability of the atmosphere and have important implications in terms of weather and climate-related risks.  We adapt an analysis pipeline inspired by Markov State Modelling and detect in an unsupervised manner the dominant winter mid-latitude Northern Hemisphere weather patterns in the Atlantic and Pacific sectors. \MeliVale{The daily 500 hPa geopotential height fields are first classified in $\sim 200$ microstates. The weather dynamics is then represented in the basis of these microstates and the slowest decaying modes are identified from the spectral properties of the transition probability matrix.} \MeliVale{These modes are defined on the basis of the nonlinear dynamical processes of the system and not as tentative metastable states as often done in Markov state analysis.} 
When focusing on a longitudinal window of 60$^\circ$, we recognise a longitude-dependent estimate of the longest relaxation time, \MeliVale{which is smaller where stronger baroclinic activity is found. In the Atlantic and Pacific sectors slow relaxation processes are mainly related to transitions between blocked regimes and zonal flow. We also find strong evidence of a dynamical regime associated with the simultaneous Atlantic-Pacific blocking. 
When the analysis is performed in a broader geographical region of the Atlantic sector, we discover that the slowest relaxation modes of the system are associated with transitions between dynamical regimes that resemble teleconnection patterns like the North Atlantic Oscillation and \textcolor{black}{weather regimes} like the Scandinavian and Greenland blocking, yet have a much stronger dynamical foundation than classical methods based e.g. on EOF analysis.  
Our method clarifies that, as a result  of the lack of a time-scale separation in the atmospheric variability of the mid-latitudes, there is no clear-cut way to represent the atmospheric dynamics in terms of few, well-defined  modes of variability.} 
The approach  proposed here can be seamlessly applied across different regions of the globe for detecting regional modes of variability, and has a great potential for intercomparing climate models and for assessing the impact of climate change on the low-frequency variability of the atmosphere.}
\maketitle

\section{\label{sec:Intro}Introduction}

Low-frequency variability in meteorology refers to a broad spectrum of atmospheric processes occurring over time scales ranging from approximately one week to about a month. Despite extensive research efforts, a comprehensive understanding of its underlying nature remains an ongoing challenge \cite{Speranza1983,Hannachi2017}. From a practical perspective, achieving efficient \DONE{medium and extended range intra-seasonal forecast, beyond the horizon of deterministic predictability of} 
typical 7-10 days in mid-latitudes poses significant difficulties. Furthermore, the climatic impacts of low-frequency  \DONE{phenomena are} substantial \cite{Ghil2002}, also in terms of \DONE{modelling} 
changing climatic conditions \cite{Masato2013}.  
Long time scales are typically entangled with large spatial scales \cite{Fraedrich1978,Speranza1983,DellAquila2005}. Large scale weather \ROUNDtwo{configurations} in the mid-latitudes hold significant importance due to their impact on weather conditions over vast regions. Understanding low frequency variability, in space and time,  is vital as it has implications for weather forecasting, climate studies, and disaster preparedness. 

One of the critical aspects of mid-latitude low-frequency variability is the transitions between two distinct flow regimes: blockings and zonal flow \cite{Benzi1986,Mo.Ghil.1987,Ruti2006,Woollings2010}. \textcolor{black}{Blockings} are synoptic features that are characterized by persistent high-pressure systems in the mid-latitudes, leading to a disruption of the typical west-to-east flow of the jet stream \cite{Rex1950,Rossby1951,Hoskins1987}.  Blocking \textcolor{black}{events} typically manifest in either the Atlantic or Pacific sectors and, more rarely, in both concurrently \cite{Woollings2008}. The lifespans of blocking events can span several days to a few weeks, resulting in the emergence of extreme and \ROUNDtwo{prolonged} weather anomalies with significant regional implications \cite{Tibaldi2018}. Depending on geographic location, season, and pre-existing conditions, blockings can induce diverse weather phenomena such as heat waves, cold spells, extensive droughts leading to wildfires, and floods  \cite{Dole2011,Xoplaki2012,Lau2012,Buehler2011,Hoskins2015,Galfi2019,Kautz2022,Lucarini2023}. Predicting the onset and the decay of blockings is rather elusive even for modern weather prediction systems \cite{Ferranti2015,Lupo2020,Kautz2022}
, and, similarly, modern climate models face difficulties in achieving a correct representation of their statistics \cite{Lucarini2007,DiBiagio2014,Davini2016}. Indeed, large uncertainties exist regarding the  response of the statistics of blocking events to climate change \cite{Woolings2018}, even if recent studies indicate the possibility of an overall increase in their size \cite{Nabizadeh2019} and intensity \cite{Steinfeld_2022}\MeliVale{, with a possible reduction in their persistence in the European sector \cite{Josh22}. 
Nonetheless, as confirmed in the most recent IPCC report, results are not entirely conclusive in terms of the impact of climate change on blockings, as there is at best a medium confidence in \textit{a projected decrease in the frequency of atmospheric blocking over Greenland and the North Pacific for the late \ROUNDtwo{21$^{st}$} century as compared to the reference climatology} \cite{seneviratne2021weather}.} These difficulties \MeliVale{possibly} reflect the fact that the blockings  are associated with atypical and anomalously unstable conditions of the atmosphere \cite{Vannitsem2001,Schubert2016,Faranda2017}, leading to serious fundamental implications in terms of overall model accuracy in describing their dynamics and statistical properties \cite{Lucarini20}.

Accurate identification of blockings is usually achieved through various 
\DONE{indices} based on the geopotential height \cite{Tibaldi1990} or potential vorticity (PV) fields \cite{Pelly2003a}. More recently, identification of blockings has been \ROUNDtwo{performed} using statistical indicators relying on Empirical Orthogonal Functions (EOFs) \cite{Barriopedro2006} as well as multidimensional methodologies \cite{Davini2012}; see \cite{Pinheiro2019} for a recent survey on the topic. 

The low-frequency variability of the northern hemisphere mid-latitudes features \textcolor{black}{also atmospheric configurations} having spatial and temporal scales larger than blockings, \textcolor{black}{exhibiting links between weather phenomena affecting far-away regions on  Earth}. These are \textcolor{black}{called} teleconnection patterns, and are responsible for driving the dynamics of the atmosphere at planetary scale and for facilitating the coupling between atmospheric and oceanic processes. Particularly relevant are the so-called  Pacific-North American teleconnection pattern (PNA) \cite{Wallace1981} and the North Atlantic Oscillation (NAO) \cite{Hurrell1995}. The PNA pattern covers a substantial portion of the North Pacific and North America and  exerts influence on synoptic activity and the subtropical jet stream over the North Pacific, leading to climate impacts over extensive regions of the North American Continent. The NAO is a dipolar pattern of mean sea level pressure over the North Atlantic, extending from the subtropical (Azores high) to sub-Arctic latitudes (Icelandic low). It significantly affects the variability of westerly winds over eastern north Atlantic and Western Europe, making it a crucial factor in shaping Europe's winter climate \cite{Hurrell2003}. Whereas the PNA and the NAO are considered to be large-scale but still regional teleconnections patterns, hemispheric teleconnections have been detected as well, such as the Northern Annular Mode (NAM, named also  Arctic Oscillation \cite{Thompson1998}) and the circumglobal teleconnection pattern \cite{Branstator2002}. These have nontrivial connections with both the NAO and the PNA \cite{Woollings2008,Harnik2016}. See \cite{croci2007,Athanasiadis2010,feldstein_franzke_2017,Hannachi2017} for a detailed discussion of the dynamics and detection methods of teleconnection patterns and off their nontrivial link with blockings.

Besides the above mentioned teleconnection patterns or main modes of variability, other recurring circulation patterns have been found relevant to describe in a coarse grained sense the weather evolution over Europe and North America. These go under the umbrella of the so-called weather regimes (see for example \cite{Baur1944,Vautard1990,Madonna2017}). They are more persistent and larger than synoptic scale weather systems, and \ROUNDtwo{often associated} with surface extremes over vast regions \cite{Franzke2013,DeLuca2019,Madonna2021,Galfi2023}. \ROUNDtwo{In many cases}, the dominant feature of North-Atlantic weather regimes is a persistent ridge or blocking (over the North-Atlantic, Greenland, central Europe or Scandinavia), but some of them are predominantly cyclonic regimes \cite{Madonna2017,Fabiano2020}. 

A rather popular way, \DONE{among others}, to describe in a coarse-grained sense the evolution of the weather in the mid-latitudes amounts to constructing via data-driven methods finite state Markov models, where each state \MeliVale{is a statistical cluster associated with a weather pattern}, 
and then investigate the properties of the resulting Markov chain \cite{Vannitsem2001,Franzke2008,Kwasniok2014,Tantet2015,Detring2021}. Recently, more advanced machine learning-based methods have been used to detect recurrent mid-latitude atmospheric regimes in a simple model of the atmosphere \cite{Mukhin2022}. In this work we wish to advance this perspective and propose a procedure to detect in an unsupervised manner anomalously persistent and recurrent weather patterns. The procedure is routed in Markov State modelling \DONE{(MSM)}, an approach originally introduced for studying metastability  in molecular systems \cite{Noe11,Brooke18}. The key idea in this approach is first classifying all the data - e.g. the molecular configurations, or, in the context of this work, the daily averaged 500 hPa geopotential height \ROUNDtwo{(Z500)} fields - in a  finite set of \emph{microstates}. \ROUNDtwo{Each microstate contains rather similar data.} One then empirically estimates the transition probability, at a fixed time lag, between the microstates.  The eigenvalues of this transition matrix  allow estimating the relaxation times of the system, \DONE{in other words how fast the system is changing}. If a gap in the spectrum exists, say after the $n^{th}$ eigenvalue, one can represent the dynamics as a Markov process between n states, which can be considered metastable on the time scale  defined by the $(n+1)^{th}$ eigenvalue. Such states are identified by the sign pattern of the eigenvectors associated to the \textit{slow} eigenvalues. 

\DONE{A key difference to standard k-means clustering analysis is that in our approach we use clusters (microstates) as a basis to represent dynamics, rather than as {\color{black}potential} metastable configurations  \MeliVale{based on the analysis of the realised configurations of the flow.}  
We choose the number of clusters by using a criterion that is not typical in ordinary clustering approaches: microstates should be as many as possible to provide a suitable partition of the data space.
In this approach, low-frequency modes emerge from the spectral analysis of the transition probability matrix among the microstates and are vectors in the microstate space, not clusters of Z500 patterns.  }

This procedure, if applied out-of-the-box to the analysis of weather data, provides poor results for two reasons. First, the \DONE{Z500 patterns} on the northern hemisphere, as a result of atmospheric turbulence, are extremely variable. Since the number of available observations in our dataset is of the order of a few thousands - see details below in Sect. \ref{sec:MD} -,  
the set of \DONE{Z500 patterns} cannot be divided in microstates which are at the same time sufficiently populated, and which contain sufficiently similar patterns. \MeliVale{Indeed it is known since long time that true atmospheric analogues are extremely rare \cite{Lorenz1969} (yet they can be useful \cite{Faranda2022}).}  
In order to address the first problem we here propose to perform MSM restricted to a longitude window. If one considers the \DONE{Z500 patterns} in a longitudinal window of, e.g., 60$^\circ$, 
 it will be more likely to find coherent patterns which are similar in the window, even if they can in principle differ outside it. We will nonetheless show below that, in agreement with the idea of atmospheric teleconnections mentioned above, coherence between the patterns is found also outside the window used for clustering purposes. In this manner, we are able to classify the \DONE{Z500 patterns} in microstates which, at the same time, include a sufficient number of observations, and include \DONE{Z500 patterns} which are qualitatively similar. Such microstates will be used as a basis to represent the dynamics, as in standard MSM. 
One can use this scheme to perform an analysis by using a window of fixed width, which moves on the longitudinal axe, covering in this manner the whole globe. 

A second problem is that  a genuine dimensional reduction using MSM can be performed only if one observes a significant gap between consecutive relaxation times, this being able to define a cutoff. Such a gap, \DONE{as shown in Supplementary Fig. 4 included in the Supp. Inf.}, is typically not  present. This, strictly speaking, implies that weather dynamics cannot be \ROUNDtwo{unequivocally} described by a Markov chain between \DONE{a small number of states} associated with different weather patterns. 
\DONE{Nevertheless, all the relaxation times are associated with an eigenvector in the space of the microstates. 
Though these Z500 patterns may not meet the criteria for genuine metastable states in the MSM lexicon, they represent Z500 configurations that tend to be persistent and, therefore, meaningful for a coarse-grained description of weather dynamics.} 

\DONE{In this work, we exploit the approach's ability to summarize robustly localized temporal variabilities to analyze these patterns in specific geographic locations and show that the patterns are akin to persistent atmospheric states reported in the literature, such as blocking, zonal flow, and weather regimes.
Moreover, the approach could also be used to study and compare dynamics predictability in different models and climate data reanalysis by inspecting their spectral properties, as an increase in the spectral gap should indicate reduced turbulence that should lead to increased predictability.
This last aspect has not been elaborated further in this work as we leave it \ROUNDtwo{as future} work.}

\begin{figure}
\includegraphics[width=\textwidth]{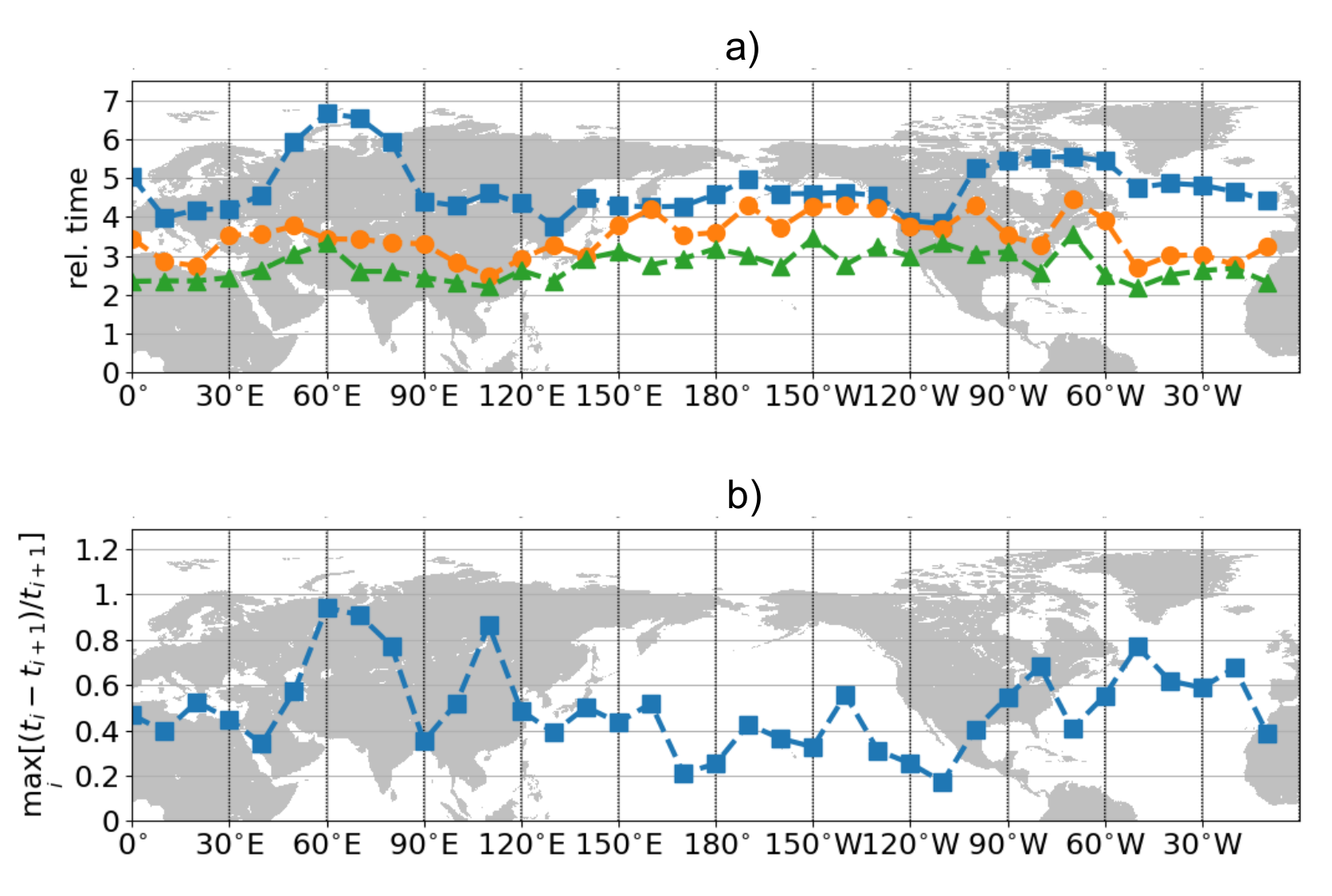}
\caption{ \textbf{Kinetic Analysis.} Kinetic analysis using a moving window with kernel width $\Phi=30^{\circ}$. Panel a): First three distinct relaxation times \DONE{(days)} for the analyzed windows. Panel b): Maximum relative increment for the first ten consecutive relaxation times.}
\label{fig:RTandB}
\end{figure}

\section{\label{sec:RESULTS}RESULTS}

In what follows, we investigate the characteristic time scales Markov chains constructed according to the procedure detailed in the previous section, which amounts to the so-called kinetic analysis of the system \cite{Pande2010,Glielmo2021} . \Cref{fig:RTandB} shows the main findings obtained on our dataset.

In Fig. \ref{fig:RTandB}a we show the first three largest relaxation times of the MSM as a function of the longitude  $\phi_0$, for a kernel width of $\Phi=30^{\circ}$. 
\DONE{The largest relaxation time (blue)} fluctuates longitudinally between approximately 4  and 7 days, with higher values near the centre of the Eurasian and American continental masses. Instead, lower values are found over the oceans, near the Pacific west coast, and over central Europe. Such a geographical pattern follows to a good degree of approximation the patterns of synoptic variability \cite{Hoskins2019}, with smaller relaxation times being associated with more intense synoptic disturbances. This fits with the concept of baroclinic instability as being the catalyser of transitions between competing states of the low-frequency variability of the atmosphere  \cite{Benzi1986,Ruti2006}. \MeliVale{The results are basically unchanged if one chooses a longer time lag, see Supplementary Fig. 3.}

\begin{figure}
\includegraphics[width=\textwidth]{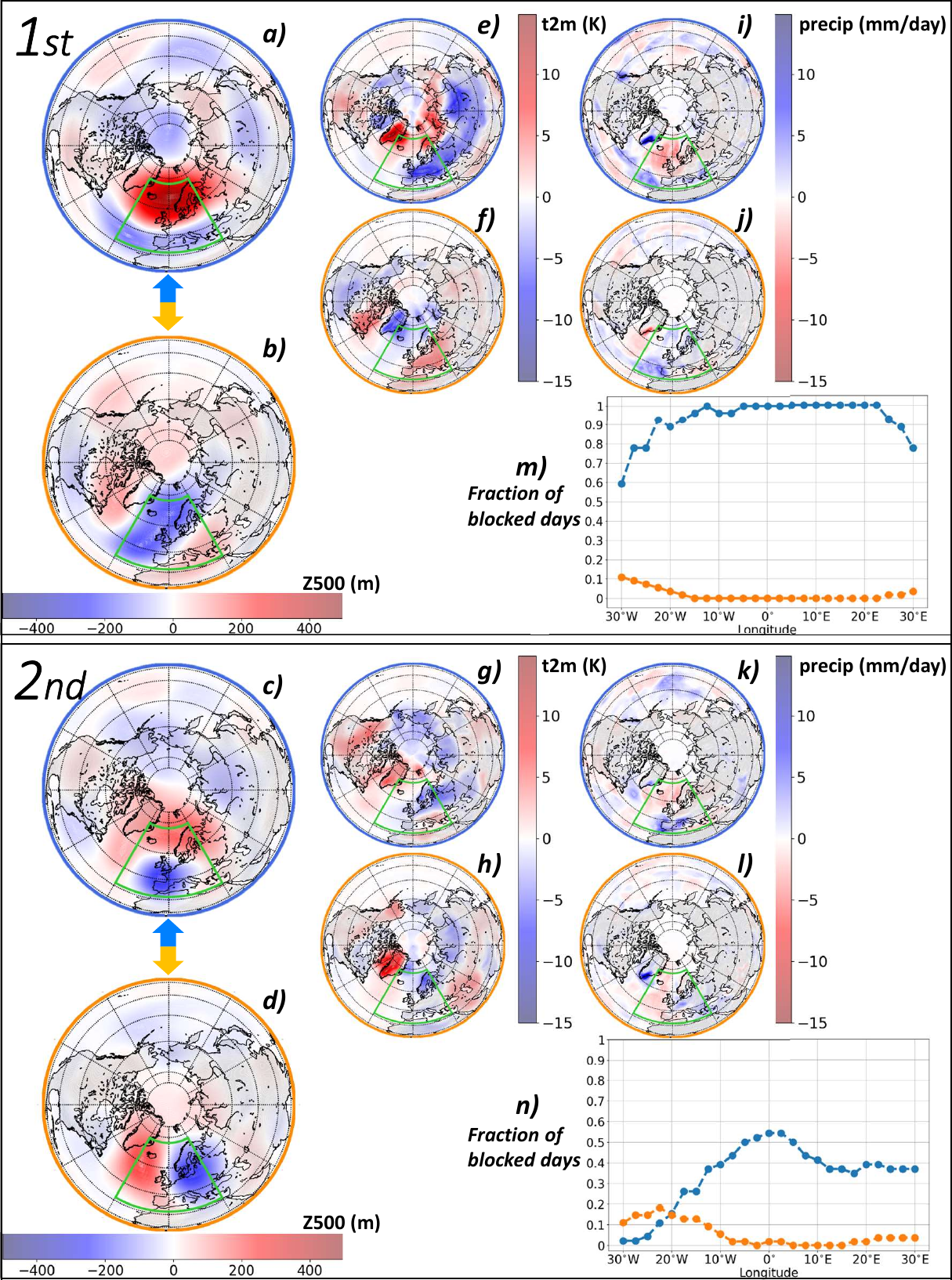}
\caption{\textbf{Modes of Variability of the Atlantic Sector.} Analysis performed by considering Z500 data from  the Atlantic sector ($[ 30^{\circ}W, 30^{\circ}E ]\times[32^\circ N,72^\circ N]$).  
Counterposed microstates mean field Z500 anomalies for the slowest relaxation mode of the system ($t_1=5.1 days$), and  the second slowest relaxation mode of the system ($t_2=3.5 days$). The green window indicates the area considered for clustering the data. Slowest relaxation model, panel a): Z500 [m] anomaly for the microstate featuring more blocked days. Panels e) and i): same as a), for the  t2m [K] and precip [mm/day] anomalies, respectively. Panels b), f), and j): same as a), e), and i), respectively, for the microstate with less blocked days. Panel m): fraction of blocked days for each microstate at the longitudes used to cluster the data; compare the orange vs. blue color code. The corresponding panels are reported for the second slowest relaxation mode in the lower section of the figure. The full fields, given by the sum of the climatological and of the anomaly patterns shown here, are portrayed in Supplementary Fig. 13a,b.}
\label{fig:Atl60}
\end{figure}

In Fig. \ref{fig:RTandB}b, we show  the maximum relative gap between two successive relaxation times at the different longitudes  for  the first ten relaxation times. The gap is at most of order one, which implies that the $i^{th}$ relaxation time, $t_i$ is at most twice  larger than the $(i+1)^{th}$. 
The lack of any significant gap in the relaxation times implies that atmospheric dynamics cannot be meaningfully described  as a Markov process between a small number of metastable states. Hence, there is always a certain degree of subjectivity in performing a cutoff in order to define a discrete, reduced order model. In other words, the relaxation to the steady state involves a mixture of processes at different time scales which are all interlaced. This observation is \DONE{not} surprising: indeed it confirms the lack of temporal (and spatial) scale separation that is very apparent from  spectral analyses of the mid-latitude atmospheric variability \cite{Fraedrich1978,Speranza1983,DellAquila2005}.

\begin{figure}
\includegraphics[width=\textwidth]{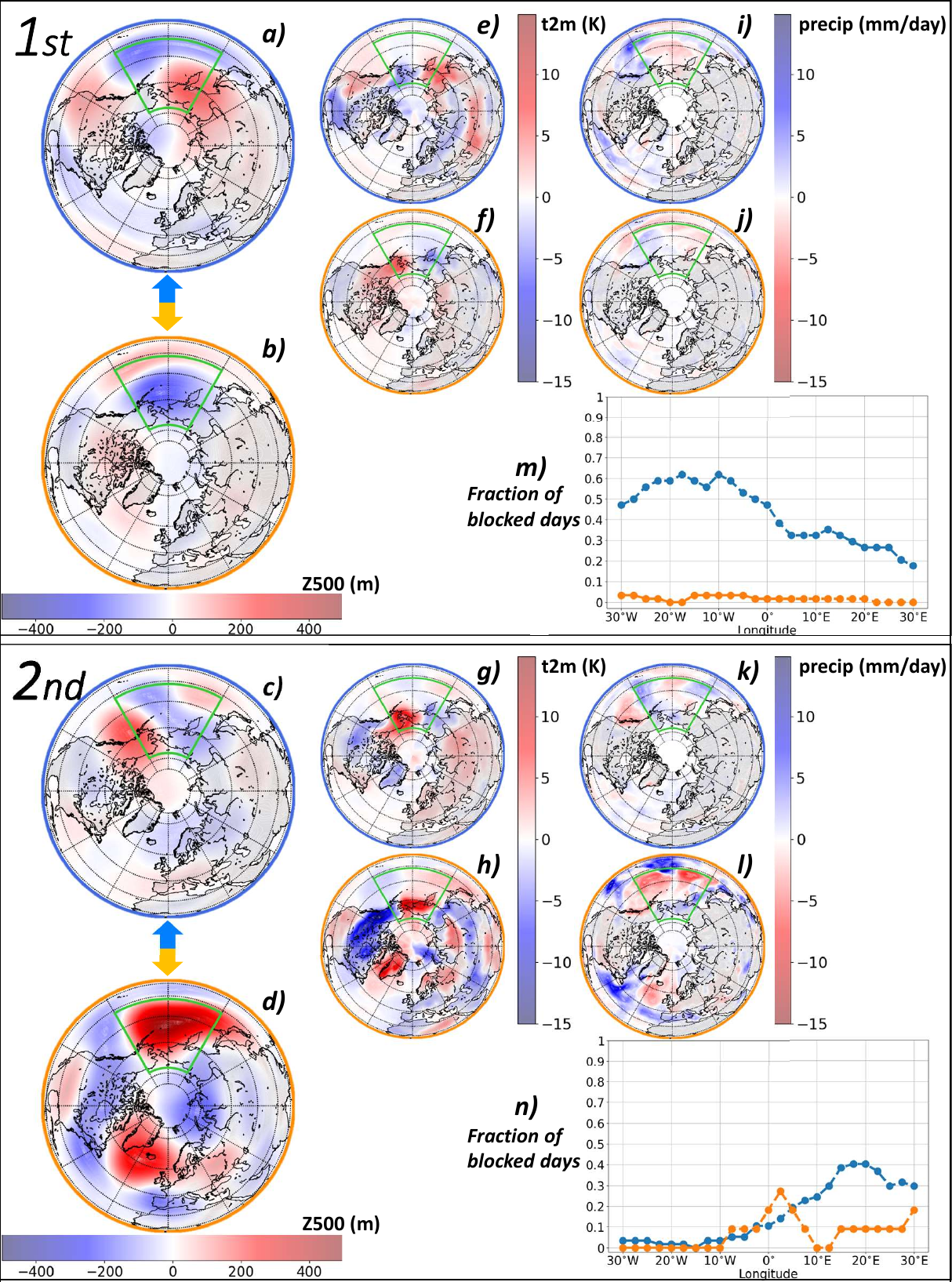}
\caption{\textbf{Modes of Variability of the Pacific Sector.} Same as Fig. \ref{fig:Atl60}, but with the analysis performed considering Z500 [m] data from  the Pacific sector ($[150^{\circ}W,150^{\circ}E ]\times[32^\circ N,72^\circ N]$). The two slowest relaxation modes correspond to $t_1=4.8$ $days$ and $t_2=3.6$ $days$, respectively.
\MeliVale{Full fields, given by the sum of the climatological and of the anomaly patterns shown here, are portrayed in Supplementary Figs. 13c,d.}
}
\label{fig:Pac60}
\end{figure}

\begin{figure}
\includegraphics[width=\textwidth]{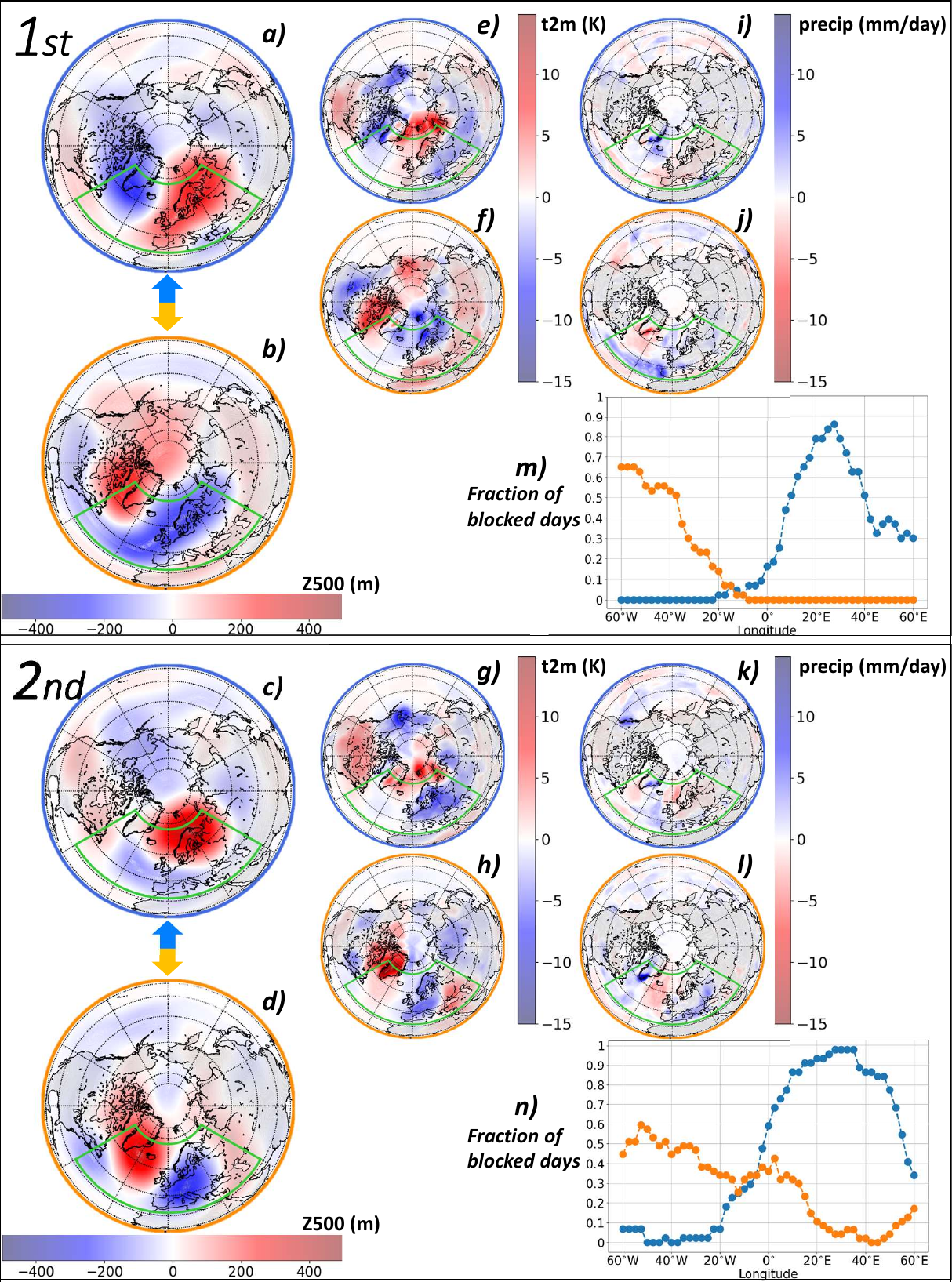}
\caption{\textbf{Modes of Variability of the Extended Atlantic Sector.} Same as Fig. \ref{fig:Atl60}, but with the analysis performed considering Z500 data for the  
extended Atlantic sector ($[60^{\circ}W, 60^{\circ}E ]\times[32^\circ N,72^\circ N]$). 
The two slowest relaxation modes correspond to $t_1=3.7$ $days$ and $t_2=3.5$ $days$, respectively. \MeliVale{Full fields, given by the sum of the climatological and of the anomaly patterns shown here, are portrayed in Supplementary Figs. 14a,b.} 
}
\label{fig:Atl120}
\end{figure}

\subsection{\ROUNDtwo{Alternation} between Blockings and Zonal Flow}
In the following we describe in detail the eigenvectors associated to the two slowest  relaxation times for windows centered around two longitudes: 
$0^{\circ}W$, for the Atlantic region, and $180^{\circ}W$, for the Pacific region\MeliVale{, choosing in first instance a window width of $60^{\circ}$.} 
The eigenvectors are visualized  by plotting  the average \DONE{Z500}, t2m, and precip fields of each the two microstates, which have been chosen according to the rationale discussed in the Sect. Methods. 
\ROUNDtwo{The average is computed over all the fields belonging to a given microstate.}

\ROUNDtwo{If the system possessed a gap in the relaxation times, each of the microstates above could be rigorously interpreted as configurations succinctly describing Markov states. Hence, the dynamics could be described at a coarse grained level as a trapping around a microstate, a rapid transition, a trapping around another microstate, and so on \cite{Glielmo2021}. Since we lack such a time scale separation, the interpretation above is valid only at a qualitative level.}

\ROUNDtwo{Nonetheless,} the eigenvectors provide, by definition, the best possible description of the transition matrix $T$ in a span of dimension 3 (the span includes also the eigenvector that is associated to the unitary eigenvalue, which corresponds to the  
stationary state). Hence, while the  patterns associated to these eigenvectors cannot be  considered genuine metastable states, they do reflect  dominant modes of variability associated with persistent  configurations. Indeed, as we will see,  they bear similarities to persistent atmospheric states previously described in the literature, such as \ROUNDtwo{blockings, zonal flow patterns, and larger scale weather regimes associated with teleconnections}.
 
Results for the Atlantic and Pacific sectors are presented in Fig. \ref{fig:Atl60} and Fig. \ref{fig:Pac60}, respectively. 
\MeliVale{Both figures have dedicated sections on the top for the first subdominant eigenvector and on the bottom for the second subdominant one. \ROUNDtwo{The arrows included in each section illustrate the fact that the two depicted microstates are, in the sense described above, counterposed and that the variability associated to each eigenvector is related to the transitions between the corresponding microstates.} In each section, the graph in the bottom right corner shows the percentage of blocking days estimated using the Tibaldi-Molteni (TM) index \cite{Tibaldi1990} as a function of the longitude} for the two opposing microstates obtained through the pipeline procedure. \ROUNDtwo{Specifically, the TM index is first computed for each longitude for all the fields belonging to the microstate. If blocked conditions are realized at the longitude of reference and in the two neighbouring longitudes and persist for at least three days, the longitude is defined as blocked and assigned unitary value. Secondly, the percentage is obtained for each latitude by averaging over all the fields in the microstate. Note that the patterns associated with higher (lower) occurrence of blockings are indicated by a blue (orange) band around the equator. We stick to this convention for the rest of the paper and in the Supp. Inf.}  


The graphs concerning the first eigenvector show that \ROUNDtwo{the two dominant weather patterns are blocked and non-blocked atmospheric states for both the Atlantic (Fig. \ref{fig:Atl60}a-b) and the Pacific sectors (Fig. \ref{fig:Pac60}a-b). While the dichotomy is strikingly accurate for the Atlantic sector (see Fig. \ref{fig:Atl60}m), over the Pacific the percentage of blocked days achieved by the state we can synoptically associated with blocking is somewhat lower (see Fig. \ref{fig:Pac60}m).} \MeliVale{Here, the core of the blockings is geographically shifted towards the western part of the selected region, near Siberia, in agreement with \cite{Gao2020a,Gao2020b}. We also notice that the \DONE{(predominantly)} zonal flow pattern over the Atlantic associate with the blockings-free microstate features the well-known meridionally tilted shape of the North-Atlantic jet stream \cite{Woolings2011}. } 
The \MeliVale{relaxation time} 
of the first eigenvectors for the Atlantic and Pacific sectors \ROUNDtwo{(4.8 and 3.7 $days$, respectively)} matches well the time scales of mid-latitude synoptic variability associated with baroclinic disturbances, which are responsible for the transitions between blocked and zonal states \cite{Benzi1986,Ruti2006}.

\DONE{Looking at the temperature and precipitation anomalies in the Atlantic sector composites \ROUNDtwo{(Fig. \ref{fig:Atl60}e,f,i,j)}, we find confirmation of the impact of blocking events on such meteorological fields, leading to surface extremes \cite{Kautz2022}. \ROUNDtwo{In what follows we focus on the impacts over land.} In agreement with the original observations by Rex \cite{Rex1951}: blockings in the Atlantic sector  are associated with a dipole of warm anomalies in northern vs cold anomalies in central and southern regions, 
whereas dry anomalies prevail almost everywhere, except in the southernmost regions. Note also the wet anomaly in South Greenland, which is presumably due to diverted storm tracks. On the other hand, zonal flow is characterized by moist air masses moving towards Europe, resulting in stronger precipitation. The enhanced eastward advection also leads to warmer conditions in the central and southern regions.}

\DONE{Changes in the advection due to the occurrence of blockings are important \ROUNDtwo{for determining anomalies in the surface fields} in the Pacific sector too (Figs. \ref{fig:Pac60}e,f,,i,j). Zonal conditions are associated with strong cold anomalies in the western sector, as a result of enhanced advection from the cold continental mass, whereas the eastern sector features warm anomalies; compare with \cite{Breeden2020,Nabizadeh2021}. The occurrence of blockings leads almost everywhere within the region of interest to reduced precipitation, as a result of prevailing anticyclonic conditions. Notably, one finds increased precipitation near the western North American coast \cite{Carrera2004}.}

\MeliVale{In order to illustrate the representativity of most extreme microstates shown in Figs. \ref{fig:Atl60} and \ref{fig:Pac60}, we show in \MeliVale{Supplementary Figs. 6 and 7} of the Supp. Inf. the Z500 anomaly fields corresponding \ROUNDtwo{to two additional pairs of microstates of the slowest relaxation mode for the Atlantic and Pacific sectors, respectively. These pairs are just slightly less far apart from each other in the complex plane of the components of the corresponding eigenvector than those depicted in Figs. \ref{fig:Atl60} and \ref{fig:Pac60}.} For both sectors, there is a close alignments among the three pairs. 
In all three cases, the pairs clearly describe transitions between blocked states and zonal flow. Nonetheless, one finds differences (to be expected) among variants of  blocking and zonal flow configurations.  Blocking can either substantially suppress the zonal flow over the mid-latitudes (e.g.  \MeliVale{Supplementary Fig. 6  states 57, 170, Supplementary Fig. 7 state 80}) or lead merely to its southward displacement. This is a characteristic of high-latitude blocking \MeliVale{(Supplementary Fig. 6  state 89, Supplementary Fig. 7  state 7)}. The zonal flow exhibits a substantial variability as well, especially over the Atlantic region, where the jet can be either more tilted than usually affecting primarily the British Isles and Scandinavia \MeliVale{(Supplementary Fig 6  state 41)} or have a pronounced zonal orientation affecting mainly the Mediterranean region (linked to a deep trough over Europe, \MeliVale{Supplementary Fig. 6  state 96)}.}

\subsection{Transitions between different blocking patterns}
\MeliVale{In the Atlantic sector (Fig. \ref{fig:Atl60}, lower \ROUNDtwo{panel}) the Z500 maps related to the second eigenvector contain features associated with transitions between different blocking structures, whereby one microstate features a high-latitude structure (Fig. \ref{fig:Atl60}c) and the other one a smaller scale  positive anomaly centered in the Atlantic ocean (Fig. \ref{fig:Atl60}d). Quite interestingly, one extreme microstate of the second eigenvector computed for the Pacific sector (Fig. \ref{fig:Pac60}d) is associated with the simultaneous presence of a western North Pacific block and of a \ROUNDtwo{Greenland}  block. The fact that a specific and rather relevant microstate capture\ROUNDtwo{s} such a configuration is quite interesting. 
In \cite{Woolings2008} it was shown that  simultaneous Atlantic-Pacific blocking   occur more frequently than what would be the case if the two sectoral blockings were wholly independent. In \cite{Lucarini20} it was shown that such a pattern is associated with rather special - and anomalously unstable - configuration of the large scale flow. Additionally, \cite{Woolings2008} and \cite{Thompson2001} connected - yet not in the strongest terms - the simultaneous Atlantic-Pacific blockings to the negative phase of the NAM.  Also in our analysis such a connection does not appear to be very strong, even if the corresponding Z500 anomaly pattern shows a disrupted polar vortex, which is the main feature of the negative NAM phase. A recent study \cite{messoridorrington2023} analysing co-occurring North American and Euro-Atlantic weather regimes identified patterns of simultaneous blocking in the two oceanic regions as well.
\ROUNDtwo{While we find stronger signature of the simultaneous blockings when targeting our analysis to the Pacific sector, a weaker global teleconnection signal emerges also when targeting the analysis to the region around Greenland, see Supplementary Figs. 15 and 16, pointing to an asymmetry in the link between the two regions.}
The fact that our method individuates in an unsupervised manner simultaneous Atlantic-Pacific blockings - targeting only one of the two sectors - further reinforces the idea of a physical link between the Pacific and Atlantic patterns \cite{Woollings2008,Lucarini20}. }

\MeliVale{Considering the remarkable nature of the simultaneous Atlantic-Pacific blocking pattern, it is worth looking at the three pairs of most extreme microstates of the second eigenvector for the Pacific sector (Supplementary Fig. 8). Indeed the simultaneous Atlantic-Pacific blocking pattern is very different compared to the other microstates associated with the occurrence of blockings in the Pacific sector. The atypical character of the simultaneous blocking is underpinned by the large distance w.r.t. the second and third most extreme eigenvalues (Supplementary Fig. 8, bottom right panel).}

\subsection{Exploring Teleconnection Patterns}
 The analysis of the $60^{\circ}$ windows \ROUNDtwo{captures rather well that the alternation between  blocked and zonal flow states is the dominant feature of atmospheric dynamics on that scale, but also points  to the relevance of considering} different blocking structures and of global teleconnections. In order to be able to better explore  these latter aspects, we increase the window size to $120^{\circ}$ in longitude. Here, we restrict ourselves to the extended  Atlantic region defined by $[60^{\circ}W, 60^{\circ}E ]\times[32^\circ N,72^\circ N]$. Our main findings are portrayed in Fig. \ref{fig:Atl120}, which is structured along the same lines as Fig. \ref{fig:Atl60}.  

\ROUNDtwo{Figure \ref{fig:Atl120} shows that t}he Z500 fields of the two extreme microstates that are associated with the first eigenvector feature very large structures with anomalies covering the whole hemisphere, pointing to the transition between positive and negative NAO phases (NAO+ and NAO-, corresponding to Figs. \ref{fig:Atl120}a and \ref{fig:Atl120}b, respectively). This is supported by the temperature and precipitation anomaly composites, shown in Figs. \ref{fig:Atl120}e,f,i,j; compare with \cite{Fabiano2020}. NAO+ is related to a strongly tilted jet and, consequently, a strong ridge over Europe leading to warm and dry anomalies over the western and northern part of the continent. Instead, NAO- is related to an anomalously zonal and southern jet, which brings abundant precipitation over the Mediterranean region, while cold arctic air masses can reach northern and central Europe. Note  that the relaxation time associated with the first eigenvector - 3.7 $days$ - \MeliVale{is in good agreement with the decorrelation time - about 5 $days$ - of the NAO index time series \cite{Onskog2018}. Note also that the  Z500 fields of the two extreme microstates are in close agreement on a global scale with the first two EOFs of the DJF Z500 fields for the whole $20^\circ N-90^\circ N$ sector (not shown), thus supporting the large-scale nature of the modes detected through our method.} 

The Z500 anomaly fields related to the second eigenvector show similar structures, \textcolor{black}{but rather than affecting the whole hemisphere, they are more confined to the North Atlantic and European regions. The dominant feature of one of the extreme microstates is a strong, meridionally restricted high Z500 anomaly over Northern Europe, representing a typical Scandinavian blocking pattern (\ROUNDtwo{Fig.\ref{fig:Atl120}c}). The opposite extreme microstate shows a \ROUNDtwo{dipole} with a strong blocking over Greenland and Iceland, and a trough over continental Europe (\ROUNDtwo{panel d}). Thus, the variability associated with the second eigenvector is related to transitions between Scandinavian blocking and Greenland blocking (or North-Atlantic ridge)}. Both circulation anomalies lead to anomalously cold air over most of the European continent, however the origin of the cold air differs. The Scandinavian blocking favours cold air advection from Siberia, whereas the deep trough next to the Greenland blocking, allows Arctic airmasses to reach all the way south to the Mediterranean. This transition between circulation patterns connected to anomalously cold conditions brings prolonged periods of cold winter weather over large parts of the continent. Both circulation structures are related to mainly wet anomalies over the Mediterranean, suggesting an anomalously southern location of the jet stream.

\MeliVale{To obtain a clearer picture about how precisely  the extreme microstates are defined, we show in Supplementary Fig. 9  the Z500 anomaly fields corresponding to the three most distant pairs of microstates  of the first eigenvector for the $120^{\circ}$ window in the Atlantic sector. \ROUNDtwo{We find that the NAO- pattern (Supplementary Fig. 9b) is  close to two patterns (Supplementary Fig. 9d,f) that describe the occurrence of Greenland blocking. This supports previous findings that reveal a close link between  NAO- and Greenland blocking, and clarifies the fundamental difficulty in distinguishing between these patterns \cite{Woollings2008b,Davini2012,Li2015}. Additionally, we find that the NAO+ pattern (Supplementary Fig. 9a) is extremely close to two patterns (Supplementary Fig. 9c,e) depicting the Scandinavian blocking. This is an agreement with recent studies suggesting that these two patterns are closely dynamically associated \cite{Li2015,Vihma2019}.}
We remark that this connection does not mean that the NAO is in a positive phase whenever Scandinavian blocking occurs, but points instead towards the dynamical similarity between the two weather patterns that would be classified as different weather configurations based on a clustering analysis or on the NAO index. Finally, Supplementary Fig. 9d reveals a clear connection between the first and second eigenvector; compare with Fig. \ref{fig:Atl120}b.}

\section{\label{sec:Conc}Conclusions}

\REV{We have analysed the low-frequency variability of the Northern Hemisphere winter mid-latitude atmosphere based on  Markov state modelling.} \DONE{Compared to similar studies in the literature, we do not choose the number of \ROUNDtwo{potential metastable states}  a-priori.} \MeliVale{Instead, we follow a two-step procedure that allows us to distillate the main feature of the dynamics of the system, going beyond a diagnostic analysis. 
Following a methodology originally devised for studying molecular dynamical processes, we first partition the phase space in a number of microstates that should be as high as made possible by the available data. In our case, this number ends up to be $O(100)$, and is the only free parameter of the procedure.} 
\MeliVale{We then construct the Markov matrix describing the transition between the microstates; and, using standard methods, end up identifying the invariant measure corresponding to the stationary probability as well as the relaxation modes, which can be ordered according to their corresponding relaxation time. Those with longest relaxation times correspond to the slowest decaying modes.}

\MeliVale{Seeking truly global modes is unfeasible because the data is insufficient. Hence, we resort to studying regional features by performing our analysis on longitudinal windows of 60$^\circ$ and 120$^\circ$, respectively. We have the freedom to shift seamlessly our domain, but we mostly focus on the Atlantic and Pacific sectors, considering the high frequency of winter blockings in these two regions \cite{Woolings2018}.}

\MeliVale{For the $60^\circ$ longitudinal window, our results show  that, in both the Atlantic and the Pacific sector, the slowest modes are associated with transitions  between blocking states and zonal flow, with the expected associated Z500, surface air temperature, and precipitation anomalies. Additionally, we find strong evidence of a mode of variability - almost a statistical outlier yet clearly relevant for the dynamics -  strongly associated with the simultaneous occurrence of blockings in both sectors; see Fig. \ref{fig:Pac60}, Supplementary Figs. 15 and 16. This reinforces the conclusions of previous studies indicating the presence of a physical link between the two blockings \cite{Woollings2008} and of a rather special dynamical configuration associated with such a regime \cite{Lucarini20}. 
Many aspects of simultaneous Atlantic - Pacific blockings are not well understood, pointing towards the necessity of future research efforts.}

Considering a $120^\circ$ longitudinal window for the Atlantic sector leads as well to the detection of blockings and zonal flows as important features. Nonetheless, the slowest transitions are in this case related \MeliVale{to large-scale dynamics, associated with teleconnections and the transition between different} blocking structures. 
\REV{We find that the slowest and second slowest relaxation modes are associated with the transition between patterns that are closely reminiscient of the NAO+ and NAO- states, and of the Scandinavian blocking and Greenland blocking, respectively.} \ROUNDtwo{A preliminary analysis of the $120^\circ$ longitudinal window centered over the Pacific is  encouraging as our method allows to identify large scale patterns that resemble both phases of the much-studied Pacific modes of atmospheric variability, namely the Pacific North America teleconnection pattern \cite{Linkin2008} and the North Pacific Oscillation/West Pacific teleconnection pattern \cite{Aru2022}, see Supplementary Figs. 11a,b,c,d. Clearly this aspect deserves further investigation on its own.}

\MeliVale{The patterns we find are to some degree similar to familiar patterns obtained with other methods, such as EOF, or weather regimes based on clustering analysis, which usually leads to considering 4-8 clusters, see e.g. \cite{Fabiano2020,Hochman2021,Messori2023}. 
However, due to the large number of microstates in our analysis, our patterns resemble more accurately observed daily fields because less information is filtered out. Furthermore, they are dynamically more meaningful, being defined based on slow modes of variability, as mentioned above. Given the dynamical nature and the high degree of granularity of these patterns, they might be promising candidates to support medium-range weather prediction, along the lines of the so-called weather types \cite{Neal2016}.} 

\MeliVale{Furthermore, our analysis underlines the complexity of atmospheric dynamics by revealing dynamical connections between different weather regimes.} Our results also confirm the difficulty in separating clearly some of these weather patterns, as in the case of the 
Greenland blocking and NAO- \cite{Davini2012} as well as Scandinavian blocking and NAO+. In \cite{Woollings2008b} it is shown that the NAO is strongly related to variations in occurrence of high latitude blocking, and can be interpreted from a blocking perspective.


An important outcome of our analysis -- and certainly related to the similarity of different weather patterns discussed above -- is that we do not find a significant gap between relaxation times of the various decaying modes. Hence, we cannot claim that a handful of weather patterns  \ROUNDtwo{unequivocally} describe the relevant large-scale atmospheric dynamics. This underlines the irreducible complexity of atmospheric dynamics of the mid-latitudes, \DONE{confirming} 
that there is always a certain degree of arbitrariness when selecting dominant modes and excluding the rest. The lack of such a gap, which is apparent when performing a spectral analysis of the mid-latitude atmospheric variability \cite{Fraedrich1978,Speranza1983,DellAquila2005}, can also be seen as the reason behind the well-studied presence of deviations from exponential behaviour of the autocorrelation of the NAO index for long time lags \cite{Feldstein2000,Onskog2018}.

In this work, we have chosen to focus on the winter season \MeliVale{considering that blockings are more prevalent in winter compared to summer \cite{Woolings2018}.} However, our method is able to detect slow modes of variability in summer as well. \MeliVale{We have performed a preliminary analysis on the summer season (June-July-August, JJA) for the Euro-Atlantic sector, where the longitudinal window has been shifted eastwards by 30$^\circ$ in order to take into account that the peak of blocking occurrence is at $\approx30^\circ E$ \cite{Drouard2018} When considering the $60^\circ$ longitudinal window, one of the identified dominant patterns of variability captures the occurrence of blockings in the region and is very similar to what \ROUNDtwo{was} shown in previous investigations \cite{Drouard2018}. When considering the extended $120^\circ$ longitudinal window, the two dominant patterns of variability have a close resemblance over Eurasia to the double-ridge and double-trough patterns presented in \cite{Yang2022}. See Supplementary Fig. 12 for details.}

Even if we discuss in detail only the results in two specific, much studied regions of the Northern Hemisphere, the methodology discussed in this contribution can be adapted seamlessly for analysing dominant weather patterns in other areas of the planet that have been much less extensively studied according to this angle, as in the case of the tropics \cite{Pope2009,Hassim2019}, or the mid-latitudes \cite{Solman2003,Wilson2013,Arizmendi2022,Loikith2019} and polar regions \cite{Pohl2019} of the Southern Hemisphere. 
Our method has the potential of revealing new and previously unknown weather patterns that are relevant in such regions, thus potentially advancing our knowledge of regional climate and facilitating the evaluation of the impacts of climate variability on human and environmental welfare. 
 
Additionally, our approach has a great potential for comparing climate models, for testing their realism, and for investigating the impact of climate change on the large scale variability of the atmosphere. \ROUNDtwo{In} this regard, we plan to take advantage of the  available data of the climate models intercomparison project \cite{Eyring2016}. The availability of much longer time series, as a result of large ensemble climate model simulation strategy  \cite{Maher2021} might allow for extending this analysis in such a way that global atmospheric patterns can be directly studied, without the need to resort to longitudinal windows as done here. This might provide a key advancement for understanding the link between synoptic scale variability and global teleconnections and be particularly useful when trying to address the dynamical reasons behind the occurrence of persistent extremes like heatwaves and cold spells \cite{Kornhuber2019,Kornhuber2021,Galfi2021,RagoneBouchet2021,Lucarini2023}

\section{Methods}\label{sec:MD}

\subsection{Meteorological Data and Definition of the Microstates}
Our study focuses on the winter low-frequency variability in the mid-latitudes of the Northern Hemisphere. Hence, following \cite{DellAquila2005}, we examined the 1950-2022 December-January-February (DJF) daily averaged Z500 field extracted from the $2.5^\circ$ resolution NCEP Renalysis 1 \cite{NCEP} for the latitudinal band $[32^\circ N,72^\circ N]$.
\DONE{To ensure that the studied signal meets the basic requirements needed to be able to adopt the analysis method we intend to use, namely that it is possible to define an equilibrium measure, we have taken a necessary preprocessing step of eliminating trends and seasonal variations from the data. In particular, we have detrended and deseasonalized the data by subtracting the DJF average for each year separately before removing the seasonal cycle constructed by computing the average for each day over all years in a pointwise manner, namely, separately for each grid point.}

On these preprocessed data, we performed Markov state modelling (MSM). 
In short, MSM  involves \DONE{$(1)$} dividing the data into a large set of discrete states called \emph{microstates}. The data belonging to a microstate are assumed to be similar (see below for a detailed discussion on this point). One then \DONE{$(2)$} estimates the  transition probability, in a given time, between the microstates. The dynamics of the system is \DONE{$(3)$} described in the basis of the eigenvectors of the transition probability matrix, with the eigenvalues providing information on the relaxation times. 

More specifically, let's denote the data points by $X_t$, where $t=1,\dots,N$ labels the different data. In the case of this work, $X_t$ is the Z500 pattern, recorded \DONE{on a geographical region} defined below on a latitude/longitude grid, and  $t$ labels the day in which a specific  pattern has been recorded.
The data are then divided in microstates using  a K-means clustering algorithm  \cite{Lloyd82}. Specifically, one chooses a number K of microstates, and, by K-means, one finds, for the \DONE{Z500 pattern} $X_t$ an integer cluster label $c_t \in \{1,\dots,K \}$. 

To decide the value of $K$  one should consider two different factors: (i)  the minimum number of elements in each microstate should be large enough for a reliable estimation of the transition probability matrix and (ii)  the Z500 patterns assigned to a microstate should be (qualitatively) similar. 

Condition (i) is pretty simple to satisfy: since the transition probability in MSM is typically sparse, it is sufficient to choose K in such a way that each cluster is visited a sufficient number of times.
Since the number of available days is approximately 6400, we opted for 180 microstates, which implies that each microstate will contain, on average, $\sim$ 35 \ROUNDtwo{fields, each associated with one day}, \MeliVale{with values ranging between $\sim$ 10 and $\sim$ 70.} The viability of this choice is demonstrated by dedicated experiments on artificial data generated by the \MeliVale{state-of-the-art MPI-ESM-LR version 1.2 Earth system model \cite{MPI-M} \DONE{(Supplementary Figs. 2 and 4)}.  {\color{black}This is among the best performing ESMs belonging to the CMIP6 class \cite{Bock2020} and, additionally, its skill in representing the atmospheric variability of the Northern Hemisphere mid-latitudes compares rather well with its higher resolution version \cite{Muller0219}.}
We have considered the same dataset - \textit{piControl} run featuring pre-industrial conditions in terms of atmospheric composition and land-use \cite{MPIdata} - used in \cite{Galfi2021,Lucarini2023} to investigate heatwaves and cold spells associated with blockings.}  \ROUNDtwo{In the Supp. Inf. we show - see Supplementary Figs. 4a,b - that the characteristic time scales of the leading modes are basically unchanged if one performs the analysis on 72 vs 1000 winters choosing 180 microstates. Similarly, small changes in the leading characteristic time scales are found if one chooses 180 vs 360 microstates to analyse the 1000 winters dataset. Moreover, Supplementary Fig. 5 shows that the actual patterns of the two most extreme microstates associated with the first eigenvector are robust with respect to the number of years used in their evaluation (72 vs 1000 years).} 

Satisfying condition (ii) turned out to be much more difficult. If one applies K-means clustering with K of order 100 on the Z500 patterns across all longitudes in the latitudinal band of interest, one finds that patterns which are qualitatively different are assigned to the same cluster. 
As an example, it is possible to observe a specific and recognizable \DONE{pattern} in the Atlantic sector (say an omega-block) and, simultaneously, a variety of different \DONE{patterns} in the Pacific sector (not shown). This observation prompted us to develop a variant of MSM, specifically designed to keep into account the fact that \DONE{patterns} can be strongly correlated for spatial scales corresponding to zonal wavenumbers up to $\approx 3$ but - in general - only weakly correlated on a truly global scale \cite{DellAquila2005}. 
In particular, to weight appropriately local correlations we multiply  the \DONE{patterns} by a Gaussian kernel, centered on a specific longitude value $\phi_0$, with standard deviation $\Phi$ of order of some tenths of degrees (see below). Therefore, the distance between \DONE{patterns} $X$  and \DONE{Z500 patterns} $X'$  is given by

\begin{equation}
    d(X,X'| \phi_0)=\sum_{\phi,\lambda} a_{\phi,\lambda} \left( X(\phi,\lambda)-X'(\phi,\lambda)\right )^2 
    e^{-\frac{(\phi-\phi_0)^2}{2\Phi^2}}
    \label{eq:distance}
\end{equation}
where $a_{\phi,\lambda}$ is the area of the grid patch of latitude $\lambda$ and longitude $\phi$, and the sum over $\lambda$ runs between \DONE{$32^\circ N$ and  $72^\circ N$}. \ROUNDtwo{Clearly, changing the latitudinal band defining the mid-latitudes has an impact on the final outcome of the procedure. Nonetheless, we have verified that the results discussed below are robust with respect to small changes in the spatial domain of references, see Supplementary Figs. 9 and 10 for an example.}

Choosing the kernel width is a crucial decision as it determines the scale of climatic events we focus on. 
We present the results for two kernel \DONE{widths}, $\Phi=30^{\circ}$ and $\Phi=60^{\circ}$, qualitatively corresponding to windows spanning a longitude of $60^{\circ}$, $120^{\circ}$. \ROUNDtwo{The former window is appropriate to detect features such as blockings occurring within the target region. The latter window is better suited to capture larger teleconnections and possibly identify global patterns of weather variability; see results below. We remark that the choice of the centre of the window is extremely relevant because the prevalence of blockings, larger scale teleconnection, and the characteristics of the jet stream in the mid-latitudes are strongly dependent on the local geographical and orographical features  \cite{Hannachi2017,Woollings2010,Hoskins2015,Woolings2018}.} \MeliVale{Furthermore, this selection of the effective window size is also roughly consistent with the advection length scale: taking $\approx10$ \ROUNDtwo{m s}$^{-1}$ as reference value for a westerly wind at 500 hPa in the region of interest \cite{Whittleston2018}, one obtains an advective length scale over 5 days of $\approx4.3\times10^3$ \ROUNDtwo{km}, which corresponds to about 50$^\circ$ longitude in the midlatitudes.}

\ROUNDtwo{We remark that in this approach the K microstates should not be considered in any manner metastable states, but just a basis to represent the dynamics. Metastability, if present, emerges from the spectral properties of the transition probability between these microstates. This leads us to detailing the next steps in our data analysis protocol.}

\subsection{Transition Probability Matrix}
After dividing the data into microstates, 
we estimate the $K \times K$ transition probability matrix $T$ of the process from the data. Each row element of the matrix represents the index of the \textit{departure} microstate, while each column element represents the index of the \textit{arrival} microstate. The element $T_{i,j}$ is the estimated probability of moving from the ith microstate to the jth microstate after the time lag $\tau$, namely $T_{i,j}=P_r(X_t + \tau = j | X_t = i)$. By construction, all the matrix elements are non-negative and the rows of the transition probability matrix sum to 1, namely $T$ is a stochastic matrix \cite{ANDRILLI2016513}.
In our specific case, we choose $\tau= 1$ day. In the Supp. Inf. we show that the slowest relaxation times remain very similar if one chooses $\tau= 2$ days, see Supplementary Fig. 3.  $T$ is estimated using only consecutive days, separately on each winter, and averaged over all the available winters in our dataset. 


Also following the standard MSM pipeline, we then  perform an eigen-decomposition of the transition probability matrix $T$, finding its \MeliVale{eigenvectors $v_i$ and corresponding eigenvalues $\lambda_i$, so that $Tv_i=\lambda_jv_i$.}
Since \DONE{$T$} is a stochastic matrix, its first eigenvalue equals one, while the other eigenvalues are complex numbers with norms smaller than one. The components of the eigenvector related to the first eigenvalue - which is unique in the non-degenerate case, as found here - are proportional to the long-term probability of each microstate, thus defining the so-called invariant measure. The eigenvectors of $T$ linked to the eigenvalues with largest norms below 1 correspond to the slow decaying modes of the system. 
\DONE{In the event that an eigenvalue is complex, its conjugate will also be found in the spectrum. This implies that any perturbation to the stationary state described by the $i^{th}$ eigenvector will decay to zero with an exponentially damped sinusoid. The time constant of the exponential, denoted by $t_{i}$, and the frequency of the oscillation, denoted by $\omega_i$, are given respectively by:}

\begin{equation}
\begin{split}
&t_{i}=\frac{-\tau}{\log(\left|\lambda_{i}\right|)} \\
&\omega_i=\frac{2\pi}{\left|\tan^{-1}\left(\frac{Im(\lambda_{i})}{Re(\lambda_{i})}\right)\right|}
\end{split}
\end{equation}
\MeliVale{In our case the complex eigenvalues correspond to modes where the oscillatory behaviour is much slower than the decaying one - say with timescale of one month vs. one week, so that at all practical purposes the phenomenology of relaxation is dominated by exponential damping.}

\MeliVale{It is  challenging to find an efficient way to portray an eigenvector given by a combination of microstates, each representing a Z500 pattern. In the case of the subdominant eigenvectors, which correspond to $|\lambda_i|<1$, we proceed as follows. We consider the two microstates that are \ROUNDtwo{as far apart as possible} in the complex} \ROUNDtwo{plane} defined by the component of the eigenvector. 
Indeed, these two microstates represent the dominant components of the anomaly associated with the eigenvector.  and they appear always in counterphase. 
\ROUNDtwo{In this sense, the two microstates can be seen in opposition and the eigenvenctor can be interpreted as describing an oscillation between these two microstates, which can be seen as weather regimes. 
We then visualise the \DONE{Z500} field associated with each microstate by plotting the average of the Z500 fields of all the  days assigned to the microstate. The same procedure can then repeated for other meteorological fields of interest, see below. 
See the Supp. Inf. for further details and clarifying examples.} 

To check whether the relaxation dynamics has a clear signal on variables which are not explicitly controlled in the MSM procedure, we also  visualise the  average 2-meter air temperature field  (t2m) and the precipitation rate field (precip).
To have a clearer picture and show large-scale patterns, we always visualise these fields over the whole northern hemisphere, highlighting the region used for deriving the MSM by a  green frame. 
In \cref{alg:1}, we provide a concise overview of the whole analysis pipeline.
\ROUNDtwo{Clearly, changing the latitudinal band defining the mid-latitudes has an impact on the final outcome of the procedure. Nonetheless, we have verified that our results are robust with respect to small changes in the spatial domain of references, see Supplementary Figs. 9 and 10 for an example.}

\begin{algorithm}
\SetAlgoLined
\caption{Pipeline summary }
\label{alg:1}
         \SetKwInOut{Input}{input}
        \SetKwInOut{Output}{output}
        \Input{ Daily average reanalysis at 500hPa for winter months of the Northern Hemisphere }
        \Output{Partition of the data in microstates, and label of microstates associated with the slowest mode of the system.}
    \SetKwBlock{Beginn}{beginn}{ende}
            Select the window central longitude $\phi_0$ and the standard deviation $\Phi$ of the Gaussian kernel; \\
            Select the number of microstates K;\\
    \Begin{

            Partition the data into K microstates by the K-means clustering algorithm;\\
            Compute the transition probability matrix T with time lag $\tau=1$;\\
            Eigendecomposition of T;\\
            Check for a possible spectral gap;\\
            Select the first $k$ eigenvectors of interest;\\
        \For{$j =1$ to $k$}{
               Find the indices of the two most extreme components of the j-th eigenvector ( Euclidean distance on both Real and Imaginary parts ) and assign them to $index_A[j]$ and $index_B[j]$
        }
        \Return data-set partition, $index_A[:], \ index_B[:]$
    }
\end{algorithm}

\bmhead{Acknowledgments}
VL acknowledges the support received from the Horizon 2020 Project TiPES (Grant No. No. 820970), from the Marie Curie ITN and CriticalEarth (Grant Agreement No. 956170), from EPSRC through the grant EP/T018178/1, and from the University of Reading's RETF project CROPS. The authors acknowledge scientific exchanges with D. Faranda, M. Ghil, A. Gritsun, and G. Messori. 

{\color{black}\bmhead{Competing Interests}
The Authors declare no Competing Financial or Non-Financial Interests

\bmhead{Data Availability}
The NCEP-NCAR Reanalysis 1 data are freely available and have been obtained from the NOAA PSL, Boulder, Colorado, USA (\url{https://psl.noaa.gov}). The data \cite{MPIdata} for the standard pre-industrial control run (piControl) of the MPI-ESM-LR version 1.2 ESM-LR model \cite{MPI-M} can be obtained from the CMIP6 website (\url{https://pcmdi.llnl.gov/CMIP6/}).
The data that support the findings of this study are publicly available at \url{https://doi.org/10.5281/zenodo.11068800}.

\bmhead{Code Availability}
The code used in this paper is inspired by \cite{Noe11,Brooke18} and is publicly available at \url{https://github.com/sspring137/npj2024.git}.

\bmhead{Author Contributions}
 V.L. and A.L. conceived the study. S.S. performed the data analysis and produced the figures. V.M.G. led the analysis of the obtained meteorological fields. All authors discussed the results and contributed to the final manuscript.

\providecommand{\noopsort}[1]{}\providecommand{\singleletter}[1]{#1}

\section{Figure Captions}
\textbf{Figure 1.} \textbf{Kinetic Analysis.} Kinetic analysis using a moving window with kernel width $\Phi=30^{\circ}$. Panel a): First three distinct relaxation times (days) for the analyzed windows. Panel b): Maximum relative increment for the first ten consecutive relaxation times.

\noindent
\textbf{Figure 2.} \textbf{Modes of Variability of the Atlantic Sector.} Analysis performed by considering Z500 data from  the Atlantic sector ($[ 30^{\circ}W, 30^{\circ}E ]\times[32^\circ N,72^\circ N]$).  
Counterposed microstates mean field Z500 anomalies for the slowest relaxation mode of the system ($t_1=5.1 days$), and  the second slowest relaxation mode of the system ($t_2=3.5 days$). The green window indicates the area considered for clustering the data. Slowest relaxation model, panel a): Z500 [m] anomaly for the microstate featuring more blocked days. Panels e) and i): same as a), for the  t2m [K] and precip [mm/day] anomalies, respectively. Panels b), f), and j): same as a), e), and i), respectively, for the microstate with less blocked days. Panel m): fraction of blocked days for each microstate at the longitudes used to cluster the data; compare the orange vs. blue color code. The corresponding panels are reported for the second slowest relaxation mode in the lower section of the figure. The full fields, given by the sum of the climatological and of the anomaly patterns shown here, are portrayed in Supplementary Figs. 13a,b.

\noindent
\textbf{Figure 3.} \textbf{Modes of Variability of the Pacific Sector.} Same as Fig. \ref{fig:Atl60}, but with the analysis performed considering Z500 [m] data from  the Pacific sector ($[150^{\circ}W,150^{\circ}E ]\times[32^\circ N,72^\circ N]$). The two slowest relaxation modes correspond to $t_1=4.8$ $days$ and $t_2=3.6$ $days$, respectively. Full fields, given by the sum of the climatological and of the anomaly patterns shown here, are portrayed in Supplementary Figs. 13c,d.

\noindent
\textbf{Figure 4.} \textbf{Modes of Variability of the Extended Atlantic Sector.}  Same as Fig. \ref{fig:Atl60}, but with the analysis performed considering Z500 data for the  extended Atlantic sector ($[60^{\circ}W, 60^{\circ}E ]\times[32^\circ N,72^\circ N]$). 
The two slowest relaxation modes correspond to $t_1=3.7$ $days$ and $t_2=3.5$ $days$, respectively. Full fields, given by the sum of the climatological and of the anomaly patterns shown here, are portrayed in Supplementary Figs. 14a,b. 


\begin{thebibliography}{114}
\ifx \bisbn   \undefined \def \bisbn  #1{ISBN #1}\fi
\ifx \binits  \undefined \def \binits#1{#1}\fi
\ifx \bauthor  \undefined \def \bauthor#1{#1}\fi
\ifx \batitle  \undefined \def \batitle#1{#1}\fi
\ifx \bjtitle  \undefined \def \bjtitle#1{#1}\fi
\ifx \bvolume  \undefined \def \bvolume#1{\textbf{#1}}\fi
\ifx \byear  \undefined \def \byear#1{#1}\fi
\ifx \bissue  \undefined \def \bissue#1{#1}\fi
\ifx \bfpage  \undefined \def \bfpage#1{#1}\fi
\ifx \blpage  \undefined \def \blpage #1{#1}\fi
\ifx \burl  \undefined \def \burl#1{\textsf{#1}}\fi
\ifx \doiurl  \undefined \def \doiurl#1{\url{https://doi.org/#1}}\fi
\ifx \betal  \undefined \def \betal{\textit{et al.}}\fi
\ifx \binstitute  \undefined \def \binstitute#1{#1}\fi
\ifx \binstitutionaled  \undefined \def \binstitutionaled#1{#1}\fi
\ifx \bctitle  \undefined \def \bctitle#1{#1}\fi
\ifx \beditor  \undefined \def \beditor#1{#1}\fi
\ifx \bpublisher  \undefined \def \bpublisher#1{#1}\fi
\ifx \bbtitle  \undefined \def \bbtitle#1{#1}\fi
\ifx \bedition  \undefined \def \bedition#1{#1}\fi
\ifx \bseriesno  \undefined \def \bseriesno#1{#1}\fi
\ifx \blocation  \undefined \def \blocation#1{#1}\fi
\ifx \bsertitle  \undefined \def \bsertitle#1{#1}\fi
\ifx \bsnm \undefined \def \bsnm#1{#1}\fi
\ifx \bsuffix \undefined \def \bsuffix#1{#1}\fi
\ifx \bparticle \undefined \def \bparticle#1{#1}\fi
\ifx \barticle \undefined \def \barticle#1{#1}\fi
\bibcommenthead
\ifx \bconfdate \undefined \def \bconfdate #1{#1}\fi
\ifx \botherref \undefined \def \botherref #1{#1}\fi
\ifx \url \undefined \def \url#1{\textsf{#1}}\fi
\ifx \bchapter \undefined \def \bchapter#1{#1}\fi
\ifx \bbook \undefined \def \bbook#1{#1}\fi
\ifx \bcomment \undefined \def \bcomment#1{#1}\fi
\ifx \oauthor \undefined \def \oauthor#1{#1}\fi
\ifx \citeauthoryear \undefined \def \citeauthoryear#1{#1}\fi
\ifx \endbibitem  \undefined \def \endbibitem {}\fi
\ifx \bconflocation  \undefined \def \bconflocation#1{#1}\fi
\ifx \arxivurl  \undefined \def \arxivurl#1{\textsf{#1}}\fi
\csname PreBibitemsHook\endcsname

\bibitem[\protect\citeauthoryear{Speranza}{1983}]{Speranza1983}
\begin{barticle}
\bauthor{\bsnm{Speranza}, \binits{A.}}:
\batitle{Deterministic and statistical properties of the westerlies}.
\bjtitle{Paleogeophysics}
\bvolume{121},
\bfpage{511}--\blpage{562}
(\byear{1983})
\end{barticle}
\endbibitem

\bibitem[\protect\citeauthoryear{Hannachi et~al.}{2017}]{Hannachi2017}
\begin{barticle}
\bauthor{\bsnm{Hannachi}, \binits{A.}},
\bauthor{\bsnm{Straus}, \binits{D.M.}},
\bauthor{\bsnm{Franzke}, \binits{C.L.E.}},
\bauthor{\bsnm{Corti}, \binits{S.}},
\bauthor{\bsnm{Woollings}, \binits{T.}}:
\batitle{{Low-frequency nonlinearity and regime behavior in the Northern
  Hemisphere extratropical atmosphere}}.
\bjtitle{Reviews of Geophysics}
\bvolume{55}(\bissue{1}),
\bfpage{199}--\blpage{234}
(\byear{2017})
\doiurl{10.1002/2015RG000509}
{\href{https://arxiv.org/abs/https://agupubs.onlinelibrary.wiley.com/doi/pdf/10.1002/2015RG000509}{{https://agupubs.onlinelibrary.wiley.com/doi/pdf/10.1002/2015RG000509}}}
\end{barticle}
\endbibitem

\bibitem[\protect\citeauthoryear{Ghil and Robertson}{2002}]{Ghil2002}
\begin{barticle}
\bauthor{\bsnm{Ghil}, \binits{M.}},
\bauthor{\bsnm{Robertson}, \binits{A.W.}}:
\batitle{{{\textquotedblleft}Waves{\textquotedblright} vs.
  {\textquotedblleft}particles{\textquotedblright} in the
  atmosphere{\textquoteright}s phase space: A pathway to long-range
  forecasting?}}
\bjtitle{Proceedings of the National Academy of Sciences}
\bvolume{99}(\bissue{suppl 1}),
\bfpage{2493}--\blpage{2500}
(\byear{2002})
\doiurl{10.1073/pnas.012580899}
{\href{https://arxiv.org/abs/https://www.pnas.org/content/99/suppl\_1/2493.full.pdf}{{https://www.pnas.org/content/99/suppl\_1/2493.full.pdf}}}
\end{barticle}
\endbibitem

\bibitem[\protect\citeauthoryear{Masato et~al.}{2013}]{Masato2013}
\begin{barticle}
\bauthor{\bsnm{Masato}, \binits{G.}},
\bauthor{\bsnm{Hoskins}, \binits{B.J.}},
\bauthor{\bsnm{Woollings}, \binits{T.}}:
\batitle{{Winter and Summer Northern Hemisphere Blocking in CMIP5 Models}}.
\bjtitle{Journal of Climate}
\bvolume{26}(\bissue{18}),
\bfpage{7044}--\blpage{7059}
(\byear{2013})
\doiurl{10.1175/JCLI-D-12-00466.1}
\end{barticle}
\endbibitem

\bibitem[\protect\citeauthoryear{Fraedrich and B\"ottger}{1978}]{Fraedrich1978}
\begin{barticle}
\bauthor{\bsnm{Fraedrich}, \binits{K.}},
\bauthor{\bsnm{B\"ottger}, \binits{H.}}:
\batitle{{A Wavenumber-Frequency Analysis of the 500 mb Geopotential at
  50$^\circ$ N}}.
\bjtitle{Journal of Atmospheric Sciences}
\bvolume{35}(\bissue{4}),
\bfpage{745}--\blpage{750}
(\byear{1978})
\doiurl{10.1175/1520-0469(1978)035<0745:AWFAOT>2.0.CO;2}
\end{barticle}
\endbibitem

\bibitem[\protect\citeauthoryear{Dell'Aquila et~al.}{2005}]{DellAquila2005}
\begin{barticle}
\bauthor{\bsnm{Dell'Aquila}, \binits{A.}},
\bauthor{\bsnm{Lucarini}, \binits{V.}},
\bauthor{\bsnm{Ruti}, \binits{P.M.}},
\bauthor{\bsnm{Calmanti}, \binits{S.}}:
\batitle{{Hayashi spectra of the northern hemisphere mid-latitude atmospheric
  variability in the NCEP--NCAR and ECMWF reanalyses}}.
\bjtitle{Climate Dynamics}
\bvolume{25}(\bissue{6}),
\bfpage{639}--\blpage{652}
(\byear{2005})
\doiurl{10.1007/s00382-005-0048-x}
\end{barticle}
\endbibitem

\bibitem[\protect\citeauthoryear{Benzi et~al.}{1986}]{Benzi1986}
\begin{barticle}
\bauthor{\bsnm{Benzi}, \binits{R.}},
\bauthor{\bsnm{Malguzzi}, \binits{P.}},
\bauthor{\bsnm{Speranza}, \binits{A.}},
\bauthor{\bsnm{Sutera}, \binits{A.}}:
\batitle{{The statistical properties of general atmospheric circulation:
  observational evidence and a minimal theory of bimodality}}.
\bjtitle{{Q. J. R. Meteorol. Soc.}}
\bvolume{112}(\bissue{473}),
\bfpage{661}--\blpage{674}
(\byear{1986})
\doiurl{10.1256/smsqj.47305}
\end{barticle}
\endbibitem

\bibitem[\protect\citeauthoryear{Mo and Ghil}{1987}]{Mo.Ghil.1987}
\begin{barticle}
\bauthor{\bsnm{Mo}, \binits{K.C.}},
\bauthor{\bsnm{Ghil}, \binits{M.}}:
\batitle{{Statistics and Dynamics of Persistent Anomalies}}.
\bjtitle{{J. Atmos. Sci.}}
\bvolume{44}(\bissue{5}),
\bfpage{877}--\blpage{902}
(\byear{1987})
\doiurl{10.1175/1520-0469(1987)044<0877:sadopa>2.0.co;2}
\end{barticle}
\endbibitem

\bibitem[\protect\citeauthoryear{Ruti et~al.}{2006}]{Ruti2006}
\begin{botherref}
\oauthor{\bsnm{Ruti}, \binits{P.M.}},
\oauthor{\bsnm{Lucarini}, \binits{V.}},
\oauthor{\bsnm{Dell'Aquila}, \binits{A.}},
\oauthor{\bsnm{Calmanti}, \binits{S.}},
\oauthor{\bsnm{Speranza}, \binits{A.}}:
{Does the subtropical jet catalyze the midlatitude atmospheric regimes?}
Geophysical Research Letters
\textbf{33}(6)
(2006)
\doiurl{10.1029/2005GL024620}
{\href{https://arxiv.org/abs/https://agupubs.onlinelibrary.wiley.com/doi/pdf/10.1029/2005GL024620}{{https://agupubs.onlinelibrary.wiley.com/doi/pdf/10.1029/2005GL024620}}}
\end{botherref}
\endbibitem

\bibitem[\protect\citeauthoryear{Woollings et~al.}{2010}]{Woollings2010}
\begin{barticle}
\bauthor{\bsnm{Woollings}, \binits{T.}},
\bauthor{\bsnm{Hannachi}, \binits{A.}},
\bauthor{\bsnm{Hoskins}, \binits{B.}}:
\batitle{{Variability of the North Atlantic eddy-driven jet stream}}.
\bjtitle{Quarterly Journal of the Royal Meteorological Society}
\bvolume{136}(\bissue{649}),
\bfpage{856}--\blpage{868}
(\byear{2010})
\doiurl{10.1002/qj.625}
{\href{https://arxiv.org/abs/https://rmets.onlinelibrary.wiley.com/doi/pdf/10.1002/qj.625}{{https://rmets.onlinelibrary.wiley.com/doi/pdf/10.1002/qj.625}}}
\end{barticle}
\endbibitem

\bibitem[\protect\citeauthoryear{Rex}{1950}]{Rex1950}
\begin{barticle}
\bauthor{\bsnm{Rex}, \binits{D.F.}}:
\batitle{{Blocking Action in the Middle Troposphere and its Effect upon
  Regional Climate}}.
\bjtitle{Tellus}
\bvolume{2}(\bissue{3}),
\bfpage{196}--\blpage{211}
(\byear{1950})
\doiurl{10.3402/tellusa.v2i3.8546}
{\href{https://arxiv.org/abs/https://doi.org/10.3402/tellusa.v2i3.8546}{{https://doi.org/10.3402/tellusa.v2i3.8546}}}
\end{barticle}
\endbibitem

\bibitem[\protect\citeauthoryear{Rossby}{1951}]{Rossby1951}
\begin{barticle}
\bauthor{\bsnm{Rossby}, \binits{C.-G.}}:
\batitle{{On the dynamics of certain types of blocking waves}}.
\bjtitle{Journal of the Chinese Geophysical Society}
\bvolume{2},
\bfpage{1}--\blpage{13}
(\byear{1951})
\end{barticle}
\endbibitem

\bibitem[\protect\citeauthoryear{Hoskins}{1987}]{Hoskins1987}
\begin{bchapter}
\bauthor{\bsnm{Hoskins}, \binits{B.J.}}:
\bctitle{{Theories of blocking}}.
In: \bbtitle{{Seminar on the Nature and Prediction of Extra Tropical Weather
  Systems. 7-11 September 1987}},
vol. \bseriesno{II},
pp. \bfpage{1}--\blpage{10}.
\bpublisher{ECMWF},
\blocation{Shinfield Park, Reading}
(\byear{1987})
\end{bchapter}
\endbibitem

\bibitem[\protect\citeauthoryear{Woollings and Hoskins}{2008}]{Woollings2008}
\begin{barticle}
\bauthor{\bsnm{Woollings}, \binits{T.}},
\bauthor{\bsnm{Hoskins}, \binits{B.}}:
\batitle{Simultaneous atlantic–pacific blocking and the northern annular
  mode}.
\bjtitle{Quarterly Journal of the Royal Meteorological Society}
\bvolume{134}(\bissue{636}),
\bfpage{1635}--\blpage{1646}
(\byear{2008})
\doiurl{10.1002/qj.310}
{\href{https://arxiv.org/abs/https://rmets.onlinelibrary.wiley.com/doi/pdf/10.1002/qj.310}{{https://rmets.onlinelibrary.wiley.com/doi/pdf/10.1002/qj.310}}}
\end{barticle}
\endbibitem

\bibitem[\protect\citeauthoryear{Tibaldi and Molteni}{2018}]{Tibaldi2018}
\begin{botherref}
\oauthor{\bsnm{Tibaldi}, \binits{S.}},
\oauthor{\bsnm{Molteni}, \binits{F.}}:
Atmospheric Blocking in Observation and Models.
Oxford University Press
(2018).
\doiurl{10.1093/acrefore/9780190228620.013.611} .
\url{https://oxfordre.com/climatescience/view/10.1093/acrefore/9780190228620.001.0001/acrefore-9780190228620-e-611}
\end{botherref}
\endbibitem

\bibitem[\protect\citeauthoryear{Dole et~al.}{2011}]{Dole2011}
\begin{botherref}
\oauthor{\bsnm{Dole}, \binits{R.} et al.}:
{Was there a basis for anticipating the 2010 Russian heat wave?}
Geophysical Research Letters
\textbf{38}(6)
(2011)
\doiurl{10.1029/2010GL046582}
{\href{https://arxiv.org/abs/https://agupubs.onlinelibrary.wiley.com/doi/pdf/10.1029/2010GL046582}{{https://agupubs.onlinelibrary.wiley.com/doi/pdf/10.1029/2010GL046582}}}
\end{botherref}
\endbibitem

\bibitem[\protect\citeauthoryear{Xoplaki et~al.}{2012}]{Xoplaki2012}
\begin{bchapter}
\bauthor{\bsnm{Xoplaki}, \binits{E.} et al.}:
\bctitle{{Large-Scale Atmospheric Circulation Driving Extreme Climate Events in
  the Mediterranean and its Related Impacts}}.
In: \beditor{\bsnm{Lionello}, \binits{P.}} (ed.)
\bbtitle{{The Climate of the Mediterranean Region}},
pp. \bfpage{347}--\blpage{417}.
\bpublisher{Elsevier},
\blocation{Oxford}
(\byear{2012}).
\doiurl{10.1016/B978-0-12-416042-2.00006-9} .
\burl{https://www.sciencedirect.com/science/article/pii/B9780124160422000069}
\end{bchapter}
\endbibitem

\bibitem[\protect\citeauthoryear{Lau and Kim}{2012}]{Lau2012}
\begin{barticle}
\bauthor{\bsnm{Lau}, \binits{W.K.M.}},
\bauthor{\bsnm{Kim}, \binits{K.-M.}}:
\batitle{{The 2010 Pakistan Flood and Russian Heat Wave: Teleconnection of
  Hydrometeorological Extremes}}.
\bjtitle{Journal of Hydrometeorology}
\bvolume{13}(\bissue{1}),
\bfpage{392}--\blpage{403}
(\byear{2012})
\doiurl{10.1175/JHM-D-11-016.1}
{\href{https://arxiv.org/abs/https://journals.ametsoc.org/jhm/article-pdf/13/1/392/4110762/jhm-d-11-016\_1.pdf}{{https://journals.ametsoc.org/jhm/article-pdf/13/1/392/4110762/jhm-d-11-016\_1.pdf}}}
\end{barticle}
\endbibitem

\bibitem[\protect\citeauthoryear{Buehler et~al.}{2011}]{Buehler2011}
\begin{barticle}
\bauthor{\bsnm{Buehler}, \binits{T.}},
\bauthor{\bsnm{Raible}, \binits{C.C.}},
\bauthor{\bsnm{Stocker}, \binits{T.F.}}:
\batitle{{The relationship of winter season North Atlantic blocking frequencies
  to extreme cold or dry spells in the ERA-40}}.
\bjtitle{Tellus A}
\bvolume{63}(\bissue{2}),
\bfpage{212}--\blpage{222}
(\byear{2011})
\doiurl{10.1111/j.1600-0870.2010.00492.x}
{\href{https://arxiv.org/abs/https://onlinelibrary.wiley.com/doi/pdf/10.1111/j.1600-0870.2010.00492.x}{{https://onlinelibrary.wiley.com/doi/pdf/10.1111/j.1600-0870.2010.00492.x}}}
\end{barticle}
\endbibitem

\bibitem[\protect\citeauthoryear{Hoskins and Woollings}{2015}]{Hoskins2015}
\begin{barticle}
\bauthor{\bsnm{Hoskins}, \binits{B.}},
\bauthor{\bsnm{Woollings}, \binits{T.}}:
\batitle{{Persistent Extratropical Regimes and Climate Extremes}}.
\bjtitle{Curr. Clim. Change Rep.}
\bvolume{1},
\bfpage{115}--\blpage{124}
(\byear{2015})
\doiurl{10.1007/s40641-015-0020-8}
\end{barticle}
\endbibitem

\bibitem[\protect\citeauthoryear{G{\'a}lfi et~al.}{2019}]{Galfi2019}
\begin{barticle}
\bauthor{\bsnm{G{\'a}lfi}, \binits{V.M.}},
\bauthor{\bsnm{Lucarini}, \binits{V.}},
\bauthor{\bsnm{Wouters}, \binits{J.}}:
\batitle{{A large deviation theory-based analysis of heat waves and cold spells
  in a simplified model of the general circulation of the atmosphere}}.
\bjtitle{Journal of Statistical Mechanics: Theory and Experiment}
\bvolume{2019}(\bissue{3}),
\bfpage{033404}
(\byear{2019})
\end{barticle}
\endbibitem

\bibitem[\protect\citeauthoryear{Kautz et~al.}{2022}]{Kautz2022}
\begin{barticle}
\bauthor{\bsnm{Kautz}, \binits{L.-A.}},
\bauthor{\bsnm{Martius}, \binits{O.}},
\bauthor{\bsnm{Pfahl}, \binits{S.}},
\bauthor{\bsnm{Pinto}, \binits{J.G.}},
\bauthor{\bsnm{Ramos}, \binits{A.M.}},
\bauthor{\bsnm{Sousa}, \binits{P.M.}},
\bauthor{\bsnm{Woollings}, \binits{T.}}:
\batitle{{Atmospheric blocking and weather extremes over the Euro-Atlantic
  sector -- a review}}.
\bjtitle{Weather and Climate Dynamics}
\bvolume{3}(\bissue{1}),
\bfpage{305}--\blpage{336}
(\byear{2022})
\doiurl{10.5194/wcd-3-305-2022}
\end{barticle}
\endbibitem

\bibitem[\protect\citeauthoryear{Lucarini et~al.}{2023}]{Lucarini2023}
\begin{barticle}
\bauthor{\bsnm{Lucarini}, \binits{V.}},
\bauthor{\bsnm{Galfi}, \binits{V.M.}},
\bauthor{\bsnm{Riboldi}, \binits{J.}},
\bauthor{\bsnm{Messori}, \binits{G.}}:
\batitle{{Typicality of the 2021 Western North America summer heatwave}}.
\bjtitle{Environmental Research Letters}
\bvolume{18}(\bissue{1}),
\bfpage{015004}
(\byear{2023})
\doiurl{10.1088/1748-9326/acab77}
\end{barticle}
\endbibitem

\bibitem[\protect\citeauthoryear{Ferranti et~al.}{2015}]{Ferranti2015}
\begin{barticle}
\bauthor{\bsnm{Ferranti}, \binits{L.}},
\bauthor{\bsnm{Corti}, \binits{S.}},
\bauthor{\bsnm{Janousek}, \binits{M.}}:
\batitle{{Flow-dependent verification of the ECMWF ensemble over the
  Euro-Atlantic sector}}.
\bjtitle{Quarterly Journal of the Royal Meteorological Society}
\bvolume{141}(\bissue{688}),
\bfpage{916}--\blpage{924}
(\byear{2015})
\doiurl{10.1002/qj.2411}
{\href{https://arxiv.org/abs/https://rmets.onlinelibrary.wiley.com/doi/pdf/10.1002/qj.2411}{{https://rmets.onlinelibrary.wiley.com/doi/pdf/10.1002/qj.2411}}}
\end{barticle}
\endbibitem

\bibitem[\protect\citeauthoryear{Lupo}{2021}]{Lupo2020}
\begin{barticle}
\bauthor{\bsnm{Lupo}, \binits{A.R.}}:
\batitle{{Atmospheric blocking events: a review}}.
\bjtitle{Annals of the New York Academy of Sciences}
\bvolume{1504}(\bissue{1}),
\bfpage{5}--\blpage{24}
(\byear{2021})
\doiurl{10.1111/nyas.14557}
{\href{https://arxiv.org/abs/https://nyaspubs.onlinelibrary.wiley.com/doi/pdf/10.1111/nyas.14557}{{https://nyaspubs.onlinelibrary.wiley.com/doi/pdf/10.1111/nyas.14557}}}
\end{barticle}
\endbibitem

\bibitem[\protect\citeauthoryear{Lucarini et~al.}{2007}]{Lucarini2007}
\begin{barticle}
\bauthor{\bsnm{Lucarini}, \binits{V.}},
\bauthor{\bsnm{Calmanti}, \binits{S.}},
\bauthor{\bsnm{Dell'Aquila}, \binits{A.}},
\bauthor{\bsnm{Ruti}, \binits{P.M.}},
\bauthor{\bsnm{Speranza}, \binits{A.}}:
\batitle{{Intercomparison of the northern hemisphere winter mid-latitude
  atmospheric variability of the IPCC models}}.
\bjtitle{Climate Dynamics}
\bvolume{28}(\bissue{7}),
\bfpage{829}--\blpage{848}
(\byear{2007})
\doiurl{10.1007/s00382-006-0213-x}
\end{barticle}
\endbibitem

\bibitem[\protect\citeauthoryear{Di~Biagio et~al.}{2014}]{DiBiagio2014}
\begin{barticle}
\bauthor{\bsnm{Di~Biagio}, \binits{V.}},
\bauthor{\bsnm{Calmanti}, \binits{S.}},
\bauthor{\bsnm{Dell'Aquila}, \binits{A.}},
\bauthor{\bsnm{Ruti}, \binits{P.M.}}:
\batitle{{Northern Hemisphere winter midlatitude atmospheric variability in
  CMIP5 models}}.
\bjtitle{Geophysical Research Letters}
\bvolume{41}(\bissue{4}),
\bfpage{1277}--\blpage{1282}
(\byear{2014})
\doiurl{10.1002/2013GL058928}
{\href{https://arxiv.org/abs/https://agupubs.onlinelibrary.wiley.com/doi/pdf/10.1002/2013GL058928}{{https://agupubs.onlinelibrary.wiley.com/doi/pdf/10.1002/2013GL058928}}}
\end{barticle}
\endbibitem

\bibitem[\protect\citeauthoryear{Davini and D'Andrea}{2016}]{Davini2016}
\begin{barticle}
\bauthor{\bsnm{Davini}, \binits{P.}},
\bauthor{\bsnm{D'Andrea}, \binits{F.}}:
\batitle{{Northern Hemisphere Atmospheric Blocking Representation in Global
  Climate Models: Twenty Years of Improvements?}}
\bjtitle{Journal of Climate}
\bvolume{29}(\bissue{24}),
\bfpage{8823}--\blpage{8840}
(\byear{2016})
\doiurl{10.1175/JCLI-D-16-0242.1}
\end{barticle}
\endbibitem

\bibitem[\protect\citeauthoryear{Woollings et~al.}{2018}]{Woolings2018}
\begin{barticle}
\bauthor{\bsnm{Woollings}, \binits{T.}},
\bauthor{\bsnm{Barriopedro}, \binits{D.}},
\bauthor{\bsnm{Methven}, \binits{J.}},
\bauthor{\bsnm{Son}, \binits{S.-W.}},
\bauthor{\bsnm{Martius}, \binits{O.}},
\bauthor{\bsnm{Harvey}, \binits{B.}},
\bauthor{\bsnm{Sillmann}, \binits{J.}},
\bauthor{\bsnm{Lupo}, \binits{A.R.}},
\bauthor{\bsnm{Seneviratne}, \binits{S.}}:
\batitle{{Blocking and its Response to Climate Change}}.
\bjtitle{Current Climate Change Reports}
\bvolume{4}(\bissue{3}),
\bfpage{287}--\blpage{300}
(\byear{2018})
\doiurl{10.1007/s40641-018-0108-z}
\end{barticle}
\endbibitem

\bibitem[\protect\citeauthoryear{Nabizadeh et~al.}{2019}]{Nabizadeh2019}
\begin{barticle}
\bauthor{\bsnm{Nabizadeh}, \binits{E.}},
\bauthor{\bsnm{Hassanzadeh}, \binits{P.}},
\bauthor{\bsnm{Yang}, \binits{D.}},
\bauthor{\bsnm{Barnes}, \binits{E.A.}}:
\batitle{{Size of the Atmospheric Blocking Events: Scaling Law and Response to
  Climate Change}}.
\bjtitle{Geophysical Research Letters}
\bvolume{46}(\bissue{22}),
\bfpage{13488}--\blpage{13499}
(\byear{2019})
\doiurl{10.1029/2019GL084863}
{\href{https://arxiv.org/abs/https://agupubs.onlinelibrary.wiley.com/doi/pdf/10.1029/2019GL084863}{{https://agupubs.onlinelibrary.wiley.com/doi/pdf/10.1029/2019GL084863}}}
\end{barticle}
\endbibitem

\bibitem[\protect\citeauthoryear{Steinfeld et~al.}{2022}]{Steinfeld_2022}
\begin{barticle}
\bauthor{\bsnm{Steinfeld}, \binits{D.}},
\bauthor{\bsnm{Sprenger}, \binits{M.}},
\bauthor{\bsnm{Beyerle}, \binits{U.}},
\bauthor{\bsnm{Pfahl}, \binits{S.}}:
\batitle{{Response of moist and dry processes in atmospheric blocking to
  climate change}}.
\bjtitle{Environmental Research Letters}
\bvolume{17}(\bissue{8}),
\bfpage{084020}
(\byear{2022})
\doiurl{10.1088/1748-9326/ac81af}
\end{barticle}
\endbibitem

\bibitem[\protect\citeauthoryear{Dorrington et~al.}{2022}]{Josh22}
\begin{botherref}
\oauthor{\bsnm{Dorrington}, \binits{J.}},
\oauthor{\bsnm{Strommen}, \binits{K.}},
\oauthor{\bsnm{Fabiano}, \binits{F.}},
\oauthor{\bsnm{Molteni}, \binits{F.}}:
{CMIP6 Models Trend Toward Less Persistent European Blocking Regimes in a
  Warming Climate}.
Geophysical Research Letters
\textbf{49}(24)
(2022)
\doiurl{10.1029/2022GL100811}
\end{botherref}
\endbibitem

\bibitem[\protect\citeauthoryear{Seneviratne
  et~al.}{2021}]{seneviratne2021weather}
\begin{bchapter}
\bauthor{\bsnm{Seneviratne}, \binits{S.I.} et al}:
(eds.)
\bbtitle{Climate Change 2021: The Physical Science Basis. Contribution of
  Working Group I to the Sixth Assessment Report of the Intergovernmental Panel
  on Climate Change},
pp. \bfpage{1513}--\blpage{1766}.
\bpublisher{Cambridge University Press},
\blocation{Cambridge, United Kingdom and New York, NY, USA}
(\byear{2021}).
\doiurl{10.1017/9781009157896.013}
\end{bchapter}
\endbibitem

\bibitem[\protect\citeauthoryear{Vannitsem}{2001}]{Vannitsem2001}
\begin{barticle}
\bauthor{\bsnm{Vannitsem}, \binits{S.}}:
\batitle{{Toward a phase-space cartography of the short- and medium-range
  predictability of weather regimes}}.
\bjtitle{Tellus A: Dynamic Meteorology and Oceanography}
\bvolume{53}(\bissue{1}),
\bfpage{56}--\blpage{73}
(\byear{2001})
\doiurl{10.3402/tellusa.v53i1.12180}
{\href{https://arxiv.org/abs/https://doi.org/10.3402/tellusa.v53i1.12180}{{https://doi.org/10.3402/tellusa.v53i1.12180}}}
\end{barticle}
\endbibitem

\bibitem[\protect\citeauthoryear{Schubert and Lucarini}{2016}]{Schubert2016}
\begin{barticle}
\bauthor{\bsnm{Schubert}, \binits{S.}},
\bauthor{\bsnm{Lucarini}, \binits{V.}}:
\batitle{{Dynamical analysis of blocking events: spatial and temporal
  fluctuations of covariant Lyapunov vectors}}.
\bjtitle{Quarterly Journal of the Royal Meteorological Society}
\bvolume{142}(\bissue{698}),
\bfpage{2143}--\blpage{2158}
(\byear{2016})
\doiurl{10.1002/qj.2808}
{\href{https://arxiv.org/abs/https://rmets.onlinelibrary.wiley.com/doi/pdf/10.1002/qj.2808}{{https://rmets.onlinelibrary.wiley.com/doi/pdf/10.1002/qj.2808}}}
\end{barticle}
\endbibitem

\bibitem[\protect\citeauthoryear{Faranda et~al.}{2017}]{Faranda2017}
\begin{barticle}
\bauthor{\bsnm{Faranda}, \binits{D.}},
\bauthor{\bsnm{Messori}, \binits{G.}},
\bauthor{\bsnm{Yiou}, \binits{P.}}:
\batitle{{Dynamical proxies of North Atlantic predictability and extremes}}.
\bjtitle{Scientific Reports}
\bvolume{7}(\bissue{1}),
\bfpage{41278}
(\byear{2017})
\doiurl{10.1038/srep41278}
\end{barticle}
\endbibitem

\bibitem[\protect\citeauthoryear{Lucarini and Gritsun}{2020}]{Lucarini20}
\begin{botherref}
\oauthor{\bsnm{Lucarini}, \binits{V.}},
\oauthor{\bsnm{Gritsun}, \binits{A.}}:
{A new mathematical framework for atmospheric blocking events}.
Climate Dynamics
\textbf{52}
(2020)
\doiurl{10.1007/s00382-019-05018-2}
\end{botherref}
\endbibitem

\bibitem[\protect\citeauthoryear{Tibaldi and Molteni}{1990}]{Tibaldi1990}
\begin{barticle}
\bauthor{\bsnm{Tibaldi}, \binits{S.}},
\bauthor{\bsnm{Molteni}, \binits{F.}}:
\batitle{On the operational predictability of blocking}.
\bjtitle{Tellus A}
\bvolume{42}(\bissue{3}),
\bfpage{343}--\blpage{365}
(\byear{1990})
\doiurl{10.1034/j.1600-0870.1990.t01-2-00003.x}
{\href{https://arxiv.org/abs/https://onlinelibrary.wiley.com/doi/pdf/10.1034/j.1600-0870.1990.t01-2-00003.x}{{https://onlinelibrary.wiley.com/doi/pdf/10.1034/j.1600-0870.1990.t01-2-00003.x}}}
\end{barticle}
\endbibitem

\bibitem[\protect\citeauthoryear{Pelly and Hoskins}{2003}]{Pelly2003a}
\begin{barticle}
\bauthor{\bsnm{Pelly}, \binits{J.L.}},
\bauthor{\bsnm{Hoskins}, \binits{B.J.}}:
\batitle{A new perspective on blocking}.
\bjtitle{Journal of the Atmospheric Sciences}
\bvolume{60}(\bissue{5}),
\bfpage{743}--\blpage{755}
(\byear{2003})
\doiurl{10.1175/1520-0469(2003)060<0743:ANPOB>2.0.CO;2}
\end{barticle}
\endbibitem

\bibitem[\protect\citeauthoryear{Barriopedro et~al.}{2006}]{Barriopedro2006}
\begin{barticle}
\bauthor{\bsnm{Barriopedro}, \binits{D.}},
\bauthor{\bsnm{Garc\'{i}a-Herrera}, \binits{R.}},
\bauthor{\bsnm{Lupo}, \binits{A.R.}},
\bauthor{\bsnm{Hern\'{a}ndez}, \binits{E.}}:
\batitle{{A Climatology of Northern Hemisphere Blocking}}.
\bjtitle{Journal of Climate}
\bvolume{19}(\bissue{6}),
\bfpage{1042}--\blpage{1063}
(\byear{2006})
\doiurl{10.1175/JCLI3678.1}
\end{barticle}
\endbibitem

\bibitem[\protect\citeauthoryear{Davini et~al.}{2012}]{Davini2012}
\begin{barticle}
\bauthor{\bsnm{Davini}, \binits{P.}},
\bauthor{\bsnm{Cagnazzo}, \binits{C.}},
\bauthor{\bsnm{Gualdi}, \binits{S.}},
\bauthor{\bsnm{Navarra}, \binits{A.}}:
\batitle{{Bidimensional Diagnostics, Variability, and Trends of Northern
  Hemisphere Blocking}}.
\bjtitle{Journal of Climate}
\bvolume{25}(\bissue{19}),
\bfpage{6496}--\blpage{6509}
(\byear{2012})
\doiurl{10.1175/JCLI-D-12-00032.1}
\end{barticle}
\endbibitem

\bibitem[\protect\citeauthoryear{Pinheiro et~al.}{2019}]{Pinheiro2019}
\begin{barticle}
\bauthor{\bsnm{Pinheiro}, \binits{M.C.}},
\bauthor{\bsnm{Ullrich}, \binits{P.A.}},
\bauthor{\bsnm{Grotjahn}, \binits{R.}}:
\batitle{{Atmospheric blocking and intercomparison of objective detection
  methods: flow field characteristics}}.
\bjtitle{Climate Dynamics}
\bvolume{53}(\bissue{7}),
\bfpage{4189}--\blpage{4216}
(\byear{2019})
\doiurl{10.1007/s00382-019-04782-5}
\end{barticle}
\endbibitem

\bibitem[\protect\citeauthoryear{Wallace and Gutzler}{1981}]{Wallace1981}
\begin{barticle}
\bauthor{\bsnm{Wallace}, \binits{J.M.}},
\bauthor{\bsnm{Gutzler}, \binits{D.S.}}:
\batitle{{Teleconnections in the Geopotential Height Field during the Northern
  Hemisphere Winter}}.
\bjtitle{Monthly Weather Review}
\bvolume{109}(\bissue{4}),
\bfpage{784}--\blpage{812}
(\byear{1981})
\doiurl{10.1175/1520-0493(1981)109<0784:TITGHF>2.0.CO;2}
\end{barticle}
\endbibitem

\bibitem[\protect\citeauthoryear{Hurrell}{1995}]{Hurrell1995}
\begin{barticle}
\bauthor{\bsnm{Hurrell}, \binits{J.W.}}:
\batitle{Decadal trends in the north atlantic oscillation: Regional
  temperatures and precipitation}.
\bjtitle{Science}
\bvolume{269}(\bissue{5224}),
\bfpage{676}--\blpage{679}
(\byear{1995}).
Accessed 2023-07-24
\end{barticle}
\endbibitem

\bibitem[\protect\citeauthoryear{Hurrell et~al.}{2003}]{Hurrell2003}
\begin{bbook}
\bauthor{\bsnm{Hurrell}, \binits{J.W.}},
\bauthor{\bsnm{Kushnir}, \binits{Y.}},
\bauthor{\bsnm{Ottersen}, \binits{G.}},
\bauthor{\bsnm{Visbeck}, \binits{M.}}:
In: \beditor{\bsnm{Hurrell}, \binits{J.W.}},
\beditor{\bsnm{Kushnir}, \binits{Y.}},
\beditor{\bsnm{Ottersen}, \binits{G.}},
\beditor{\bsnm{Visbeck}, \binits{M.}} (eds.)
\bbtitle{{An Overview of the North Atlantic Oscillation}},
pp. \bfpage{1}--\blpage{35}.
\bpublisher{American Geophysical Union (AGU)}, 
(\byear{2003}).
\doiurl{10.1029/134GM01}
\end{bbook}
\endbibitem

\bibitem[\protect\citeauthoryear{Thompson and Wallace}{1998}]{Thompson1998}
\begin{barticle}
\bauthor{\bsnm{Thompson}, \binits{D.W.J.}},
\bauthor{\bsnm{Wallace}, \binits{J.M.}}:
\batitle{{The Arctic oscillation signature in the wintertime geopotential
  height and temperature fields}}.
\bjtitle{Geophysical Research Letters}
\bvolume{25}(\bissue{9}),
\bfpage{1297}--\blpage{1300}
(\byear{1998})
\doiurl{10.1029/98GL00950}
{\href{https://arxiv.org/abs/https://agupubs.onlinelibrary.wiley.com/doi/pdf/10.1029/98GL00950}{{https://agupubs.onlinelibrary.wiley.com/doi/pdf/10.1029/98GL00950}}}
\end{barticle}
\endbibitem

\bibitem[\protect\citeauthoryear{Branstator}{2002}]{Branstator2002}
\begin{barticle}
\bauthor{\bsnm{Branstator}, \binits{G.}}:
\batitle{{Circumglobal teleconnections, the jet stream waveguide, and the North
  Atlantic Oscillation}}.
\bjtitle{Journal of Climate}
\bvolume{15},
\bfpage{1893}--\blpage{1910}
(\byear{2002})
\doiurl{10.1175/1520-0442(2002)015<1893:CTTJSW>2.0.CO;2}
\end{barticle}
\endbibitem

\bibitem[\protect\citeauthoryear{Harnik et~al.}{2016}]{Harnik2016}
\begin{barticle}
\bauthor{\bsnm{Harnik}, \binits{N.}},
\bauthor{\bsnm{Messori}, \binits{G.}},
\bauthor{\bsnm{Caballero}, \binits{R.}},
\bauthor{\bsnm{Feldstein}, \binits{S.B.}}:
\batitle{{The Circumglobal North American wave pattern and its relation to cold
  events in eastern North America}}.
\bjtitle{Geophysical Research Letters}
\bvolume{43},
\bfpage{11015}--\blpage{11023}
(\byear{2016})
\doiurl{10.1002/2016GL070760}
\end{barticle}
\endbibitem

\bibitem[\protect\citeauthoryear{Croci-Maspoli et~al.}{2007}]{croci2007}
\begin{barticle}
\bauthor{\bsnm{Croci-Maspoli}, \binits{M.}},
\bauthor{\bsnm{Schwierz}, \binits{C.}},
\bauthor{\bsnm{Davies}, \binits{H.C.}}:
\batitle{{Atmospheric blocking: space-time links to the NAO and PNA}}.
\bjtitle{Climate Dynamics}
\bvolume{29}(\bissue{7}),
\bfpage{713}--\blpage{725}
(\byear{2007})
\doiurl{10.1007/s00382-007-0259-4}
\end{barticle}
\endbibitem

\bibitem[\protect\citeauthoryear{Athanasiadis et~al.}{2010}]{Athanasiadis2010}
\begin{barticle}
\bauthor{\bsnm{Athanasiadis}, \binits{P.J.}},
\bauthor{\bsnm{Wallace}, \binits{J.M.}},
\bauthor{\bsnm{Wettstein}, \binits{J.J.}}:
\batitle{{Patterns of Wintertime Jet Stream Variability and Their Relation to
  the Storm Tracks}}.
\bjtitle{Journal of the Atmospheric Sciences}
\bvolume{67}(\bissue{5}),
\bfpage{1361}--\blpage{1381}
(\byear{2010})
\doiurl{10.1175/2009JAS3270.1}
\end{barticle}
\endbibitem

\bibitem[\protect\citeauthoryear{Feldstein and
  Franzke}{2017}]{feldstein_franzke_2017}
\begin{bbook}
\bauthor{\bsnm{Feldstein}, \binits{S.B.}},
\bauthor{\bsnm{Franzke}, \binits{C.L.E.}}:
In: \beditor{\bsnm{Franzke}, \binits{C.L.E.}},
\beditor{\bsnm{O’Kane}, \binits{T.J.}} (eds.)
\bbtitle{Atmospheric Teleconnection Patterns},
pp. \bfpage{54}--\blpage{104}.
\bpublisher{Cambridge University Press},
\blocation{Cambridge}
(\byear{2017})
\end{bbook}
\endbibitem

\bibitem[\protect\citeauthoryear{Baur et~al.}{1944}]{Baur1944}
\begin{botherref}
\oauthor{\bsnm{Baur}, \binits{F.}},
\oauthor{\bsnm{Hess}, \binits{P.}},
\oauthor{\bsnm{Nagel}, \binits{H.}}:
Kalender der großwetterlagen europas 1881–1939.
Technical report,
DWD: Bad Homburg
(1944)
\end{botherref}
\endbibitem

\bibitem[\protect\citeauthoryear{Vautard}{1990}]{Vautard1990}
\begin{barticle}
\bauthor{\bsnm{Vautard}, \binits{R.}}:
\batitle{{Multiple Weather Regimes over the North Atlantic: Analysis of
  Precursors and Successors}}.
\bjtitle{Monthly Weather Review}
\bvolume{118}(\bissue{10}),
\bfpage{2056}--\blpage{2081}
(\byear{1990})
\doiurl{10.1175/1520-0493(1990)118<2056:MWROTN>2.0.CO;2}
\end{barticle}
\endbibitem

\bibitem[\protect\citeauthoryear{Madonna et~al.}{2017}]{Madonna2017}
\begin{barticle}
\bauthor{\bsnm{Madonna}, \binits{E.}},
\bauthor{\bsnm{Li}, \binits{C.}},
\bauthor{\bsnm{Grams}, \binits{C.M.}},
\bauthor{\bsnm{Woollings}, \binits{T.}}:
\batitle{{The link between eddy-driven jet variability and weather regimes in
  the North Atlantic-European sector}}.
\bjtitle{Quarterly Journal of the Royal Meteorological Society}
\bvolume{143},
\bfpage{2960}--\blpage{2972}
(\byear{2017})
\doiurl{10.1002/qj.3155}
\end{barticle}
\endbibitem

\bibitem[\protect\citeauthoryear{Franzke}{2013}]{Franzke2013}
\begin{botherref}
\oauthor{\bsnm{Franzke}, \binits{C.L.E.}}:
{Persistent regimes and extreme events of the North Atlantic atmospheric
  circulation}.
Philosophical Transactions of the Royal Society A: Mathematical, Physical and
  Engineering Sciences
\textbf{371}
(2013)
\doiurl{10.1098/rsta.2011.0471}
\end{botherref}
\endbibitem

\bibitem[\protect\citeauthoryear{De~Luca et~al.}{2019}]{DeLuca2019}
\begin{botherref}
\oauthor{\bsnm{De~Luca}, \binits{P.} et al.}:
{Past and Projected Weather Pattern Persistence with Associated Multi-Hazards
  in the British Isles}.
Atmosphere
\textbf{10}(10)
(2019)
\doiurl{10.3390/atmos10100577}
\end{botherref}
\endbibitem

\bibitem[\protect\citeauthoryear{Madonna et~al.}{2021}]{Madonna2021}
\begin{barticle}
\bauthor{\bsnm{Madonna}, \binits{E.}},
\bauthor{\bsnm{Battisti}, \binits{D.S.}},
\bauthor{\bsnm{Li}, \binits{C.}},
\bauthor{\bsnm{White}, \binits{R.H.}}:
\batitle{{Reconstructing winter climate anomalies in the Euro-Atlantic sector
  using circulation patterns}}.
\bjtitle{Weather and Climate Dynamics}
\bvolume{2},
\bfpage{777}--\blpage{794}
(\byear{2021})
\doiurl{10.5194/wcd-2-777-2021}
\end{barticle}
\endbibitem

\bibitem[\protect\citeauthoryear{Galfi and Messori}{2023}]{Galfi2023}
\begin{botherref}
\oauthor{\bsnm{Galfi}, \binits{V.M.}},
\oauthor{\bsnm{Messori}, \binits{G.}}:
{Persistent anomalies of the North Atlantic jet stream and associated surface
  extremes over Europe}.
Environmental Research Letters
\textbf{18}
(2023)
\doiurl{10.1088/1748-9326/acaedf}
\end{botherref}
\endbibitem

\bibitem[\protect\citeauthoryear{Fabiano et~al.}{2020}]{Fabiano2020}
\begin{barticle}
\bauthor{\bsnm{Fabiano}, \binits{F.}},
\bauthor{\bsnm{Christensen}, \binits{H.M.}},
\bauthor{\bsnm{Strommen}, \binits{K.}},
\bauthor{\bsnm{Athanasiadis}, \binits{P.}},
\bauthor{\bsnm{Baker}, \binits{A.}},
\bauthor{\bsnm{Schiemann}, \binits{R.}},
\bauthor{\bsnm{Corti}, \binits{S.}}:
\batitle{{Euro-Atlantic weather Regimes in the PRIMAVERA coupled climate
  simulations: impact of resolution and mean state biases on model
  performance}}.
\bjtitle{Climate Dynamics}
\bvolume{54}(\bissue{11}),
\bfpage{5031}--\blpage{5048}
(\byear{2020})
\doiurl{10.1007/s00382-020-05271-w}
\end{barticle}
\endbibitem

\bibitem[\protect\citeauthoryear{Franzke et~al.}{2008}]{Franzke2008}
\begin{barticle}
\bauthor{\bsnm{Franzke}, \binits{C.}},
\bauthor{\bsnm{Crommelin}, \binits{D.}},
\bauthor{\bsnm{Fischer}, \binits{A.}},
\bauthor{\bsnm{Majda}, \binits{A.J.}}:
\batitle{A hidden markov model perspective on regimes and metastability in
  atmospheric flows}.
\bjtitle{Journal of Climate}
\bvolume{21}(\bissue{8}),
\bfpage{1740}--\blpage{1757}
(\byear{2008})
\doiurl{10.1175/2007JCLI1751.1}
\end{barticle}
\endbibitem

\bibitem[\protect\citeauthoryear{Kwasniok}{2014}]{Kwasniok2014}
\begin{barticle}
\bauthor{\bsnm{Kwasniok}, \binits{F.}}:
\batitle{{Enhanced regime predictability in atmospheric low-order models due to
  stochastic forcing}}.
\bjtitle{Philosophical Transactions of the Royal Society A: Mathematical,
  Physical and Engineering Sciences}
\bvolume{372}(\bissue{2018}),
\bfpage{20130286}
(\byear{2014})
\doiurl{10.1098/rsta.2013.0286}
{\href{https://arxiv.org/abs/https://royalsocietypublishing.org/doi/pdf/10.1098/rsta.2013.0286}{{https://royalsocietypublishing.org/doi/pdf/10.1098/rsta.2013.0286}}}
\end{barticle}
\endbibitem

\bibitem[\protect\citeauthoryear{Tantet et~al.}{2015}]{Tantet2015}
\begin{barticle}
\bauthor{\bsnm{Tantet}, \binits{A.}},
\bauthor{\bsnm{Burgt}, \binits{F.R.}},
\bauthor{\bsnm{Dijkstra}, \binits{H.A.}}:
\batitle{{An early warning indicator for atmospheric blocking events using
  transfer operators}}.
\bjtitle{Chaos: An Interdisciplinary Journal of Nonlinear Science}
\bvolume{25}(\bissue{3}),
\bfpage{036406}
(\byear{2015})
\doiurl{10.1063/1.4908174}
{\href{https://arxiv.org/abs/https://pubs.aip.org/aip/cha/article-pdf/doi/10.1063/1.4908174/14610546/036406\_1\_online.pdf}{{https://pubs.aip.org/aip/cha/article-pdf/doi/10.1063/1.4908174/14610546/036406\_1\_online.pdf}}}
\end{barticle}
\endbibitem

\bibitem[\protect\citeauthoryear{Detring et~al.}{2021}]{Detring2021}
\begin{barticle}
\bauthor{\bsnm{Detring}, \binits{C.}},
\bauthor{\bsnm{M\"uller}, \binits{A.}},
\bauthor{\bsnm{Schielicke}, \binits{L.}},
\bauthor{\bsnm{N\'evir}, \binits{P.}},
\bauthor{\bsnm{Rust}, \binits{H.W.}}:
\batitle{{Occurrence and transition probabilities of omega and high-over-low
  blocking in the Euro-Atlantic region}}.
\bjtitle{Weather and Climate Dynamics}
\bvolume{2}(\bissue{4}),
\bfpage{927}--\blpage{952}
(\byear{2021})
\doiurl{10.5194/wcd-2-927-2021}
\end{barticle}
\endbibitem

\bibitem[\protect\citeauthoryear{Mukhin et~al.}{2022}]{Mukhin2022}
\begin{barticle}
\bauthor{\bsnm{Mukhin}, \binits{D.}},
\bauthor{\bsnm{Hannachi}, \binits{A.}},
\bauthor{\bsnm{Braun}, \binits{T.}},
\bauthor{\bsnm{Marwan}, \binits{N.}}:
\batitle{{Revealing recurrent regimes of mid-latitude atmospheric variability
  using novel machine learning method}}.
\bjtitle{Chaos: An Interdisciplinary Journal of Nonlinear Science}
\bvolume{32}(\bissue{11}),
\bfpage{113105}
(\byear{2022})
\doiurl{10.1063/5.0109889}
{\href{https://arxiv.org/abs/https://pubs.aip.org/aip/cha/article-pdf/doi/10.1063/5.0109889/16497678/113105\_1\_online.pdf}{{https://pubs.aip.org/aip/cha/article-pdf/doi/10.1063/5.0109889/16497678/113105\_1\_online.pdf}}}
\end{barticle}
\endbibitem

\bibitem[\protect\citeauthoryear{Prinz et~al.}{2011}]{Noe11}
\begin{barticle}
\bauthor{\bsnm{Prinz}, \binits{J.-H.} et al.}:
\batitle{{Markov models of molecular kinetics: Generation and validation}}.
\bjtitle{The Journal of Chemical Physics}
\bvolume{134}(\bissue{17}),
\bfpage{174105}
(\byear{2011})
\doiurl{10.1063/1.3565032}
{\href{https://arxiv.org/abs/https://pubs.aip.org/aip/jcp/article-pdf/doi/10.1063/1.3565032/15437339/174105\_1\_online.pdf}{{https://pubs.aip.org/aip/jcp/article-pdf/doi/10.1063/1.3565032/15437339/174105\_1\_online.pdf}}}
\end{barticle}
\endbibitem

\bibitem[\protect\citeauthoryear{Husic and Pande}{2018}]{Brooke18}
\begin{barticle}
\bauthor{\bsnm{Husic}, \binits{B.E.}},
\bauthor{\bsnm{Pande}, \binits{V.S.}}:
\batitle{{Markov State Models: From an Art to a Science}}.
\bjtitle{Journal of the American Chemical Society}
\bvolume{140}(\bissue{7}),
\bfpage{2386}--\blpage{2396}
(\byear{2018})
\doiurl{10.1021/jacs.7b12191}
{\href{https://arxiv.org/abs/https://doi.org/10.1021/jacs.7b12191}{{https://doi.org/10.1021/jacs.7b12191}}}.
\bcomment{PMID: 29323881}
\end{barticle}
\endbibitem

\bibitem[\protect\citeauthoryear{Lorenz}{1969}]{Lorenz1969}
\begin{barticle}
\bauthor{\bsnm{Lorenz}, \binits{E.N.}}:
\batitle{{Atmospheric Predictability as Revealed by Naturally Occurring
  Analogues}}.
\bjtitle{Journal of Atmospheric Sciences}
\bvolume{26}(\bissue{4}),
\bfpage{636}--\blpage{646}
(\byear{1969})
\doiurl{10.1175/1520-0469(1969)26<636:APARBN>2.0.CO;2}
\end{barticle}
\endbibitem

\bibitem[\protect\citeauthoryear{Faranda et~al.}{2022}]{Faranda2022}
\begin{barticle}
\bauthor{\bsnm{Faranda}, \binits{D.}et al.}:
\batitle{A climate-change attribution retrospective of some impactful weather
  extremes of 2021}.
\bjtitle{Weather and Climate Dynamics}
\bvolume{3}(\bissue{4}),
\bfpage{1311}--\blpage{1340}
(\byear{2022})
\doiurl{10.5194/wcd-3-1311-2022}
\end{barticle}
\endbibitem

\bibitem[\protect\citeauthoryear{Kalnay et~al.}{1996}]{NCEP}
\begin{barticle}
\bauthor{\bsnm{Kalnay}, \binits{E.} et al.}:
\batitle{{The NCEP/NCAR 40-Year Reanalysis Project}}.
\bjtitle{Bulletin of the American Meteorological Society}
\bvolume{77}(\bissue{3}),
\bfpage{437}--\blpage{472}
(\byear{1996})
\doiurl{10.1175/1520-0477(1996)077<0437:TNYRP>2.0.CO;2}
\end{barticle}
\endbibitem

\bibitem[\protect\citeauthoryear{Lloyd}{1982}]{Lloyd82}
\begin{barticle}
\bauthor{\bsnm{Lloyd}, \binits{S.}}:
\batitle{Least squares quantization in pcm}.
\bjtitle{IEEE Transactions on Information Theory}
\bvolume{28}(\bissue{2}),
\bfpage{129}--\blpage{137}
(\byear{1982})
\doiurl{10.1109/TIT.1982.1056489}
\end{barticle}
\endbibitem

\bibitem[\protect\citeauthoryear{Mauritsen et~al.}{2019}]{MPI-M}
\begin{barticle}
\bauthor{\bsnm{Mauritsen}, \binits{T.} et al.}:
\batitle{{Developments in the MPI-M Earth System Model version 1.2 (MPI-ESM1.2)
  and Its Response to Increasing CO2}}.
\bjtitle{Journal of Advances in Modeling Earth Systems}
\bvolume{11}(\bissue{4}),
\bfpage{998}--\blpage{1038}
(\byear{2019})
\doiurl{10.1029/2018MS001400}
{\href{https://arxiv.org/abs/https://agupubs.onlinelibrary.wiley.com/doi/pdf/10.1029/2018MS001400}{{https://agupubs.onlinelibrary.wiley.com/doi/pdf/10.1029/2018MS001400}}}
\end{barticle}
\endbibitem

\bibitem[\protect\citeauthoryear{Bock et~al.}{2020}]{Bock2020}
\begin{barticle}
\bauthor{\bsnm{Bock}, \binits{L.} et al.},
\batitle{{Quantifying Progress Across Different CMIP Phases With the
  ESMValTool}}.
\bjtitle{Journal of Geophysical Research: Atmospheres}
\bvolume{125}(\bissue{21}),
\bfpage{2019}--\blpage{032321}
(\byear{2020})
\doiurl{10.1029/2019JD032321}
{\href{https://arxiv.org/abs/https://agupubs.onlinelibrary.wiley.com/doi/pdf/10.1029/2019JD032321}{{https://agupubs.onlinelibrary.wiley.com/doi/pdf/10.1029/2019JD032321}}}.
\bcomment{e2019JD032321 2019JD032321}
\end{barticle}
\endbibitem

\bibitem[\protect\citeauthoryear{M\"uller et~al.}{2018}]{Muller0219}
\begin{barticle}
\bauthor{\bsnm{M\"uller}, \binits{W.A.} et al.}:
\batitle{{A Higher-resolution Version of the Max Planck Institute Earth System
  Model (MPI-ESM1.2-HR)}}.
\bjtitle{Journal of Advances in Modeling Earth Systems}
\bvolume{10}(\bissue{7}),
\bfpage{1383}--\blpage{1413}
(\byear{2018})
\doiurl{10.1029/2017MS001217}
{\href{https://arxiv.org/abs/https://agupubs.onlinelibrary.wiley.com/doi/pdf/10.1029/2017MS001217}{{https://agupubs.onlinelibrary.wiley.com/doi/pdf/10.1029/2017MS001217}}}
\end{barticle}
\endbibitem

\bibitem[\protect\citeauthoryear{Wieners et~al.}{2019}]{MPIdata}
\begin{botherref}
\oauthor{\bsnm{Wieners}, \binits{K.-H.} et al.}:
MPI-M MPI-ESM1.2-LR model output prepared for CMIP6 CMIP esm-piControl.
Earth System Grid Federation
(2019).
\doiurl{10.22033/ESGF/CMIP6.6553} .
\url{https://doi.org/10.22033/ESGF/CMIP6.6553}
\end{botherref}
\endbibitem

\bibitem[\protect\citeauthoryear{Galfi and Lucarini}{2021}]{Galfi2021}
\begin{barticle}
\bauthor{\bsnm{Galfi}, \binits{V.M.}},
\bauthor{\bsnm{Lucarini}, \binits{V.}}:
\batitle{{Fingerprinting Heatwaves and Cold Spells and Assessing Their Response
  to Climate Change Using Large Deviation Theory}}.
\bjtitle{Phys. Rev. Lett.}
\bvolume{127},
\bfpage{058701}
(\byear{2021})
\doiurl{10.1103/PhysRevLett.127.058701}
\end{barticle}
\endbibitem

\bibitem[\protect\citeauthoryear{Whittleston et~al.}{2018}]{Whittleston2018}
\begin{barticle}
\bauthor{\bsnm{Whittleston}, \binits{D.}},
\bauthor{\bsnm{McColl}, \binits{K.A.}},
\bauthor{\bsnm{Entekhabi}, \binits{D.}}:
\batitle{{Multimodel Future Projections of Wintertime North Atlantic and North
  Pacific Tropospheric Jets: A Bayesian Analysis}}.
\bjtitle{Journal of Climate}
\bvolume{31}(\bissue{6}),
\bfpage{2533}--\blpage{2545}
(\byear{2018})
\doiurl{10.1175/JCLI-D-17-0316.1}
\end{barticle}
\endbibitem

\bibitem[\protect\citeauthoryear{Andrilli and Hecker}{2016}]{ANDRILLI2016513}
\begin{bchapter}
\bauthor{\bsnm{Andrilli}, \binits{S.}},
\bauthor{\bsnm{Hecker}, \binits{D.}}:
\bctitle{Chapter 8 - additional applications}.
In: \beditor{\bsnm{Andrilli}, \binits{S.}},
\beditor{\bsnm{Hecker}, \binits{D.}} (eds.)
\bbtitle{Elementary Linear Algebra (Fifth Edition)},
\bedition{Fifth edition} edn.,
pp. \bfpage{513}--\blpage{605}.
\bpublisher{Academic Press},
\blocation{Boston}
(\byear{2016}).
\doiurl{10.1016/B978-0-12-800853-9.00008-6} .
\burl{https://www.sciencedirect.com/science/article/pii/B9780128008539000086}
\end{bchapter}
\endbibitem

\bibitem[\protect\citeauthoryear{Pande et~al.}{2010}]{Pande2010}
\begin{barticle}
\bauthor{\bsnm{Pande}, \binits{V.S.}},
\bauthor{\bsnm{Beauchamp}, \binits{K.}},
\bauthor{\bsnm{Bowman}, \binits{G.R.}}:
\batitle{{Everything you wanted to know about Markov State Models but were
  afraid to ask}}.
\bjtitle{Methods}
\bvolume{52}(\bissue{1}),
\bfpage{99}--\blpage{105}
(\byear{2010})
\doiurl{10.1016/j.ymeth.2010.06.002} .
\bcomment{Protein Folding}
\end{barticle}
\endbibitem

\bibitem[\protect\citeauthoryear{Glielmo et~al.}{2021}]{Glielmo2021}
\begin{barticle}
\bauthor{\bsnm{Glielmo}, \binits{A.} et al.}:
\batitle{{Unsupervised Learning Methods for Molecular Simulation Data}}.
\bjtitle{Chemical Reviews}
\bvolume{121}(\bissue{16}),
\bfpage{9722}--\blpage{9758}
(\byear{2021})
\doiurl{10.1021/acs.chemrev.0c01195}
\end{barticle}
\endbibitem

\bibitem[\protect\citeauthoryear{Hoskins and Hodges}{2019}]{Hoskins2019}
\begin{barticle}
\bauthor{\bsnm{Hoskins}, \binits{B.J.}},
\bauthor{\bsnm{Hodges}, \binits{K.I.}}:
\batitle{{The Annual Cycle of Northern Hemisphere Storm Tracks. Part I:
  Seasons}}.
\bjtitle{Journal of Climate}
\bvolume{32}(\bissue{6}),
\bfpage{1743}--\blpage{1760}
(\byear{2019})
\doiurl{10.1175/JCLI-D-17-0870.1}
\end{barticle}
\endbibitem

\bibitem[\protect\citeauthoryear{Gao et~al.}{2022}]{Gao2020a}
\begin{barticle}
\bauthor{\bsnm{Gao}, \binits{M.}},
\bauthor{\bsnm{Yang}, \binits{S.}},
\bauthor{\bsnm{Li}, \binits{T.}}:
\batitle{Assessments on simulation of pacific blocking frequency during boreal
  winter in cmip6 climate models}.
\bjtitle{Dynamics of Atmospheres and Oceans}
\bvolume{100},
\bfpage{101333}
(\byear{2022})
\doiurl{10.1016/j.dynatmoce.2022.101333}
\end{barticle}
\endbibitem

\bibitem[\protect\citeauthoryear{Gao et~al.}{2020}]{Gao2020b}
\begin{botherref}
\oauthor{\bsnm{Gao}, \binits{M.}},
\oauthor{\bsnm{Yang}, \binits{S.}},
\oauthor{\bsnm{Li}, \binits{T.}}:
{The Spatio-Temporal Variation of Pacific Blocking Frequency within Winter
  Months and Its Relationship with Surface Air Temperature}.
Atmosphere
\textbf{11}(9)
(2020)
\doiurl{10.3390/atmos11090960}
\end{botherref}
\endbibitem

\bibitem[\protect\citeauthoryear{Woollings and Blackburn}{2012}]{Woolings2011}
\begin{barticle}
\bauthor{\bsnm{Woollings}, \binits{T.}},
\bauthor{\bsnm{Blackburn}, \binits{M.}}:
\batitle{{The North Atlantic Jet Stream under Climate Change and Its Relation
  to the NAO and EA Patterns}}.
\bjtitle{Journal of Climate}
\bvolume{25}(\bissue{3}),
\bfpage{886}--\blpage{902}
(\byear{2012})
\doiurl{10.1175/JCLI-D-11-00087.1}
\end{barticle}
\endbibitem

\bibitem[\protect\citeauthoryear{Rex}{1951}]{Rex1951}
\begin{barticle}
\bauthor{\bsnm{Rex}, \binits{D.F.}}:
\batitle{{The Effect of Atlantic Blocking Action upon European Climate}}.
\bjtitle{Tellus}
\bvolume{3}(\bissue{2}),
\bfpage{100}--\blpage{112}
(\byear{1951})
\doiurl{10.3402/tellusa.v3i2.8617}
{\href{https://arxiv.org/abs/https://doi.org/10.3402/tellusa.v3i2.8617}{{https://doi.org/10.3402/tellusa.v3i2.8617}}}
\end{barticle}
\endbibitem

\bibitem[\protect\citeauthoryear{Breeden et~al.}{2020}]{Breeden2020}
\begin{barticle}
\bauthor{\bsnm{Breeden}, \binits{M.L.}},
\bauthor{\bsnm{Hoover}, \binits{B.T.}},
\bauthor{\bsnm{Newman}, \binits{M.}},
\bauthor{\bsnm{Vimont}, \binits{D.J.}}:
\batitle{{Optimal North Pacific Blocking Precursors and Their Deterministic
  Subseasonal Evolution during Boreal Winter}}.
\bjtitle{Monthly Weather Review}
\bvolume{148}(\bissue{2}),
\bfpage{739}--\blpage{761}
(\byear{2020})
\doiurl{10.1175/MWR-D-19-0273.1}
\end{barticle}
\endbibitem

\bibitem[\protect\citeauthoryear{Nabizadeh et~al.}{2021}]{Nabizadeh2021}
\begin{barticle}
\bauthor{\bsnm{Nabizadeh}, \binits{E.}},
\bauthor{\bsnm{Lubis}, \binits{S.W.}},
\bauthor{\bsnm{Hassanzadeh}, \binits{P.}}:
\batitle{{The 3D Structure of Northern Hemisphere Blocking Events: Climatology,
  Role of Moisture, and Response to Climate Change}}.
\bjtitle{Journal of Climate}
\bvolume{34}(\bissue{24}),
\bfpage{9837}--\blpage{9860}
(\byear{2021})
\doiurl{10.1175/JCLI-D-21-0141.1}
\end{barticle}
\endbibitem

\bibitem[\protect\citeauthoryear{Carrera et~al.}{2004}]{Carrera2004}
\begin{barticle}
\bauthor{\bsnm{Carrera}, \binits{M.L.}},
\bauthor{\bsnm{Higgins}, \binits{R.W.}},
\bauthor{\bsnm{Kousky}, \binits{V.E.}}:
\batitle{{Downstream Weather Impacts Associated with Atmospheric Blocking over
  the Northeast Pacific}}.
\bjtitle{Journal of Climate}
\bvolume{17}(\bissue{24}),
\bfpage{4823}--\blpage{4839}
(\byear{2004})
\doiurl{10.1175/JCLI-3237.1}
\end{barticle}
\endbibitem

\bibitem[\protect\citeauthoryear{Woollings and Hoskins}{2008}]{Woolings2008}
\begin{barticle}
\bauthor{\bsnm{Woollings}, \binits{T.}},
\bauthor{\bsnm{Hoskins}, \binits{B.}}:
\batitle{{Simultaneous Atlantic–Pacific blocking and the Northern Annular
  Mode}}.
\bjtitle{Quarterly Journal of the Royal Meteorological Society}
\bvolume{134}(\bissue{636}),
\bfpage{1635}--\blpage{1646}
(\byear{2008})
\doiurl{10.1002/qj.310}
{\href{https://arxiv.org/abs/https://rmets.onlinelibrary.wiley.com/doi/pdf/10.1002/qj.310}{{https://rmets.onlinelibrary.wiley.com/doi/pdf/10.1002/qj.310}}}
\end{barticle}
\endbibitem

\bibitem[\protect\citeauthoryear{Thompson and Wallace}{2001}]{Thompson2001}
\begin{barticle}
\bauthor{\bsnm{Thompson}, \binits{D.W.J.}},
\bauthor{\bsnm{Wallace}, \binits{J.M.}}:
\batitle{{Regional Climate Impacts of the Northern Hemisphere Annular Mode}}.
\bjtitle{Science}
\bvolume{293}(\bissue{5527}),
\bfpage{85}--\blpage{89}
(\byear{2001})
\doiurl{10.1126/science.1058958}
{\href{https://arxiv.org/abs/https://www.science.org/doi/pdf/10.1126/science.1058958}{{https://www.science.org/doi/pdf/10.1126/science.1058958}}}
\end{barticle}
\endbibitem

\bibitem[\protect\citeauthoryear{Messori and
  Dorrington}{2023}]{messoridorrington2023}
\begin{barticle}
\bauthor{\bsnm{Messori}, \binits{G.}},
\bauthor{\bsnm{Dorrington}, \binits{J.}}:
\batitle{{A Joint Perspective on North American and Euro-Atlantic Weather
  Regimes}}.
\bjtitle{Geophysical Research Letters}
\bvolume{50}(\bissue{21}),
\bfpage{2023}--\blpage{104696}
(\byear{2023})
\doiurl{10.1029/2023GL104696}
{\href{https://arxiv.org/abs/https://agupubs.onlinelibrary.wiley.com/doi/pdf/10.1029/2023GL104696}{{https://agupubs.onlinelibrary.wiley.com/doi/pdf/10.1029/2023GL104696}}}.
\bcomment{e2023GL104696 2023GL104696}
\end{barticle}
\endbibitem

\bibitem[\protect\citeauthoryear{{\"O}nskog et~al.}{2018}]{Onskog2018}
\begin{barticle}
\bauthor{\bsnm{{\"O}nskog}, \binits{T.}},
\bauthor{\bsnm{Franzke}, \binits{C.L.E.}},
\bauthor{\bsnm{Hannachi}, \binits{A.}}:
\batitle{{Predictability and Non-Gaussian Characteristics of the North Atlantic
  Oscillation}}.
\bjtitle{Journal of Climate}
\bvolume{31}(\bissue{2}),
\bfpage{537}--\blpage{554}
(\byear{2018})
\doiurl{10.1175/JCLI-D-17-0101.1}
\end{barticle}
\endbibitem

\bibitem[\protect\citeauthoryear{Woollings et~al.}{2008}]{Woollings2008b}
\begin{barticle}
\bauthor{\bsnm{Woollings}, \binits{T.J.}},
\bauthor{\bsnm{Hoskins}, \binits{B.}},
\bauthor{\bsnm{Blackburn}, \binits{M.}},
\bauthor{\bsnm{Berrisford}, \binits{P.}}:
\batitle{{A new Rossby wave-breaking interpretation of the North Atlantic
  Oscillation}}.
\bjtitle{Journal of the Atmospheric Sciences}
\bvolume{65},
\bfpage{609}--\blpage{626}
(\byear{2008})
\doiurl{10.1175/2007JAS2347.1}
\end{barticle}
\endbibitem

\bibitem[\protect\citeauthoryear{Li and Zhang}{2015}]{Li2015}
\begin{barticle}
\bauthor{\bsnm{Li}, \binits{C.}},
\bauthor{\bsnm{Zhang}, \binits{Q.}}:
\batitle{{An observed connection between wintertime temperature anomalies over
  Northwest China and weather regime transitions in North Atlantic}}.
\bjtitle{J. Meteorol. Res.}
\bvolume{29},
\bfpage{201}--\blpage{203}
(\byear{2015})
\end{barticle}
\endbibitem

\bibitem[\protect\citeauthoryear{Vihma et~al.}{2020}]{Vihma2019}
\begin{barticle}
\bauthor{\bsnm{Vihma}, \binits{T.} et al.}:
\batitle{{Effects of the tropospheric large-scale circulation on European
  winter temperatures during the period of amplified Arctic warming}}.
\bjtitle{International Journal of Climatology}
\bvolume{40}(\bissue{1}),
\bfpage{509}--\blpage{529}
(\byear{2020})
\doiurl{10.1002/joc.6225}
{\href{https://arxiv.org/abs/https://rmets.onlinelibrary.wiley.com/doi/pdf/10.1002/joc.6225}{{https://rmets.onlinelibrary.wiley.com/doi/pdf/10.1002/joc.6225}}}
\end{barticle}
\endbibitem

\bibitem[\protect\citeauthoryear{Linkin and Nigam}{2008}]{Linkin2008}
\begin{barticle}
\bauthor{\bsnm{Linkin}, \binits{M.E.}},
\bauthor{\bsnm{Nigam}, \binits{S.}}:
\batitle{{The North Pacific Oscillation–West Pacific Teleconnection Pattern:
  Mature-Phase Structure and Winter Impacts}}.
\bjtitle{Journal of Climate}
\bvolume{21}(\bissue{9}),
\bfpage{1979}--\blpage{1997}
(\byear{2008})
\doiurl{10.1175/2007JCLI2048.1}
\end{barticle}
\endbibitem

\bibitem[\protect\citeauthoryear{Aru et~al.}{2022}]{Aru2022}
\begin{barticle}
\bauthor{\bsnm{Aru}, \binits{H.}},
\bauthor{\bsnm{Chen}, \binits{S.}},
\bauthor{\bsnm{Chen}, \binits{W.}}:
\batitle{{Change in the variability in the Western Pacific pattern during
  boreal winter: roles of tropical Pacific sea surface temperature anomalies
  and North Pacific storm track activity}}.
\bjtitle{Climate Dynamics}
\bvolume{58},
\bfpage{2451}--\blpage{2468}
(\byear{2022})
\doiurl{10.1007/s00382-021-06014-1}
\end{barticle}
\endbibitem

\bibitem[\protect\citeauthoryear{Hochman et~al.}{2021}]{Hochman2021}
\begin{botherref}
\oauthor{\bsnm{Hochman}, \binits{A.}},
\oauthor{\bsnm{Messori}, \binits{G.}},
\oauthor{\bsnm{Quinting}, \binits{J.F.}},
\oauthor{\bsnm{Pinto}, \binits{J.G.}},
\oauthor{\bsnm{Grams}, \binits{C.M.}}:
{Do Atlantic-European Weather Regimes Physically Exist?}
Geophysical Research Letters
\textbf{48}
(2021)
\doiurl{10.1029/2021GL095574}
\end{botherref}
\endbibitem

\bibitem[\protect\citeauthoryear{Messori and Dorrington}{2023}]{Messori2023}
\begin{barticle}
\bauthor{\bsnm{Messori}, \binits{G.}},
\bauthor{\bsnm{Dorrington}, \binits{J.}}:
\batitle{{A Joint Perspective on North American and Euro-Atlantic Weather
  Regimes}}.
\bjtitle{Geophysical Research Letters}
\bvolume{50}(\bissue{21}),
\bfpage{2023}--\blpage{104696}
(\byear{2023})
\doiurl{10.1029/2023GL104696}
{\href{https://arxiv.org/abs/https://agupubs.onlinelibrary.wiley.com/doi/pdf/10.1029/2023GL104696}{{https://agupubs.onlinelibrary.wiley.com/doi/pdf/10.1029/2023GL104696}}}.
\bcomment{e2023GL104696 2023GL104696}
\end{barticle}
\endbibitem

\bibitem[\protect\citeauthoryear{Neal et~al.}{2016}]{Neal2016}
\begin{barticle}
\bauthor{\bsnm{Neal}, \binits{R.}},
\bauthor{\bsnm{Fereday}, \binits{D.}},
\bauthor{\bsnm{Crocker}, \binits{R.}},
\bauthor{\bsnm{Comer}, \binits{R.E.}}:
\batitle{{A flexible approach to defining weather patterns and their
  application in weather forecasting over Europe}}.
\bjtitle{Meteorological Applications}
\bvolume{23}(\bissue{3}),
\bfpage{389}--\blpage{400}
(\byear{2016})
\doiurl{10.1002/met.1563}
{\href{https://arxiv.org/abs/https://rmets.onlinelibrary.wiley.com/doi/pdf/10.1002/met.1563}{{https://rmets.onlinelibrary.wiley.com/doi/pdf/10.1002/met.1563}}}
\end{barticle}
\endbibitem

\bibitem[\protect\citeauthoryear{Feldstein}{2000}]{Feldstein2000}
\begin{barticle}
\bauthor{\bsnm{Feldstein}, \binits{S.B.}}:
\batitle{{The Timescale, Power Spectra, and Climate Noise Properties of
  Teleconnection Patterns}}.
\bjtitle{Journal of Climate}
\bvolume{13}(\bissue{24}),
\bfpage{4430}--\blpage{4440}
(\byear{2000})
\doiurl{10.1175/1520-0442(2000)013<4430:TTPSAC>2.0.CO;2}
\end{barticle}
\endbibitem

\bibitem[\protect\citeauthoryear{Drouard and Woollings}{2018}]{Drouard2018}
\begin{barticle}
\bauthor{\bsnm{Drouard}, \binits{M.}},
\bauthor{\bsnm{Woollings}, \binits{T.}}:
\batitle{{Contrasting Mechanisms of Summer Blocking Over Western Eurasia}}.
\bjtitle{Geophysical Research Letters}
\bvolume{45}(\bissue{21}),
\bfpage{12040}--\blpage{12048}
(\byear{2018})
\doiurl{10.1029/2018GL079894}
{\href{https://arxiv.org/abs/https://agupubs.onlinelibrary.wiley.com/doi/pdf/10.1029/2018GL079894}{{https://agupubs.onlinelibrary.wiley.com/doi/pdf/10.1029/2018GL079894}}}
\end{barticle}
\endbibitem

\bibitem[\protect\citeauthoryear{Yang and Wang}{2022}]{Yang2022}
\begin{botherref}
\oauthor{\bsnm{Yang}, \binits{D.}},
\oauthor{\bsnm{Wang}, \binits{L.}}:
{The Summertime Circulation Types over Eurasia and Their Connections with the
  North Atlantic Oscillation Modulated by North Atlantic SST}.
Atmosphere
\textbf{13}(12)
(2022)
\doiurl{10.3390/atmos13122093}
\end{botherref}
\endbibitem

\bibitem[\protect\citeauthoryear{Pope et~al.}{2009}]{Pope2009}
\begin{barticle}
\bauthor{\bsnm{Pope}, \binits{M.}},
\bauthor{\bsnm{Jakob}, \binits{C.}},
\bauthor{\bsnm{Reeder}, \binits{M.J.}}:
\batitle{{Regimes of the North Australian Wet Season}}.
\bjtitle{Journal of Climate}
\bvolume{22}(\bissue{24}),
\bfpage{6699}--\blpage{6715}
(\byear{2009})
\doiurl{10.1175/2009JCLI3057.1}
\end{barticle}
\endbibitem

\bibitem[\protect\citeauthoryear{Hassim and Timbal}{2019}]{Hassim2019}
\begin{barticle}
\bauthor{\bsnm{Hassim}, \binits{M.E.E.}},
\bauthor{\bsnm{Timbal}, \binits{B.}}:
\batitle{{Observed Rainfall Trends over Singapore and the Maritime Continent
  from the Perspective of Regional-Scale Weather Regimes}}.
\bjtitle{Journal of Applied Meteorology and Climatology}
\bvolume{58}(\bissue{2}),
\bfpage{365}--\blpage{384}
(\byear{2019})
\doiurl{10.1175/JAMC-D-18-0136.1}
\end{barticle}
\endbibitem

\bibitem[\protect\citeauthoryear{Solman and Men{\'e}ndez}{2003}]{Solman2003}
\begin{barticle}
\bauthor{\bsnm{Solman}, \binits{S.A.}},
\bauthor{\bsnm{Men{\'e}ndez}, \binits{C.G.}}:
\batitle{{Weather regimes in the South American sector and neighbouring oceans
  during winter}}.
\bjtitle{Climate Dynamics}
\bvolume{21}(\bissue{1}),
\bfpage{91}--\blpage{104}
(\byear{2003})
\doiurl{10.1007/s00382-003-0320-x}
\end{barticle}
\endbibitem

\bibitem[\protect\citeauthoryear{Wilson et~al.}{2013}]{Wilson2013}
\begin{barticle}
\bauthor{\bsnm{Wilson}, \binits{L.}},
\bauthor{\bsnm{Manton}, \binits{M.J.}},
\bauthor{\bsnm{Siems}, \binits{S.T.}}:
\batitle{{Relationship between rainfall and weather regimes in south-eastern
  Queensland, Australia}}.
\bjtitle{International Journal of Climatology}
\bvolume{33}(\bissue{4}),
\bfpage{979}--\blpage{991}
(\byear{2013})
\doiurl{10.1002/joc.3484}
{\href{https://arxiv.org/abs/https://rmets.onlinelibrary.wiley.com/doi/pdf/10.1002/joc.3484}{{https://rmets.onlinelibrary.wiley.com/doi/pdf/10.1002/joc.3484}}}
\end{barticle}
\endbibitem

\bibitem[\protect\citeauthoryear{Arizmendi et~al.}{2022}]{Arizmendi2022}
\begin{barticle}
\bauthor{\bsnm{Arizmendi}, \binits{F.}},
\bauthor{\bsnm{Trinchin}, \binits{R.}},
\bauthor{\bsnm{Barreiro}, \binits{M.}}:
\batitle{{Weather regimes in subtropical South America and their impacts over
  Uruguay}}.
\bjtitle{International Journal of Climatology}
\bvolume{42}(\bissue{16}),
\bfpage{9253}--\blpage{9270}
(\byear{2022})
\doiurl{10.1002/joc.7816}
{\href{https://arxiv.org/abs/https://rmets.onlinelibrary.wiley.com/doi/pdf/10.1002/joc.7816}{{https://rmets.onlinelibrary.wiley.com/doi/pdf/10.1002/joc.7816}}}
\end{barticle}
\endbibitem

\bibitem[\protect\citeauthoryear{Loikith et~al.}{2019}]{Loikith2019}
\begin{barticle}
\bauthor{\bsnm{Loikith}, \binits{P.C.} et al.}:
\batitle{{A climatology of daily synoptic circulation patterns and associated
  surface meteorology over southern South America}}.
\bjtitle{Climate Dynamics}
\bvolume{53}(\bissue{7}),
\bfpage{4019}--\blpage{4035}
(\byear{2019})
\doiurl{10.1007/s00382-019-04768-3}
\end{barticle}
\endbibitem

\bibitem[\protect\citeauthoryear{Pohl et~al.}{2021}]{Pohl2019}
\begin{barticle}
\bauthor{\bsnm{Pohl}, \binits{B.} et al.}:
\batitle{{Relationship Between Weather Regimes and Atmospheric Rivers in East
  Antarctica}}.
\bjtitle{Journal of Geophysical Research: Atmospheres}
\bvolume{126}(\bissue{24}),
\bfpage{2021}--\blpage{035294}
(\byear{2021})
\doiurl{10.1029/2021JD035294}
{\href{https://arxiv.org/abs/https://agupubs.onlinelibrary.wiley.com/doi/pdf/10.1029/2021JD035294}{{https://agupubs.onlinelibrary.wiley.com/doi/pdf/10.1029/2021JD035294}}}.
\bcomment{e2021JD035294 2021JD035294}
\end{barticle}
\endbibitem

\bibitem[\protect\citeauthoryear{Eyring et~al.}{2016}]{Eyring2016}
\begin{barticle}
\bauthor{\bsnm{Eyring}, \binits{V.} et al.}:
\batitle{{Overview of the Coupled Model Intercomparison Project Phase 6 (CMIP6)
  experimental design and organization}}.
\bjtitle{Geoscientific Model Development}
\bvolume{9},
\bfpage{10539}--\blpage{10583}
(\byear{2016})
\end{barticle}
\endbibitem

\bibitem[\protect\citeauthoryear{Maher et~al.}{2021}]{Maher2021}
\begin{barticle}
\bauthor{\bsnm{Maher}, \binits{N.}},
\bauthor{\bsnm{Milinski}, \binits{S.}},
\bauthor{\bsnm{Ludwig}, \binits{R.}}:
\batitle{{Large ensemble climate model simulations: introduction, overview, and
  future prospects for utilising multiple types of large ensemble}}.
\bjtitle{Earth System Dynamics}
\bvolume{12}(\bissue{2}),
\bfpage{401}--\blpage{418}
(\byear{2021})
\doiurl{10.5194/esd-12-401-2021}
\end{barticle}
\endbibitem

\bibitem[\protect\citeauthoryear{Kornhuber et~al.}{2019}]{Kornhuber2019}
\begin{barticle}
\bauthor{\bsnm{Kornhuber}, \binits{K.} et al.}:
\batitle{{Extreme weather events in early summer 2018 connected by a recurrent
  hemispheric wave-7 pattern}}.
\bjtitle{Environmental Research Letters}
\bvolume{14}(\bissue{5}),
\bfpage{054002}
(\byear{2019})
\doiurl{10.1088/1748-9326/ab13bf}
\end{barticle}
\endbibitem

\bibitem[\protect\citeauthoryear{Kornhuber and
  Tamarin-Brodsky}{2021}]{Kornhuber2021}
\begin{barticle}
\bauthor{\bsnm{Kornhuber}, \binits{K.}},
\bauthor{\bsnm{Tamarin-Brodsky}, \binits{T.}}:
\batitle{{Future changes in Northern Hemisphere summer weather persistence
  linked to projected Arctic warming.}}
\bjtitle{Geophysical Research Letters}
\bvolume{48}(\bissue{4}),
\bfpage{2020}--\blpage{091603}
(\byear{2021})
\doiurl{10.1029/2020GL091603}
{\href{https://arxiv.org/abs/https://agupubs.onlinelibrary.wiley.com/doi/pdf/10.1029/2020GL091603}{{https://agupubs.onlinelibrary.wiley.com/doi/pdf/10.1029/2020GL091603}}}
\end{barticle}
\endbibitem

\bibitem[\protect\citeauthoryear{Ragone and Bouchet}{2021}]{RagoneBouchet2021}
\begin{barticle}
\bauthor{\bsnm{Ragone}, \binits{F.}},
\bauthor{\bsnm{Bouchet}, \binits{F.}}:
\batitle{{Rare Event Algorithm Study of Extreme Warm Summers and Heatwaves Over
  Europe}}.
\bjtitle{Geophysical Research Letters}
\bvolume{48}(\bissue{12}),
\bfpage{2020}--\blpage{091197}
(\byear{2021})
\doiurl{10.1029/2020GL091197}
{\href{https://arxiv.org/abs/https://agupubs.onlinelibrary.wiley.com/doi/pdf/10.1029/2020GL091197}{{https://agupubs.onlinelibrary.wiley.com/doi/pdf/10.1029/2020GL091197}}}.
\bcomment{e2020GL091197 2020GL091197}
\end{barticle}
\endbibitem

\end{thebibliography}
\end{document}



\newcommand{\DONE}[1]{\textcolor{black}{#1}}
\newcommand{\FIXME}[1]{\textcolor{black}{#1}}

\newcommand{\SebaAle}[1]{\textcolor{black}{#1}}
\newcommand{\MeliVale}[1]{\textcolor{black}{#1}}

\renewcommand{\thefigure}{S\arabic{figure}}

\setcounter{figure}{0}

\title{Supplementary Information - 
\MeliVale{Unsupervised detection of Large-scale Weather Patterns in the Northern Hemisphere via Markov State Modelling: from Blockings to Teleconnections}}


\author[1]{\fnm{Sebastian} \sur{Springer}}\email{sebastian.springer@sissa.it}
\equalcont{These authors contributed equally to this work.}
\author*[1,5]{\fnm{Alessandro} \sur{Laio}}\email{laio@sissa.it}
\equalcont{These authors contributed equally to this work.}

\author[2]{\fnm{Vera } \sur{Melinda Galfi}}\email{v.m.galfi@vu.nl}
\equalcont{These authors contributed equally to this work.}

\author*[3,4]{\fnm{Valerio } \sur{Lucarini}}\email{v.lucarini@reading.ac.uk}
\equalcont{These authors contributed equally to this work.}

\affil*[1]{\orgdiv{Physics Department}, \orgname{International School for Advanced Studies (SISSA)}, \orgaddress{\street{Via Bonomea 265}, \city{Trieste}, \postcode{34136}, \country{Italy}}}

\affil[2]{\orgdiv{ Institute for Environmental Studies}, \orgname{Vrije Universiteit Amsterdam}, \orgaddress{\street{De Boelelaan 1105}, \city{Amsterdam}, \postcode{081 HV}, \country{Holland}}}

\affil[3]{\orgdiv{School of Computing and Mathematical Sciences}, \orgname{University of Leicester}, \orgaddress{\street{University Road}, \city{Leicester}, \postcode{LE17RH}, \country{UK}}}

\affil[4]{\orgdiv{Department of Mathematics and Statistics \& Centre for the Mathematics of Planet Earth}, \orgname{University of Reading}, \orgaddress{\street{Whiteknights Campus}, \city{Reading}, \postcode{RG66AX}, \country{UK}}}

\affil[5]{\orgname{International Center of Theoretical Physics (ICTP)}, \orgaddress{\street{Viale Miramare 11}, \city{Trieste}, \postcode{34151}, \country{Italy}}}

\maketitle

\newpage
\begin{figure*}[ht!]
\includegraphics[width=.8\textwidth]{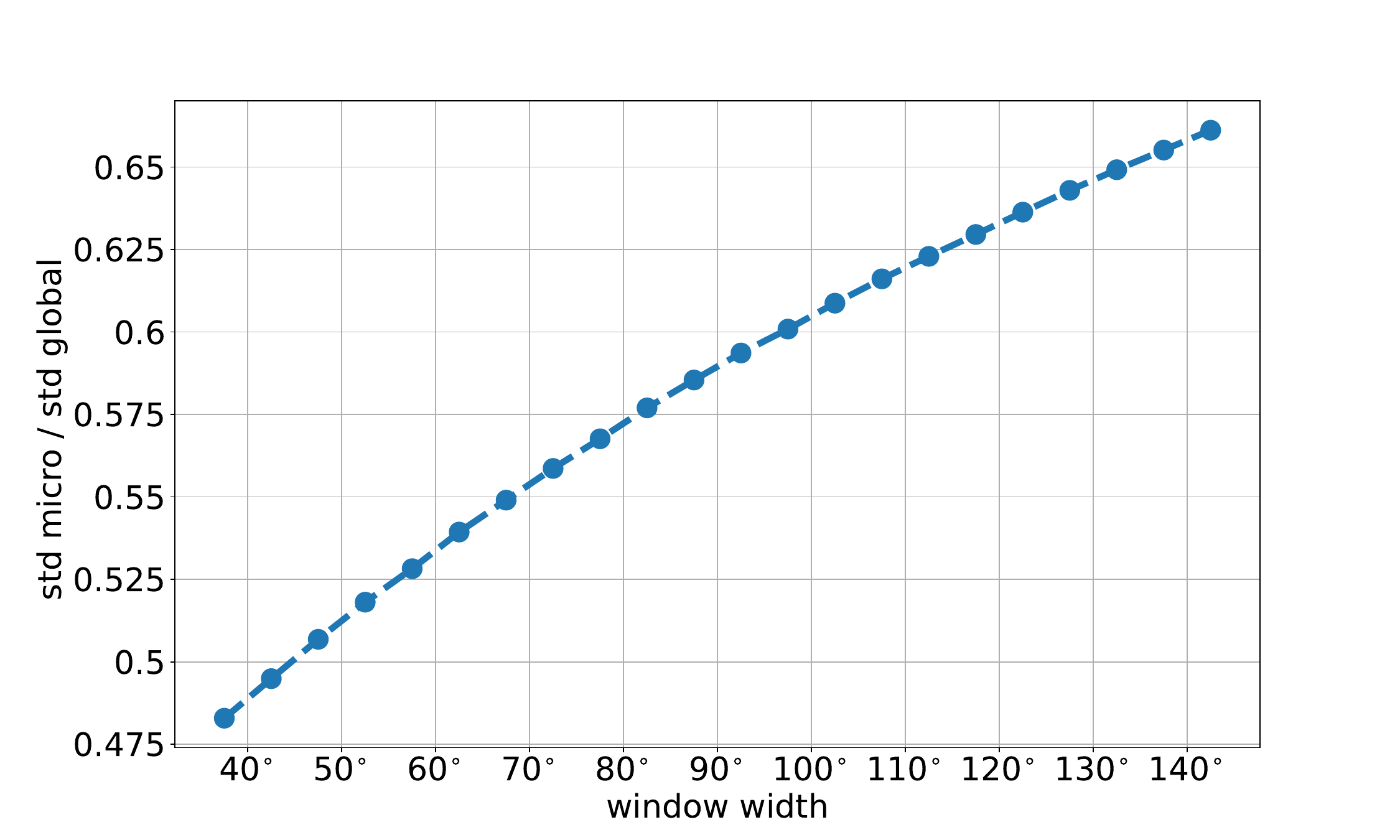}

\textbf{Supplementary Figure 1.} {Impact of increasing the width of the longitudinal window (x-axis) centered at the longitude $\phi_0=0^{\circ}$ on the homogeneity of the microstates. The number of  microstates is $K=180$. We plot (y-axis) the  ratio between the average  (computed over all the microstates) of standard deviation within each microstate and the standard deviation over all the data of the sample.}
\end{figure*}

\begin{figure*}[ht!]
\includegraphics[width=.8\textwidth]{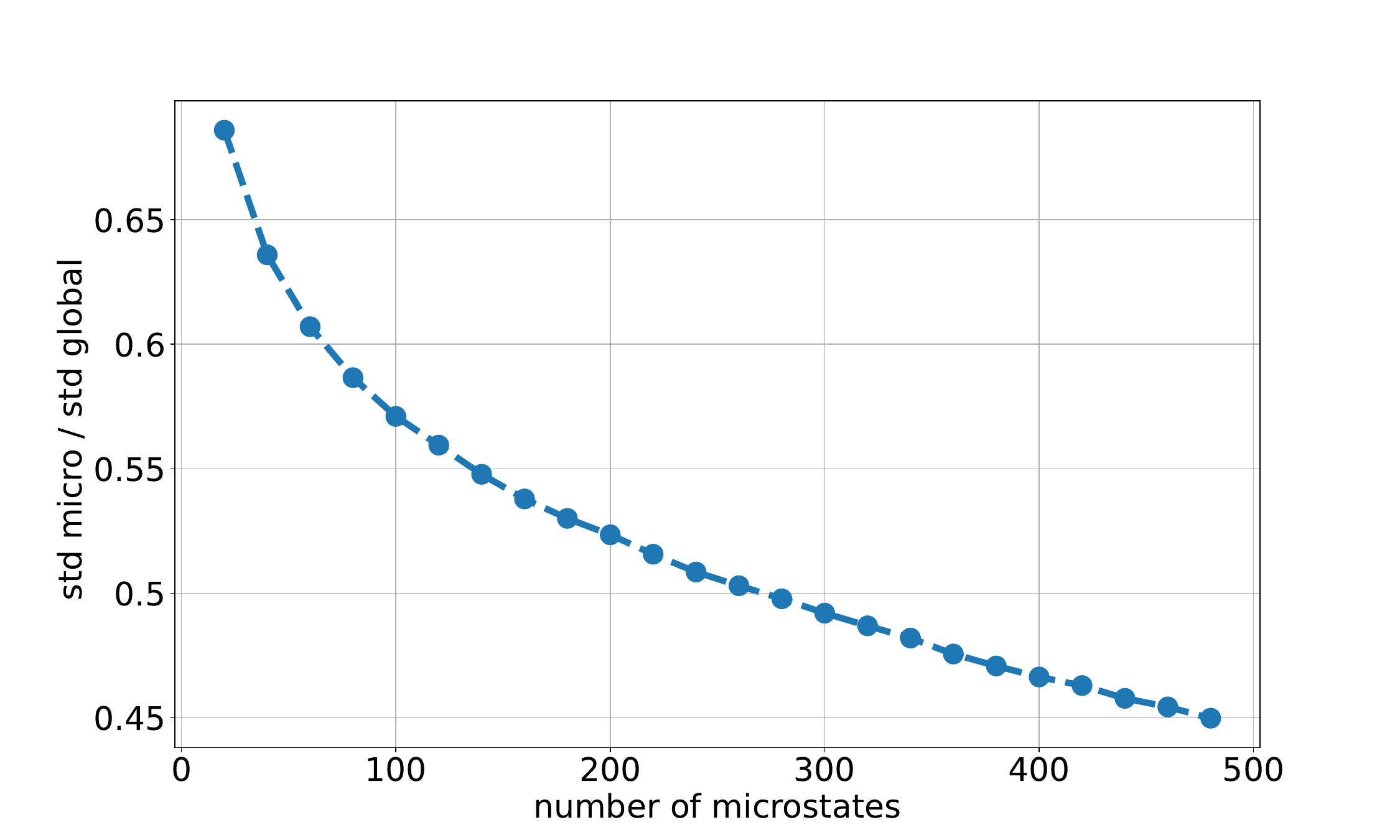}

\textbf{Supplementary Figure 2}. \MeliVale{Same as Supplementary Fig. 1, but here the width of the longitudinal window is fixed ($\Phi=30^{\circ}$)  and the number of microstates varies (x-axis).}
\end{figure*}

\begin{figure}[ht!]
\includegraphics[width=\textwidth]{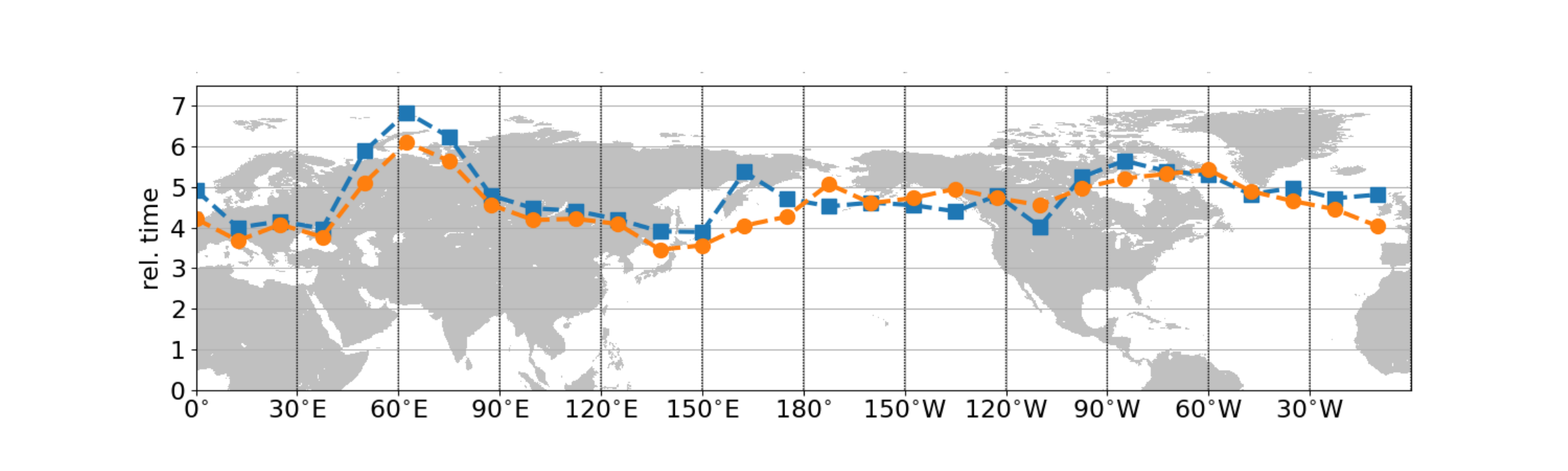}

\textbf{Supplementary Figure 3.} Kinetic analysis using a moving longitudinal window with kernel width $\Phi=30^{\circ}$. Each dot refers to the center of the window. First relaxation times when $\tau=1 \ day$ (blue line); $\tau=2 \ days$ (orange line) are plotted.
\end{figure}

\begin{figure}[ht!]
\includegraphics[width=\textwidth]{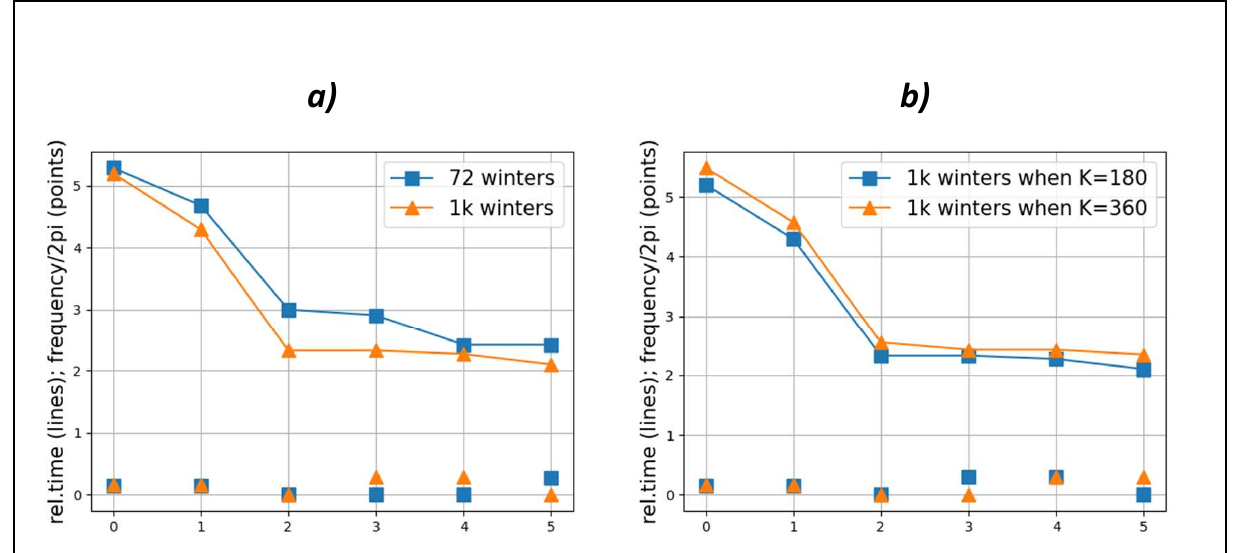}

\textbf{Supplementary Figure 4.} Robustness of the Markov state modelling pipeline used in this work.  Panel a): results are compared in the case where data from 72 winters (blue line) and 1000 winters (orange line) are used and the number of microstates is $K=180$. Panel b): results are compared in the case where  1000 winters are considered and  two choice - $K=180$ (blue line) and $K=360$ (orange line) - are adopted for the the number of microstates.  Analysis performed by considering the Z500 data from the Atlantic sector ($[30^{\circ}W, 30^{\circ}E ]\times[32^\circ N,72^\circ N]$) produced by the MPI-ESM-LR version 1.2 \cite{MPI-M} model, same data used as in \cite{Galfi2021}.
\end{figure}

\begin{figure*}[ht!]
\includegraphics[width=\textwidth]{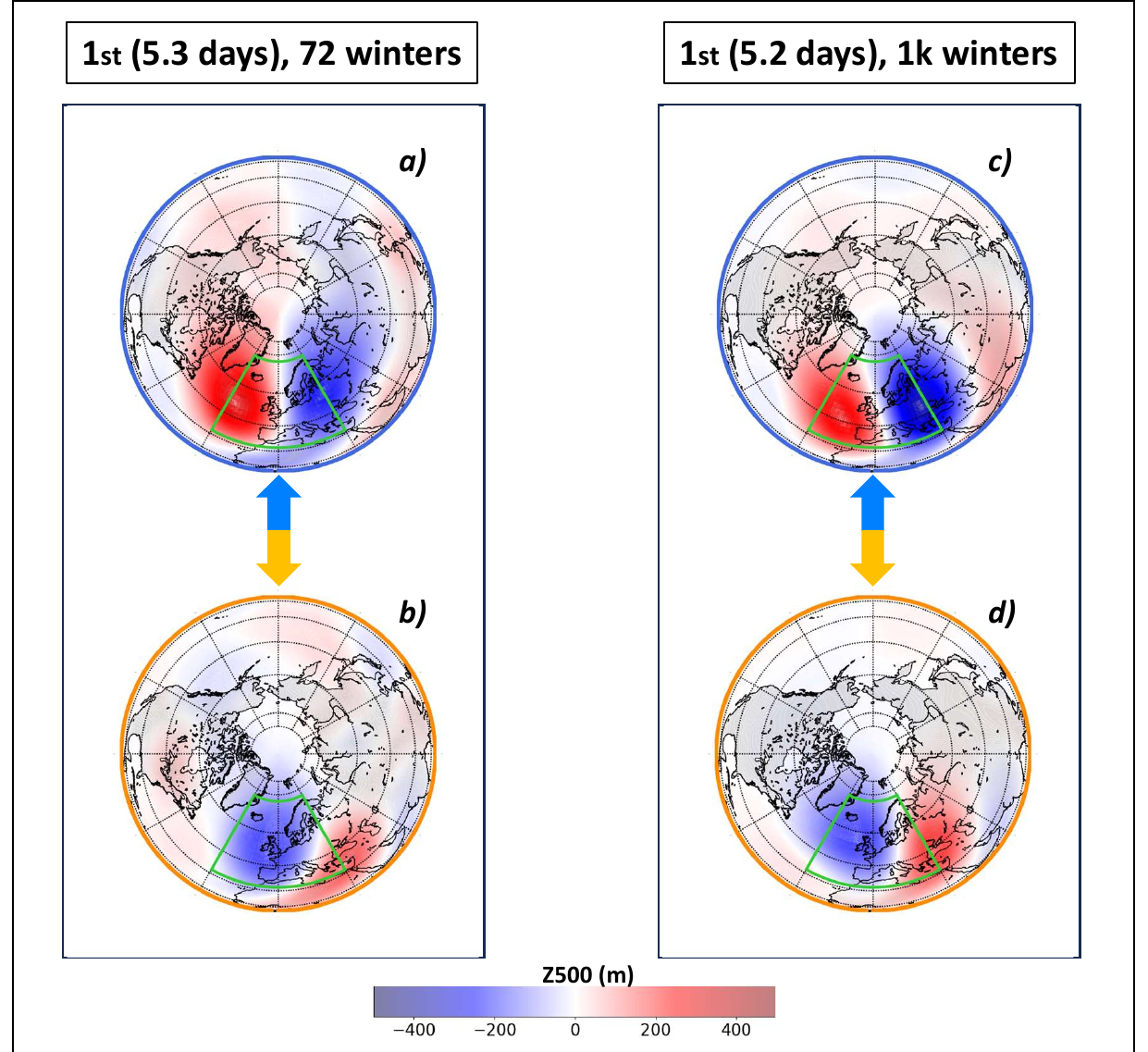}
\textbf{Supplementary Figure 5.} \MeliVale{Same data as in Supplementary Fig. 4a. 
Counterposed microstates mean field anomalies for the slowest relaxation mode of the system estimated using DJF data from 72 years (panels a and b) and 1000 years (panels c and d).}  
The green window indicates the area considered for clustering the data. 

\label{fig:Atl60_model}
\end{figure*}

\newpage

\begin{figure*}[ht!]
\includegraphics[width=\textwidth]{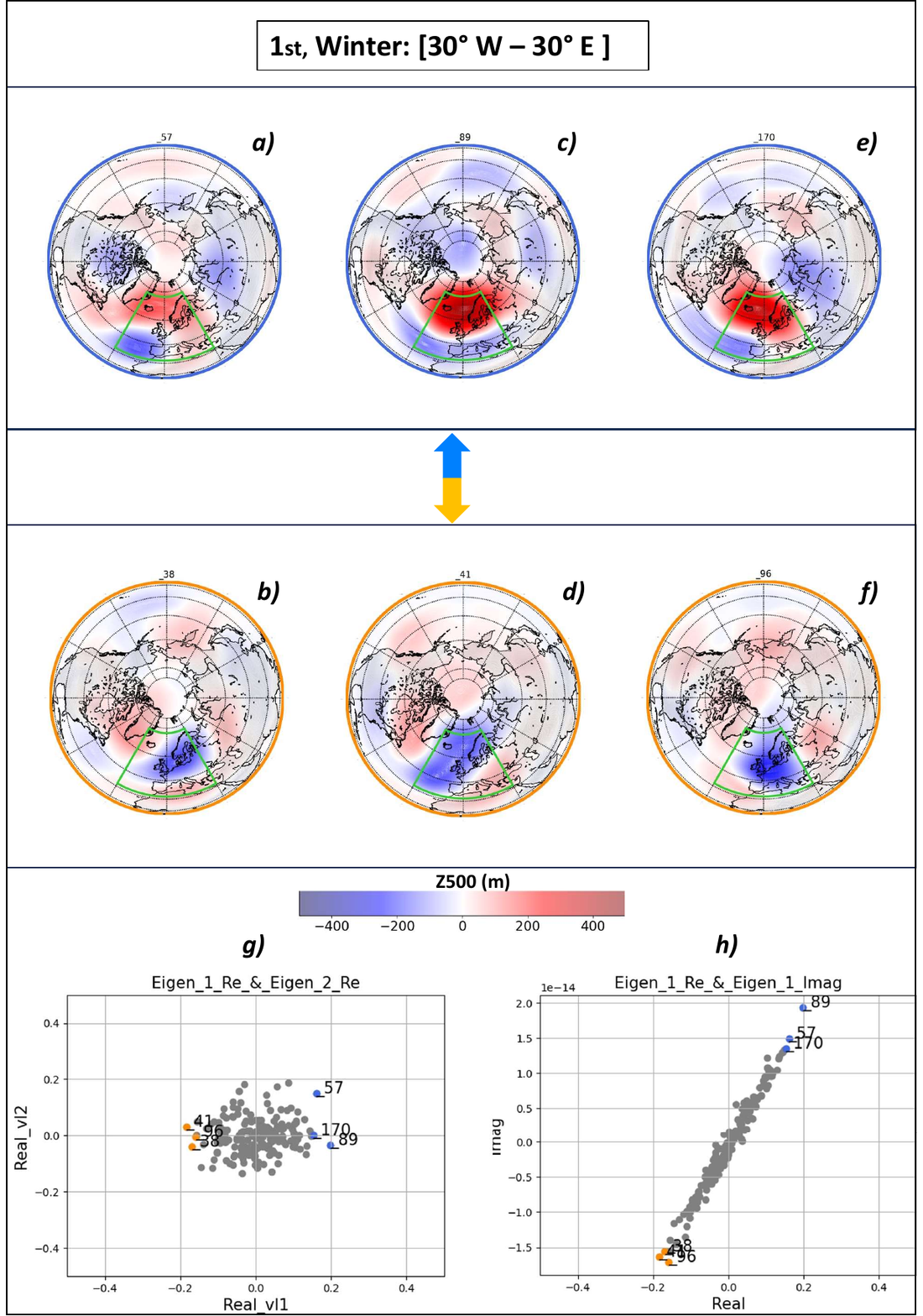}

\textbf{Supplementary Figure 6.} \MeliVale{Upper panel: Counterposed microstates mean field anomalies for the slowest relaxation mode of the system. The three pairs of most extreme states are portrayed (index values are included). Lower panel: Projections of the six chosen microstates on first and second leading eigenvectors. Analysis performed by considering NCEP/NCAR renalysis 1950-2022 DJF Z500 data from the sector ($[30^{\circ}W,30^{\circ}E ]\times[32^\circ N,72^\circ N]$).  The green window indicates the area considered for clustering the data.}
\label{fig:3Closest_b}
\end{figure*}
\newpage

\begin{figure*}[ht!]
\includegraphics[width=\textwidth]{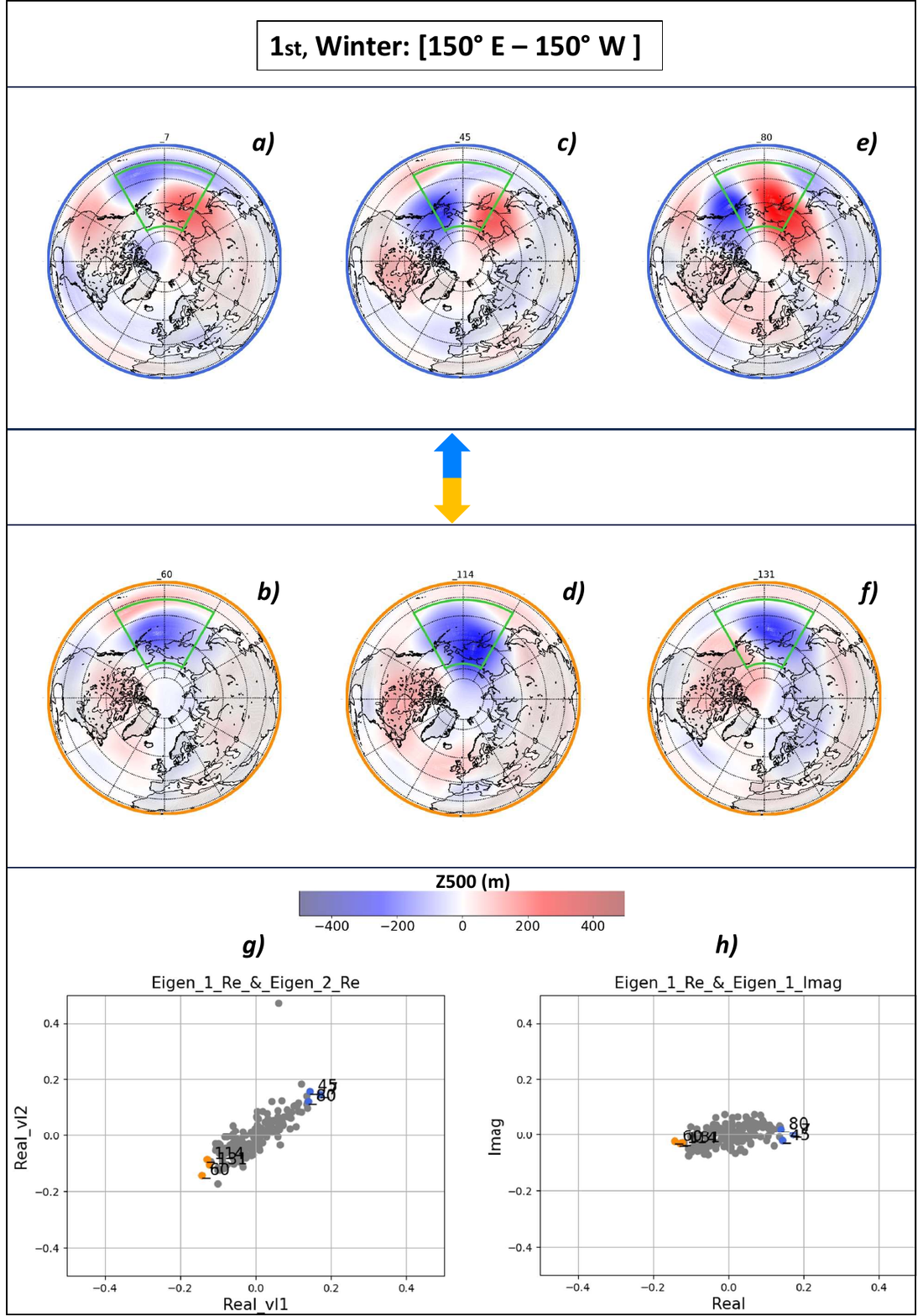}

\textbf{Supplementary Figure 7.} \MeliVale{Same as Supplementary Fig. 6, but for the  sector ($[150^{\circ}E,150^{\circ}W ]\times[32^\circ N,72^\circ N]$).}

\label{fig:3Closest_d}
\end{figure*}

\begin{figure*}[ht!]
\includegraphics[width=\textwidth]{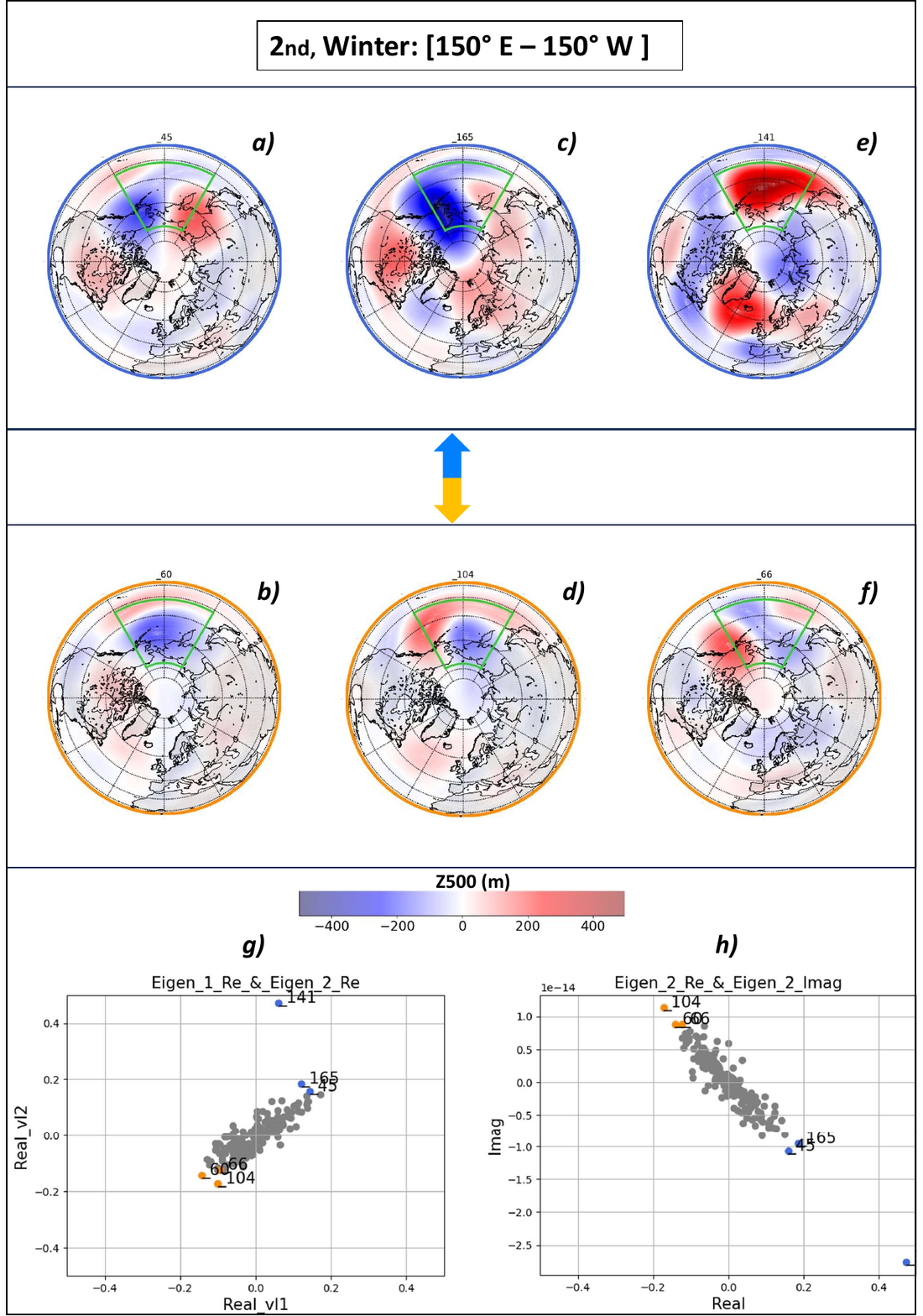}

\textbf{Supplementary Figure 8.} \MeliVale{Same as Supplementary Fig. 7, but for the second slowest relaxation mode of the system.} 

\label{fig:3Closest_c}
\end{figure*}

\begin{figure*}[ht!]
\includegraphics[width=\textwidth]{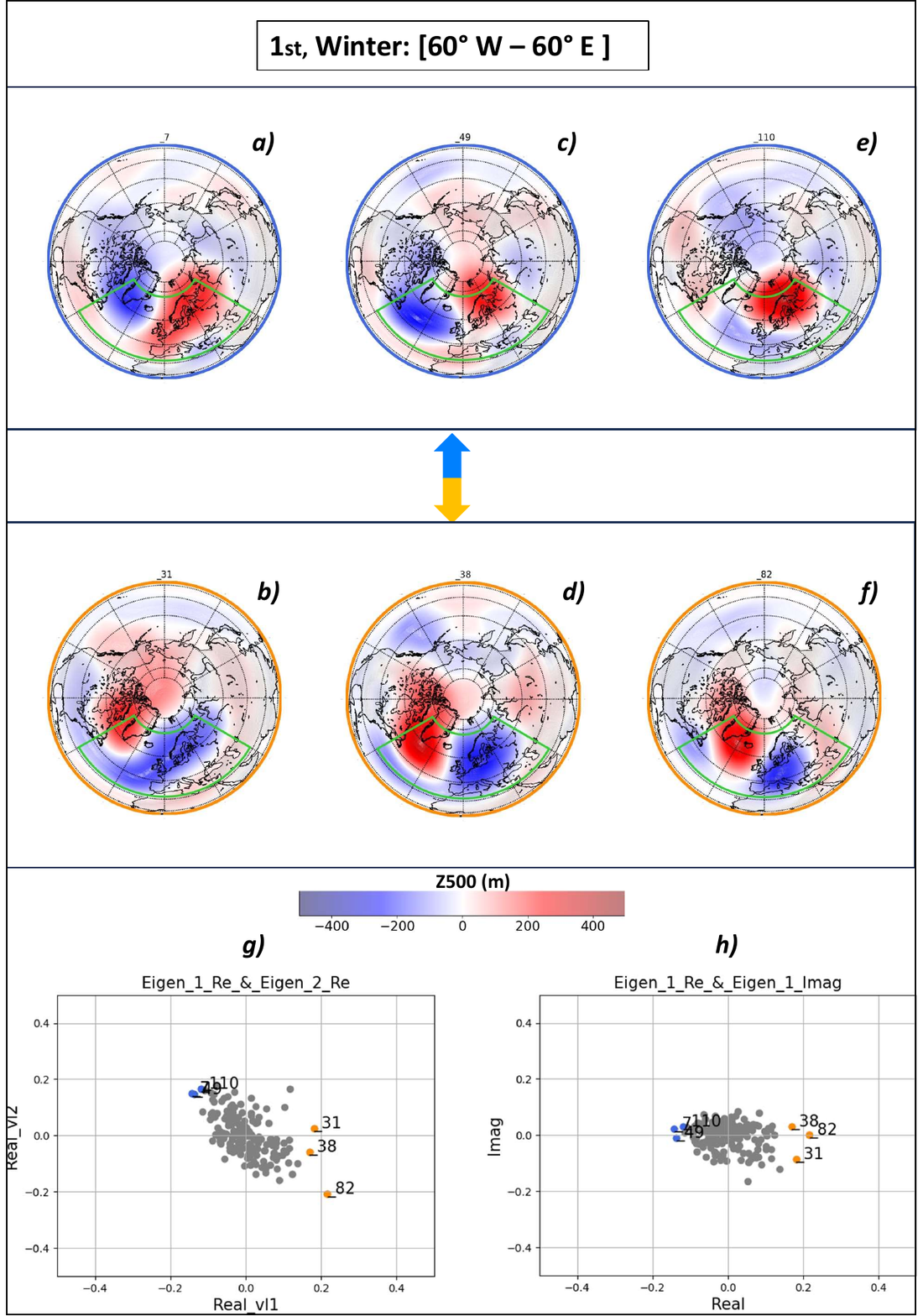}

\textbf{Supplementary Figure 9.}  \MeliVale{Same as Supplementary Fig. 6, but for the sector $[60^{\circ}W,60^{\circ}W ]\times[32^\circ N,72^\circ N]$.}
\label{fig:3Closest_e}
\end{figure*}

\begin{figure*}[ht!]
\includegraphics[width=\textwidth]{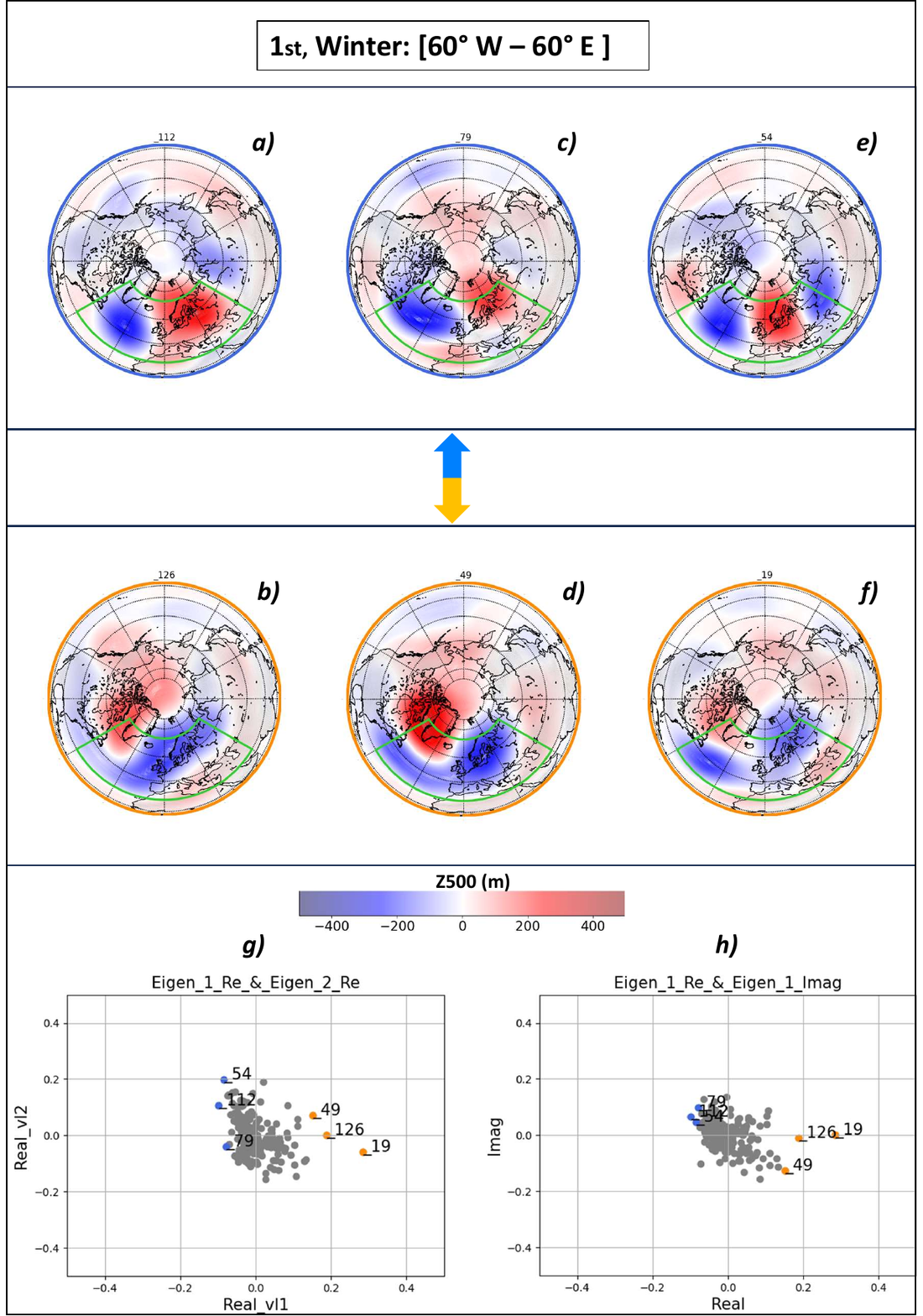}
\textbf{Supplementary Figure 10.} \MeliVale{Same as Supplementary Fig. 6, but for the sector $[60^{\circ}W,60^{\circ}W ]\times[30^\circ N,70^\circ N]$.}
\label{fig:3Closest_e}
\end{figure*}

\begin{figure*}[ht!]
\includegraphics[width=\textwidth]{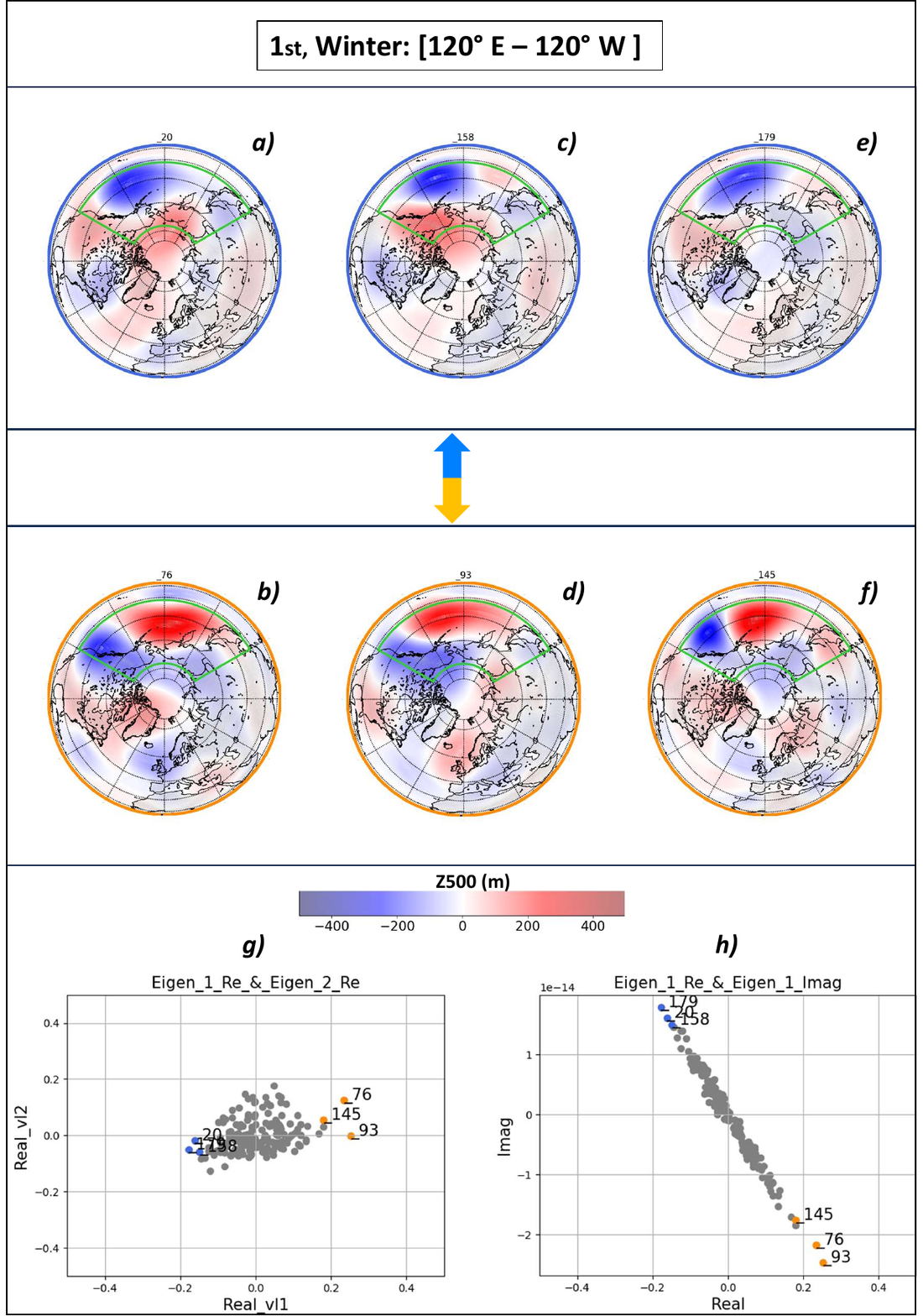}

\textbf{Supplementary Figure 11.} \MeliVale{Same as Supplementary Fig. 6, but for the  sector ($[120^{\circ}E,120^{\circ}W ]\times[32^\circ N,72^\circ N]$).}
 Counterposed microstates mean field anomalies for the slowest relaxation mode of the system when the dataset contains Winters [1950-2022]. The green window indicates the area considered for clustering the data.

\label{fig:3Closest_d}
\end{figure*}

\begin{figure*}[ht!]
\includegraphics[width=\textwidth]{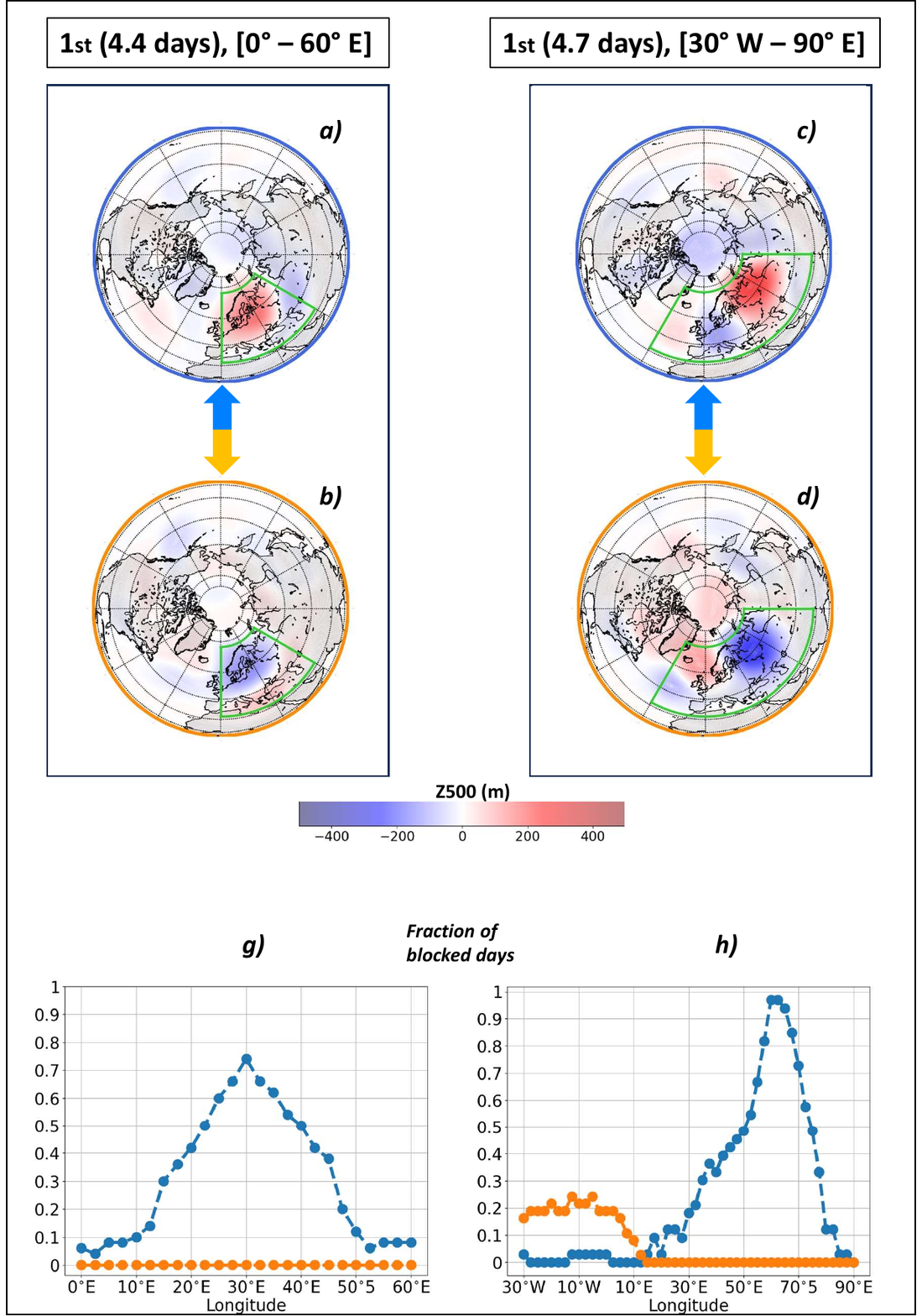}

\textbf{Supplementary Figure 12.} \MeliVale{Counterposed microstates mean field anomalies for the slowest relaxation mode of the system. Analysis performed on the NCEP/NCAR renalysis 1950-2022 JJA Z500 data from the sector $[0^{\circ},60^{\circ}E ]\times[32^\circ N,72^\circ N]$ (left) and from the sector $[30^{\circ}W,90^{\circ}E ]\times[32^\circ N,72^\circ N]$ (right).}  The green window indicates the area considered for clustering the data. 
\label{fig:StPet}
\end{figure*}

\newpage

\begin{figure*}[ht!]
\includegraphics[width=\textwidth]{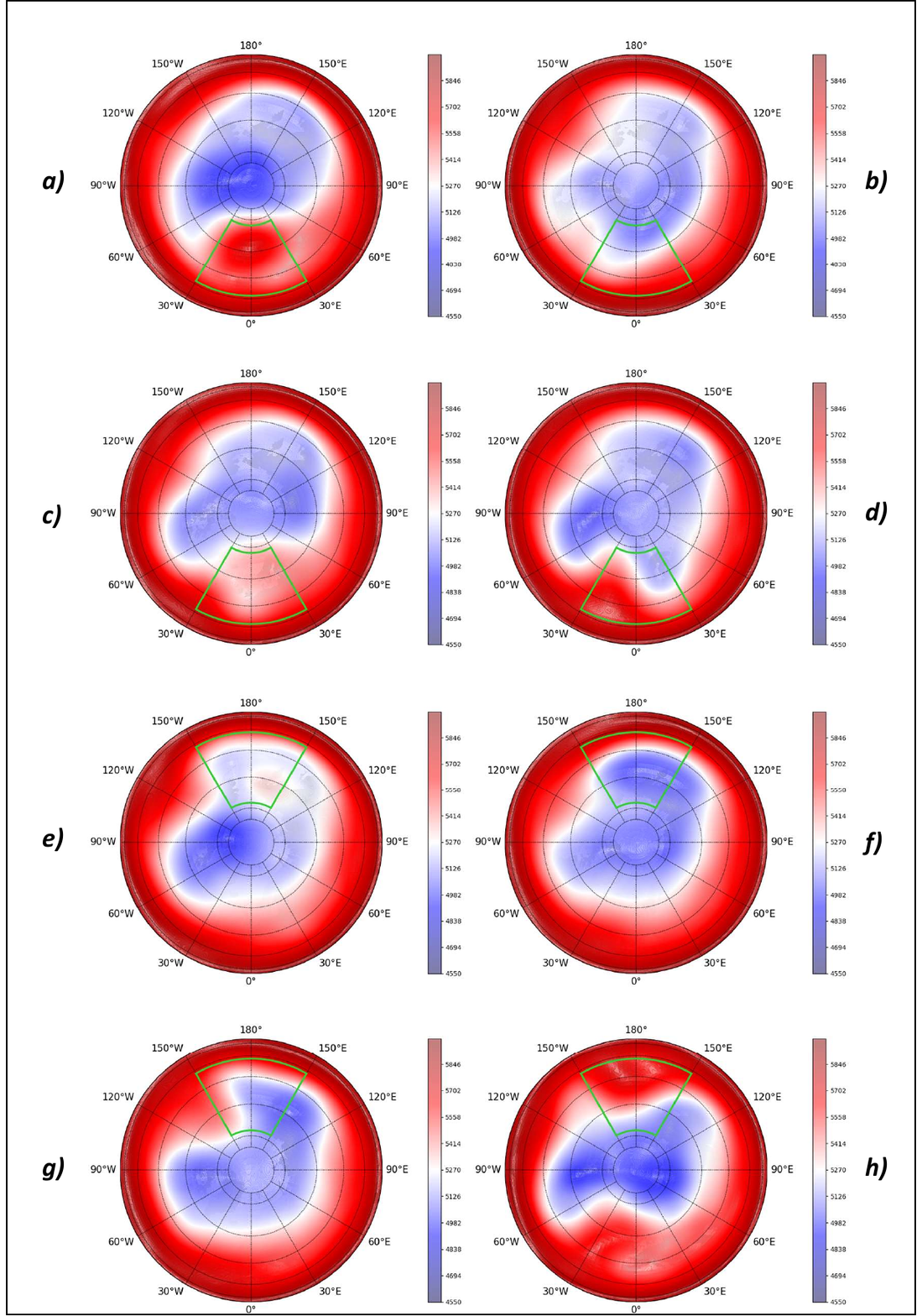}

\textbf{Supplementary Figure 13.} Z500 counterposed microstates mean fields for the slowest relaxation mode of the system. The green window indicates the area considered for clustering the data. The anomalies coincide with the detrended data described in the Section Methods and used to cluster the data.
Window $[ 30^{\circ}W, 30^{\circ}E ]$ in green. 
Z500 mean fields. Panel a): first eigenvector extreme microstate 1 (having high $\%$ of Blocked days). Panel b) first eigenvector extreme microstate 2 (having low $\%$ of Blocked days). Panels c)-d): same as panels a)-b), but for the second eigenvector.
Window $[ 150^{\circ}E, 150^{\circ}W ]$ in green. Z500 mean fields. Panel e): first eigenvector extreme microstate 1 (having high $\%$ of Blocked days). Panel f):  first eigenvector extreme microstate 2 (having low $\%$ of Blocked days). Panels g)-h): same as panels e)-f), but for the second eigenvector.

\label{fig:GHP60_raw}
\end{figure*}

\begin{figure*}[ht!]
\includegraphics[width=\textwidth]{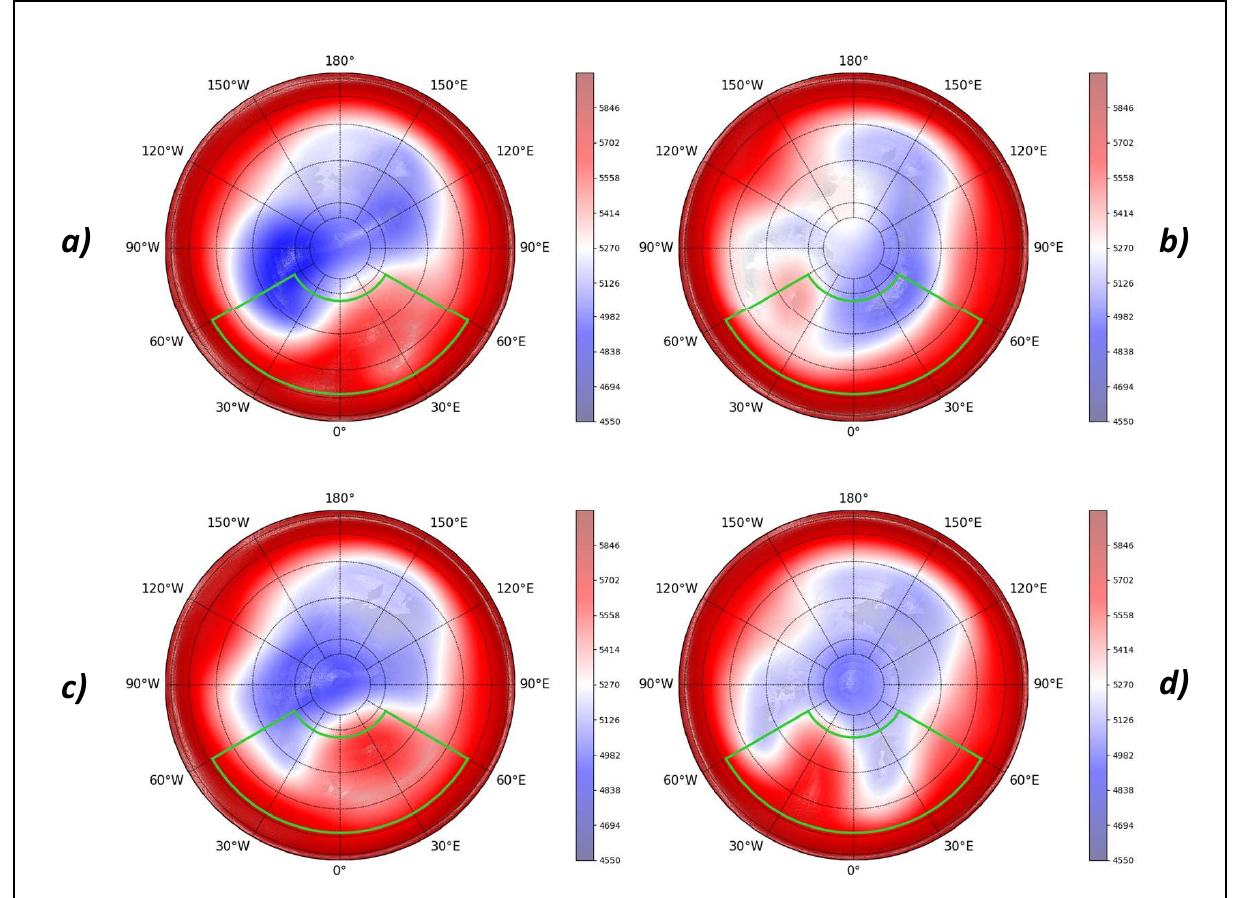}

\textbf{Supplementary Figure 14.} Counterposed microstates mean fields for the first two slowest relaxation mode of the system in the $[ 60^{\circ}W, 60^{\circ}E ]$ sector. The green window indicates the area considered for clustering the data. The Z500 anomalies coincide with the detrended data described in the Section Methods and Data and used to cluster the data.Window $[ 60^{\circ}W, 60^{\circ}E ]$ in green. Z500  mean fields. Panel a): first eigenvector extreme microstate 1 (having high $\%$ of Blocked days on the Eastern side of the window). Panel b): first eigenvector extreme microstate 2 (having high $\%$ of Blocked days on the Western side of the window).Panels c)-d): same as panels a)-b), but for the second eigenvector.

\label{fig:GPH120_Eigen1_raw}
\end{figure*}

   

\begin{figure*}[ht!]
\includegraphics[width=\textwidth]{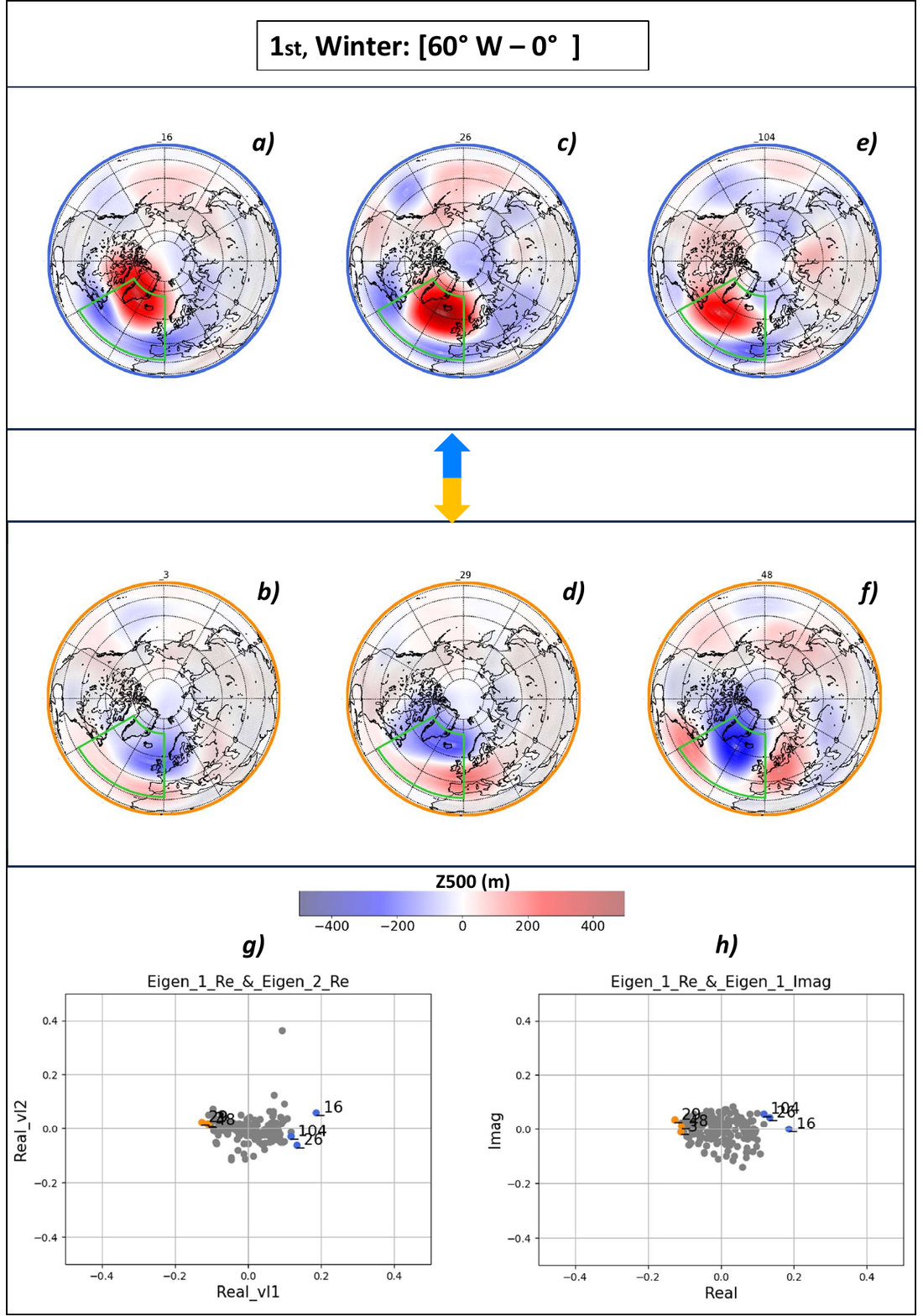}

\textbf{Supplementary Figure 15.} \MeliVale{Same as Supplementary Fig. 6, but for the  sector ($[60^{\circ}W,0^{\circ} ]\times[32^\circ N,72^\circ N]$).}

\label{fig:3Closest_c}
\end{figure*}

\begin{figure*}[ht!]
\includegraphics[width=\textwidth]{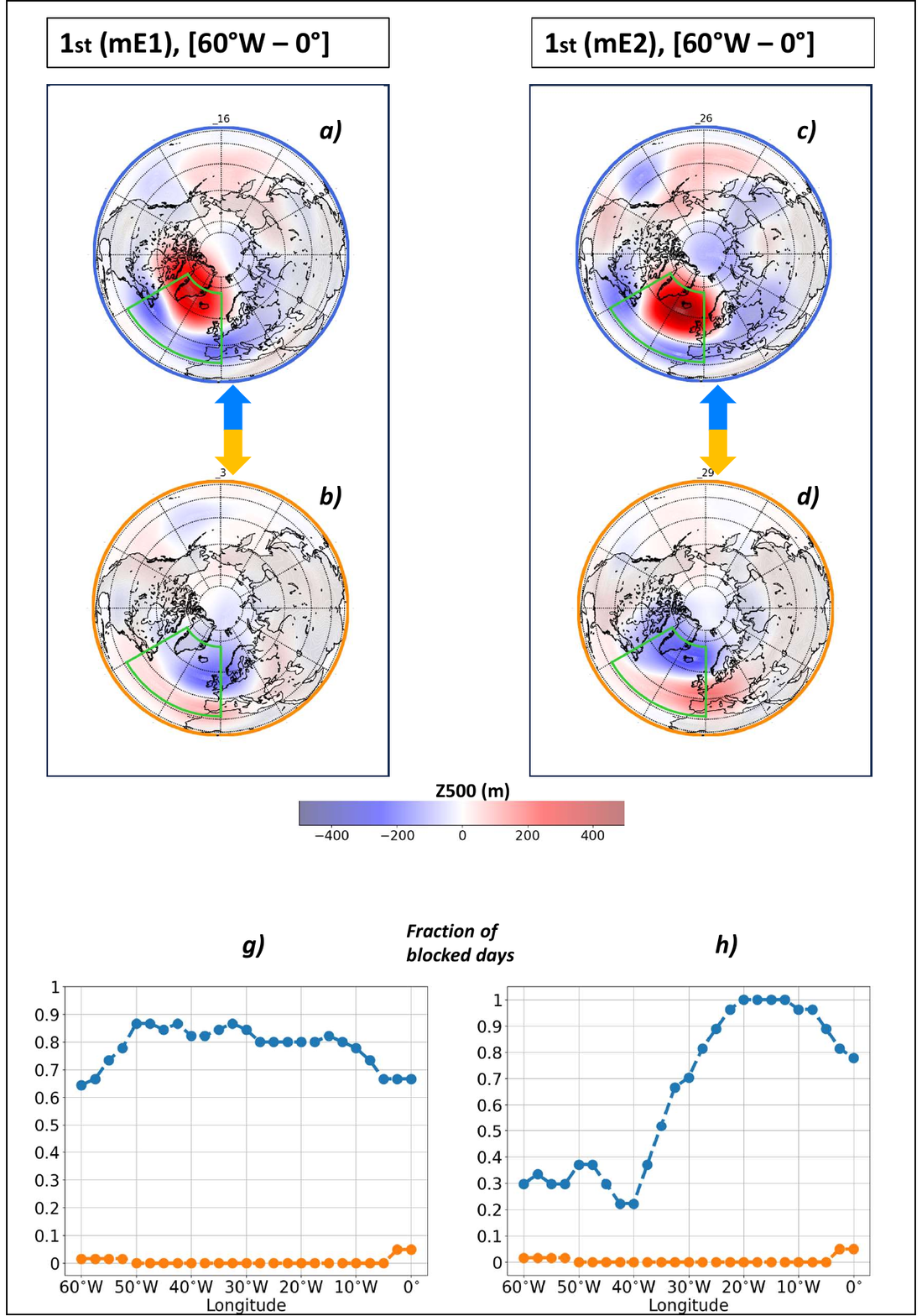}

\textbf{Supplementary Figure 16.} \MeliVale{Counterposed microstates mean field anomalies for the slowest relaxation mode of the system. Analysis performed on the NCEP/NCAR renalysis 1950-2022 JJA Z500 data from the sector $[60^{\circ}W,0^{\circ} ]\times[32^\circ N,72^\circ N]$. Panels a)-b): Most extreme microstates (mE1). Panels c)-d): second most extreme microstates (mE2).The green window indicates the area considered for clustering the data.  }

\label{fig:StPet}
\end{figure*}

\clearpage
\newpage

\providecommand{\noopsort}[1]{}\providecommand{\singleletter}[1]{#1}